%% file: main.tex
\pdfoutput=1
\PassOptionsToPackage{table}{xcolor}
\documentclass[logo]{sofa_lab_2026}

\graphicspath{{./}}

\usepackage[numbers,sort&compress]{natbib}
\usepackage{bm}
\usepackage{nicefrac}
\input{math_commands.tex}

\usepackage{float}
\usepackage{wrapfig}
\usepackage{subfig}
\usepackage{tikz}
\usepackage{fontawesome5}
\usetikzlibrary{arrows.meta,positioning,fit,backgrounds,calc,shapes.geometric,shadows,decorations.pathmorphing}
\usepackage{pgfplots}
\pgfplotsset{compat=1.18}

\usepackage{siunitx}
\usepackage{multirow}
\usepackage{adjustbox}
\usepackage{placeins}
\usepackage{arydshln}
\usepackage[normalem]{ulem}
\useunder{\uline}{\ul}{}
\usepackage{xspace}
\usepackage{xparse}
\usepackage{varwidth}
\usepackage[scaled=0.9]{helvet}   %
\usepackage{flafter}

\usepackage{listings}
\usepackage{algorithm}
\usepackage{algpseudocode}

\definecolor{ink}{HTML}{1A1D23}
\definecolor{ink2}{HTML}{5B616B}
\colorlet{clay}{BrandAccent}
\colorlet{claydark}{BrandPrimary}
\definecolor{sky}{HTML}{40627F}
\definecolor{sage}{HTML}{5E7D70}
\definecolor{heather}{HTML}{6C6A88}
\colorlet{kraft}{BrandAccent}
\definecolor{ivory}{HTML}{F3F5F7}
\definecolor{ivoryline}{HTML}{D4DAE0}
\definecolor{oursbg}{HTML}{DCE5EE}
\definecolor{famSolo}{HTML}{9AA3AB}\definecolor{famArgue}{HTML}{A79BB5}\definecolor{famFact}{HTML}{5E8574}
\definecolor{famSelf}{HTML}{8B3344}\definecolor{famNotes}{HTML}{6E8597}\definecolor{famDesign}{HTML}{B8955A}\definecolor{famTrain}{HTML}{8BAA98}

\newcommand{\famtag}[1]{\textcolor{fam#1}{\textsf{\textbf{\csname famname#1\endcsname}}}}
\DeclareRobustCommand{\mkFact}{\tikz[baseline=-0.6ex]\fill[famFact] (0,0) circle (2.1pt);}
\DeclareRobustCommand{\mkDesign}{\tikz[baseline=-0.6ex]\fill[famDesign] (-2pt,-2pt) rectangle (2pt,2pt);}
\DeclareRobustCommand{\mkNotes}{\tikz[baseline=-0.6ex]\fill[famNotes] (0,-2.6pt) -- (2.6pt,0) -- (0,2.6pt) -- (-2.6pt,0) -- cycle;}
\DeclareRobustCommand{\mkSelf}{\tikz[baseline=-0.6ex]\fill[famSelf] (-2.3pt,-1.8pt) -- (2.3pt,-1.8pt) -- (0,2.4pt) -- cycle;}
\colorlet{navy}{BrandLink}
\definecolor{pend}{HTML}{8A8F98}
\definecolor{chanprice}{HTML}{B5651D}
\colorlet{rule}{BrandRule}
\definecolor{backcolour}{rgb}{0.95,0.95,0.92}
\definecolor{codegreen}{rgb}{0,0.6,0}
\definecolor{codegray}{rgb}{0.5,0.5,0.5}
\definecolor{codepurple}{rgb}{0.58,0,0.82}
\definecolor{methodblue}{HTML}{3D5F7E}
\definecolor{takeawayblue}{HTML}{40627F}
\definecolor{mnDarkGray}{HTML}{4D4D4D}
\definecolor{tableInk}{HTML}{263B4A}
\definecolor{tableRule}{HTML}{74828B}
\definecolor{tableHeader}{HTML}{DCE7ED}
\definecolor{tableSubhead}{HTML}{EDF3F6}
\definecolor{tableGroup}{HTML}{F2F5F6}
\definecolor{tableBand}{HTML}{F7F9FA}
\definecolor{tableOurs}{HTML}{E2EDF3}
\definecolor{takeawayink}{HTML}{2F4456}
\definecolor{takeawayline}{HTML}{9CAEBB}
\definecolor{takeawayback}{HTML}{ECEFF2}
\definecolor{takeawaytitleback}{HTML}{DCE2E7}

\arrayrulecolor{rule}
\newcommand{\tsz}{\footnotesize}
\newcommand{\tgroup}[2]{\multicolumn{#1}{@{\hspace{2pt}}l}{\cellcolor{tableGroup}\textcolor{tableInk}{\textit{#2}}}}   %
\newcommand{\Lv}[1]{\ifnum#1=3\textbf{\textcolor{methodblue}{3}}\else\textcolor{ink2}{#1}\fi}   %
\newcommand{\viaacc}{\textcolor{famFact}{\textsf{\scriptsize acc.}}}\newcommand{\viacost}{\textcolor{chanprice}{\textsf{\scriptsize acc.\,+\,cost}}}\newcommand{\vianone}{\textcolor{pend}{\textsf{\scriptsize none}}}
\newcommand{\oursrow}{\rowcolor{oursbg}}
\newcommand{\refrow}{\rowcolor{ivory}}

\newcommand{\blank}{\textcolor{pend}{--}}

\definecolor{seblue}{HTML}{2F6DB5}
\newcommand{\se}[1]{{\scriptsize\textcolor{seblue}{\,$\pm$#1}}}   %

\newcommand{\cred}{r}
\newcommand{\ours}[1]{\textcolor{methodblue}{#1}}

\theoremstyle{plain}
\newtheorem{proposition}{Proposition}

\algrenewcommand\algorithmicrequire{\textbf{Input:}}
\algrenewcommand\algorithmicensure{\textbf{Output:}}
\algrenewcommand\algorithmiccomment[1]{\hfill\textcolor{gray}{$\triangleright$ \footnotesize #1}}

\algnewcommand{\LineComment}[1]{\State \textcolor{gray}{$\triangleright$ \footnotesize #1}}

\tcbset{
  takeawaybox/.style={
    width=\linewidth, enhanced, breakable,
    rounded corners, arc=2.8mm, outer arc=2.8mm,
    boxrule=0.7pt,
    top=5pt, bottom=5pt, left=7pt, right=7pt,
    before skip=7pt, after skip=7pt,
    colback=takeawayback, colframe=takeawayline, coltext=black,
    colbacktitle=takeawaytitleback, coltitle=black,
    fontupper=\itshape, fonttitle=\bfseries\itshape\small,
    titlerule=0.45pt, toptitle=4pt, bottomtitle=4pt,
    lefttitle=7pt, righttitle=7pt,
  }
}
\newtcolorbox{takeawaybox}[2][]{takeawaybox, title=#2, #1}
\definecolor{keyhead}{HTML}{505095}
\definecolor{keyback}{HTML}{F7F7FF}
\definecolor{keyframe}{HTML}{C2C2D6}
\newtcolorbox{keypoint}[1]{enhanced, before skip=14pt, after skip=9pt,
  colback=keyback, colframe=keyframe, boxrule=0.3pt, arc=2pt, outer arc=2pt,
  borderline west={4pt}{0pt}{keyhead},
  left=10pt, right=8pt, top=8pt, bottom=5pt,
  attach boxed title to top left={xshift=6pt, yshift=-5.5pt, yshifttext=-3pt},
  boxed title style={enhanced, colback=keyhead, colframe=keyhead, boxrule=0pt, arc=3pt, outer arc=3pt, boxsep=0pt,
    left=6pt, right=7pt, top=3.4pt, bottom=3.4pt},
  varwidth boxed title*=-14pt,
  title={\faLightbulb[regular]\enspace #1}, fonttitle=\bfseries\small, coltitle=white, fontupper=\normalsize}
\newtcolorbox{qualpanel}[3][]{enhanced, colback=#2!8!white, colframe=#2!75!white, boxrule=0.9pt, arc=4pt, outer arc=4pt,
  left=6pt, right=6pt, top=3.5pt, bottom=4pt, before skip=0pt, after skip=3pt, title={#3}, attach title to upper,
  after title={\par\smallskip}, coltitle=black, fonttitle=\sffamily\bfseries\footnotesize, fontupper=\sffamily\scriptsize, #1}
\newtcolorbox{qualrow}[2][0.27\linewidth]{enhanced, frame hidden, interior hidden, boxsep=0pt, left=0pt, right=0pt,
  top=0pt, bottom=0pt, before skip=0pt, after skip=0pt, sidebyside, sidebyside align=top seam, sidebyside gap=12pt,
  lefthand width=#1, fontupper=\sffamily\scriptsize, fontlower=\sffamily\scriptsize,
  segmentation style={draw=#2!75!white, line width=0.6pt, dash pattern=on 4pt off 1.5pt}}
\newcommand{\msg}[3][]{\par\hangindent=1.2em\hangafter=1\noindent{\bfseries #2:}\enspace\textit{#3}\ifx&#1&\else\,#1\fi\par\vspace{2pt}}

\definecolor{groupbg}{RGB}{232,232,232}
\ExplSyntaxOn
\NewDocumentCommand \hname { m }
  {
    \group_begin:
      \tl_set:Nn \l_tmpa_tl {#1}
      \tl_replace_all:Nnn \l_tmpa_tl { - } { - \allowbreak }
      \cs_set:Npn \_ { \textunderscore \allowbreak }
      \ttfamily
      \tl_use:N \l_tmpa_tl
    \group_end:
  }
\ExplSyntaxOff

\lstdefinestyle{codebase}{
  basicstyle=\ttfamily\scriptsize\color{ink},
  backgroundcolor=\color{ivory},
  frame=l, framerule=1.6pt, framesep=6pt, rulecolor=\color{ivoryline},
  xleftmargin=7.6pt, xrightmargin=0pt,
  aboveskip=9pt, belowskip=7pt,
  captionpos=t, abovecaptionskip=2pt, belowcaptionskip=5pt,
  columns=fullflexible, keepspaces=true,
  breaklines=true, breakatwhitespace=false,
  postbreak=\mbox{\textcolor{ink2}{$\hookrightarrow$}\space},
  showstringspaces=false, upquote=true, tabsize=2,
}
\lstdefinestyle{codetext}{style=codebase,
  rulecolor=\color{BrandPrimary},
  moredelim=[l][\color{BrandPrimary}\bfseries]{\#},
  moredelim=[s][\bfseries]{**}{**},
}
\lstdefinestyle{codeyaml}{style=codebase,
  rulecolor=\color{kraft},
    morestring=[b]", stringstyle=\color{sage!80!black},
  morekeywords={true,false,null}, keywordstyle=\color{navy},
}
\lstdefinestyle{codepy}{style=codebase,
  language=Python,
  rulecolor=\color{BrandTint},
  keywordstyle=\color{navy}\bfseries,
  morekeywords={True,False,None},
  commentstyle=\color{ink2},
  stringstyle=\color{sage!80!black},
}

\title{Recursive Collective Intelligence: Emergence on an Equal-Budget Ladder in Multi-Agent LLM Systems}

\definecolor{linkicon}{RGB}{75,107,132}
\newcommand{\linkitem}[3]{\href{#2}{\makebox[1.1em][c]{\textcolor{linkicon}{#1}}\hspace{0.3em}\texttt{#3}}}
\makeatletter
\gdef\@author{%
  \begin{tabular}{@{}c@{}}
  {\Authfont Lixuan He$^{1}$\quad Shengqu Cai$^{1}$\quad Fang Wu$^{1}$\quad Shikang Zheng$^{2}$\quad Liwei Jiang$^{3}$\quad Manling Li$^{4}$}\\
  {\Authfont Yi Ma$^{2}$\quad Li Fei-Fei$^{1}$\quad Leonidas Guibas$^{1}$\quad Gordon Wetzstein$^{1}$\quad Yejin Choi$^{1}$\quad Yining Hong$^{1}$}\\[4pt]
  {\Affilfont $^{1}$Stanford University\quad $^{2}$University of California, Berkeley\quad $^{3}$NVIDIA\quad $^{4}$Northwestern University}\\[4pt]
  {\normalfont\footnotesize\setlength{\tabcolsep}{0pt}\renewcommand{\arraystretch}{1.12}%
   \begin{tabular}{@{}l@{\hspace{2.4em}}l@{}}
   \linkitem{\faGlobe}{https://lixuan27.github.io/RCI/}{lixuan27.github.io/RCI} & \linkitem{\faGithub}{https://github.com/lixuan27/RCI-official}{github.com/lixuan27/RCI-official}\\
   \linkitem{\faPenNib}{https://lixuan27.github.io/RCI/blog/}{lixuan27.github.io/RCI/blog} & \linkitem{\faMousePointer}{https://lixuan27.github.io/RCI/playground/}{lixuan27.github.io/RCI/playground}
   \end{tabular}}
  \end{tabular}}
\makeatother

\teaser{%
  \centering\resizebox{0.92\linewidth}{!}{\input{figures/teaser_main}}\par
  \captionsetup{skip=4pt}%
  \captionof{figure}{\textbf{Recursive collective intelligence.} Each generation trains on its predecessor's verified outputs, without labels; diamonds mark releases, and PlanBench is drawn through G2. From G1 on, the group on PlanBench exceeds any final plan of the four base agents. On HiddenBench, it rises far above the vote of four base agents and above single-agent self-training with the same caps. On Hanabi (right axis), recursive training rises above a non-recursive control (Section~\ref{sec:exp-chain}).}%
  \label{fig:teaser}%
}

\begin{document}
\maketitle

\begin{abstract}
\input{subtex/abstract}
\end{abstract}

\clearpage
\setcounter{tocdepth}{2}
{\small\setlength{\parskip}{0pt}%
 \makeatletter\patchcmd{\l@section}{\addvspace{1.0em \@plus\p@}}{\addvspace{0.45em \@plus\p@}}{}{\GenericWarning{}{TOC spacing patch failed}}\makeatother
 \tableofcontents}
\fancypagestyle{contactstyle}{\fancyfoot[L]{\footerfont\textcolor{linkicon}{\faEnvelope[regular]}\hspace{0.4em}Email: \href{mailto:lixuanhe@stanford.edu}{\texttt{lixuanhe@stanford.edu}}}}
\thispagestyle{contactstyle}
\clearpage

\input{subtex/intro}

\input{subtex/related}
\input{subtex/framing}
\input{subtex/method}
\input{subtex/experiment}
\input{subtex/discussion}
\input{subtex/conclusion}

\clearpage
\renewcommand{\bibfont}{\small}
\setlength{\bibsep}{2pt plus 1pt}
\bibliography{references}
\bibliographystyle{plainnat}

\clearpage
\appendix
\input{subtex/appendix}

\end{document}

%% file: math_commands.tex
\usepackage{amsmath,amsfonts,bm}

\def\eqref#1{equation~\ref{#1}}

\def\1{\bm{1}}

\def\vone{{\bm{1}}}

\DeclareMathAlphabet{\mathsfit}{\encodingdefault}{\sfdefault}{m}{sl}
\SetMathAlphabet{\mathsfit}{bold}{\encodingdefault}{\sfdefault}{bx}{n}

%% file: figures/teaser_main.tex
\begingroup%
\definecolor{tzNavy}{HTML}{25374F}\definecolor{tzSage}{HTML}{4F7D69}\definecolor{tzGray}{HTML}{8A919A}%
\definecolor{tzInk}{HTML}{3A4656}\definecolor{tzRule}{HTML}{D4DAE0}\definecolor{tzPlum}{HTML}{7A5C93}\definecolor{tzCrim}{HTML}{B8323F}%
\tikzset{note/.style={draw=tzGray!60, fill=white, line width=0.45pt, rounded corners=1.5pt, inner sep=2.6pt, align=left, font=\rmfamily\bfseries\fontsize{9}{10.6}\selectfont, text=tzInk},%
         dia/.style={draw=white, line width=0.5pt, diamond, inner sep=0pt, minimum size=8.5pt}}%
\pgfplotsset{tzbase/.style={scale only axis, width=15.6cm, height=9.4cm, xmin=-0.15, xmax=3.15, xtick={0,1,2,3}, xticklabels={G0,G1,G2,G3},%
  axis line style={tzInk, line width=0.7pt}, tick style={tzGray}, major tick length=2.5pt,%
  tick label style={font=\rmfamily\bfseries\fontsize{11}{12}\selectfont, text=tzInk},%
  label style={font=\rmfamily\bfseries\itshape\fontsize{11.5}{13}\selectfont, text=tzInk}}}%
\begin{tikzpicture}
\begin{axis}[tzbase, axis y line*=left, axis x line*=bottom, ymin=0, ymax=1.22, ytick={0,0.2,0.4,0.6,0.8,1.0},
  ylabel={accuracy (PlanBench, HiddenBench)}, xlabel={generation}, grid=major, grid style={tzRule, line width=0.4pt},
  legend style={at={(0.5,0.995)}, anchor=north, font=\rmfamily\bfseries\fontsize{9.2}{10.6}\selectfont, draw=none, fill=none, inner sep=3pt, row sep=0pt, text=tzInk}, legend cell align=left, legend columns=4, transpose legend, legend style={/tikz/every even column/.append style={column sep=10pt}}, legend image code/.code={\draw[#1] (0cm,0cm) -- (0.5cm,0cm);}]
\addplot[tzSage, line width=1.3pt, densely dotted, domain=-0.15:3.15, samples=2, forget plot] {0.683};
\addplot[tzPlum, line width=1.3pt, densely dotted, domain=-0.15:3.15, samples=2, forget plot] {0.070};
\addlegendimage{empty legend}\addlegendentry{\hspace{-0.62cm}\textbf{PlanBench}}
\addplot[tzSage, line width=2.3pt, const plot] coordinates {(0,0.613) (1,0.857) (2,0.937) (3,0.937)};
\addlegendentry{collective training}
\addlegendimage{tzSage, line width=1.3pt, densely dotted}\addlegendentry{any final plan of four base agents}
\addlegendimage{empty legend}\addlegendentry{}
\addlegendimage{empty legend}\addlegendentry{\hspace{-0.62cm}\textbf{HiddenBench}}
\addplot[tzPlum, line width=2.3pt, const plot] coordinates {(0,0.509) (1,0.703) (2,0.755) (3,0.836) (3.15,0.836)};
\addlegendentry{collective training}
\addplot[tzPlum, line width=1.4pt, dashed, const plot] coordinates {(0,0.070) (1,0.396) (2,0.563) (3,0.658) (3.15,0.658)};
\addlegendentry{single-agent self-training}
\addlegendimage{tzPlum, line width=1.3pt, densely dotted}\addlegendentry{vote of four base agents}
\addlegendimage{empty legend}\addlegendentry{\hspace{-0.62cm}\textbf{Hanabi} (right axis)}
\addlegendimage{tzCrim, line width=2.3pt}\addlegendentry{recursive training}
\addlegendimage{tzCrim, line width=1.4pt, dashed}\addlegendentry{non-recursive control}
\addlegendimage{legend image code/.code={\node[dia, fill=tzInk] at (0.275cm,0cm) {};}}\addlegendentry{released generation}
\foreach \x/\y/\c in {1/0.857/tzSage, 2/0.937/tzSage, 1/0.703/tzPlum, 2/0.755/tzPlum, 3/0.836/tzPlum}{\edef\temp{\noexpand\node[dia, fill=\c] at (axis cs:\x,\y) {};}\temp}
\node[note, draw=none, fill=none, anchor=north east] (n1) at (axis cs:3.1,0.405) {collective training\\stays above single-agent\\self-training in every\\generation};
\end{axis}
\begin{axis}[tzbase, axis y line*=right, axis x line=none, ymin=0, ymax=5.6, ytick={0,1,2,3,4,5}, ylabel={Hanabi challenge score}, ylabel style={text=tzCrim}, y tick label style={text=tzCrim}]
\addplot[tzCrim, line width=2.3pt, const plot] coordinates {(0,0) (1,0.678) (2,2.953) (3,4.110) (3.15,4.110)};
\addplot[tzCrim, line width=1.4pt, dashed, const plot] coordinates {(1,0.678) (2,2.000) (3,2.907) (3.15,2.907)};
\foreach \x/\y in {1/0.678, 2/2.953, 3/4.110}{\edef\temp{\noexpand\node[dia, fill=tzCrim] at (axis cs:\x,\y) {};}\temp}
\end{axis}
\end{tikzpicture}
\endgroup

%% file: subtex/abstract.tex
Large language models increasingly work as teams of agents, and a team is expected to solve problems beyond the reach of any single member. Most evaluations credit a team for outperforming a single agent, without accounting for the additional computation: a team may consume many times the tokens of one agent, and a plain vote over independent answers already explains most of the reported gain. Two questions therefore remain open: When does interaction itself make a group smarter, and can a group improve its next generation by learning from its own work? We answer both with the \textbf{equal-budget ladder}, which grades a group against independent agents at three levels: outperforming a single agent, a compute-matched majority vote, and compute-matched pass@$n$. On this ladder we define \textbf{Recursive Collective Intelligence (RCI)} as a group whose own output, without a teacher or external labels, makes its next generation smarter at equal cost. Three findings follow. First, interaction improves collective intelligence when it moves information across agents:  forwarding split facts lifts groups above pass@$n$ at three model scales, by about a third at 27B. Second, organization emerges spontaneously: identical 27B agents form hubs and improve accuracy by 18 percentage points; encoding their organization in a prompt enables them to reach level 3 at less than half the token budget in a descriptive comparison. Third, groups recursively improve themselves: Two label-free generations progressively climb the ladder, the second 9B generation surpassing both pass@$n$ and a control trained on base-model turns; on PlanBench, HiddenBench and Hanabi, each generation outperforms its predecessor on unseen problems for four to six generations. A group therefore outperforms its members when it forwards useful information more efficiently than it spends tokens on coordination, and learns to do so from its own work. \looseness=-1

%% file: subtex/intro.tex
\section{Introduction}
\label{sec:intro}

Collective intelligence is the ability of a group to solve problems better than its members can alone, and it can be measured as a property of the group itself~\citep{woolley2010collective,hong2004diverse}. Language models now work in groups that could have this property. Copies of one model debate or build on each other's results through a common memory and return one joint answer~\citep{du2024debate,wang2024moa,qian2025macnet,han2025blackboard,mao2026delm}. These systems hold two promises: groups can solve problems beyond the reach of individual agents, and they can recursively improve by learning from their own work. \looseness=-1

Testing these promises, though, requires the right comparison. Majority voting over independent samples, each incurring its own computational cost, already improves upon a single agent~\citep{wang2023selfconsistency,li2024moreagents}, yet much of the improvement is credited to multi-agent debate~\citep{choi2025debatevote}. Thus, outperforming a single agent does not necessarily demonstrate the benefits of interaction, but may simply reflect greater computational cost. The value of interaction becomes particularly important when information is distributed across agents. Yet discussions tend to focus on information shared by all members while overlooking facts known to only one member~\citep{stasser1985pooling}, a pattern also observed in language-model groups~\citep{li2025hiddenbench}. Self-improvement methods, meanwhile, rely on external feedback and evaluate models in isolation~\citep{zelikman2022star,singh2024beyond}. Two questions therefore remain open: when does interaction itself make a group smarter than its members under an equal budget, and can a group make its next generation smarter without a teacher?

We answer both questions on one scale. Studies of human collective intelligence compare group performance against individuals working independently, with the strongest baseline counting a group as correct whenever any individual is correct, analogous to pass@$n$ in LLM evaluation~\citep{steiner1972group,lorge1955twomodels,laughlin1986demonstrability}. We extend this framework to language-model agents by additionally accounting for computational cost. The \textbf{equal-budget ladder} measures a group's performance against three baselines under the same token budget: a single agent, the majority vote, and pass@$n$ (Figure~\ref{fig:ladder-def}). Because pass@$n$ bounds any selection among independent answers, reaching level 3 demonstrates gains from interaction beyond answer selection. Decomposing the margin into accuracy and cost further reveals whether these gains come from better answers or greater token efficiency. On this ladder we further define \textbf{Recursive Collective Intelligence (RCI)}: a group's own output, without a teacher or external label, lifts its next generation. \looseness=-1

Building on this framework, our extensive experiments yield three major contributions and findings:

\begin{itemize}[leftmargin=*,itemsep=2pt]
  \item \textbf{An equal-budget ladder.}
  We introduce three levels of collective intelligence: outperforming a single agent, a compute-matched majority vote, and pass@$n$. We prove that pass@$n$ bounds independent answer selection and find that existing methods fail to beat compute-matched voting on shared-information tasks, while forwarding split facts exceeds pass@$n$ by about a third at 27B.

  \item \textbf{Emergent organization and coordination.}
  Identical agents spontaneously form hubs and relays, improving accuracy by 18 percentage points. Encoding their organization in a prompt enables groups to reach level 3 at less than half the token budget in a descriptive comparison.

  \item \textbf{Recursive Collective Intelligence.}
  On our hidden-profile benchmark SplitBench and on three further benchmarks, groups improve themselves through label-free training on their own verified outputs. On SplitBench, two generations progressively climb the ladder, with the second 9B generation surpassing both pass@$n$ and a control trained on base-model turns, and both improvements transfer to the official HiddenBench tasks. On PlanBench, HiddenBench and Hanabi, each generation outperforms its predecessor on unseen problems for four to six successive generations, and collective training outperforms single-agent self-training when information is split across agents.
\end{itemize}

%% file: subtex/related.tex
\section{Related Work}
\label{sec:related}

\paragraph{Evaluating groups of language-model agents.}\label{sec:rel-measure}
Multi-agent systems are commonly credited for outperforming a single agent. We think part of this improvement comes from additional computation alone. Majority voting over independent samples already improves upon a single agent~\citep{wang2023selfconsistency,li2024moreagents,chen2024morecalls}. The more samples are drawn, the more likely at least one of them is correct~\citep{brown2024monkeys}. Multi-agent debate, in contrast, rarely outperforms majority voting or a single agent with a good prompt~\citep{smit2024mad,wang2024rethinking,zhang2025stopmad,choi2025debatevote}. Modular AI systems raise the same question of whether many models together outperform each model alone~\citep{feng2026participation}. A fair evaluation should therefore compare a group with independent agents under the same computation. This matters most when information is distributed across agents, because the group then succeeds only when agents share their private facts, a step that language-model groups often skip~\citep{li2025hiddenbench}. Our ladder makes this comparison explicit.

\paragraph{Organizing language-model agents.}\label{sec:rel-org}
Multi-agent systems organize agents in many ways: agents debate~\citep{du2024debate}, aggregate layered proposals~\citep{wang2024moa}, or coordinate through directed graphs~\citep{qian2025macnet}, a central coordinator~\citep{ruan2026aorchestra,dang2025puppeteer} or a shared blackboard~\citep{hayesroth1985blackboard,han2025blackboard,mao2026delm}. Other work stops sampling once answers agree~\citep{aggarwal2023adaptive,li2024esc} or analyzes why multi-agent systems fail~\citep{cemri2025fail}. Related benchmarks measure collaboration through what models exchange~\citep{feng2026moco} and through long-horizon software projects~\citep{shen2026swecraft}. These systems fix their organization in advance, which leads us to ask whether identical agents can form their own.

\paragraph{Training agents and collectives.}\label{sec:rel-train}
Agentic reinforcement learning trains a single model toward verifiable outcomes with group-relative policy gradients~\citep{jin2025searchr1,zhang2025agenticrl,shao2024deepseekmath,yu2025dapo}, while distillation learns from a teacher~\citep{agarwal2024gkd,gu2024minillm} and imitation from successful runs~\citep{dong2023raft,motwani2025malt,chen2024optima}. Multi-agent extensions reward debaters with a verifier or share a team reward~\citep{park2025maporl,liu2025magrpo,zhao2025strongermas,hong2025mgrpo,li2026agentflow}. Because a shared reward is the same for every agent, it hides what each agent contributed. Cooperative reinforcement learning assigns this credit through difference rewards, value decomposition and Shapley values~\citep{wolpert2001collectives,foerster2018coma,sunehag2018vdn,rashid2018qmix,wang2020shapleyq,shapley1953value}; our Shapley credit applies the same idea to agents that share one policy. \looseness=-1

\paragraph{Recursive self-improvement.}\label{sec:rel-rsi}
Self-referential systems such as the G{\"o}del machine improve the process that improves them~\citep{schmidhuber2003godel}. Language models improve by training on their own rationales, filtered answers, self-assigned rewards or majority votes~\citep{zelikman2022star,huang2023selfimprove,singh2024beyond,yuan2024selfrewarding,zuo2025ttrl}. Agent scaffolds evolve their prompts~\citep{fernando2024promptbreeder}, their improver~\citep{zelikman2024stop} or their own code~\citep{zhang2026dgm}, and groups of models finetune on their own debates~\citep{subramaniam2025multiagentft} or successful team trajectories~\citep{zhao2025sirius}. These methods score answers with labels or verifiers and evaluate each model on its own. Our generations instead learn from turns verified by the group's own protocol, and the ladder credits an improvement to the group only when it outperforms independent agents at equal cost. \looseness=-1

%% file: subtex/framing.tex
\section{From Multi-Agent Systems to Collective Intelligence}
\label{sec:framing}

\begin{figure}[!t]
\centering
\resizebox{\linewidth}{!}{\input{figures/teaser_ladder}}
\caption{\textbf{The equal-budget ladder.} We define the equal-budget ladder by three levels: for a group whose token budget is equivalent to running $n_B$ independent agents, it reaches level 1 when it outperforms a single agent, level 2 when it outperforms the majority vote of $n_B$ independent agents, and level 3 when it exceeds their pass@$n_B$. Four Qwen3.8-27B multi-agent protocols on HiddenBench illustrate this ladder, each placed at the highest level it achieves.}
\label{fig:ladder-def}
\end{figure}
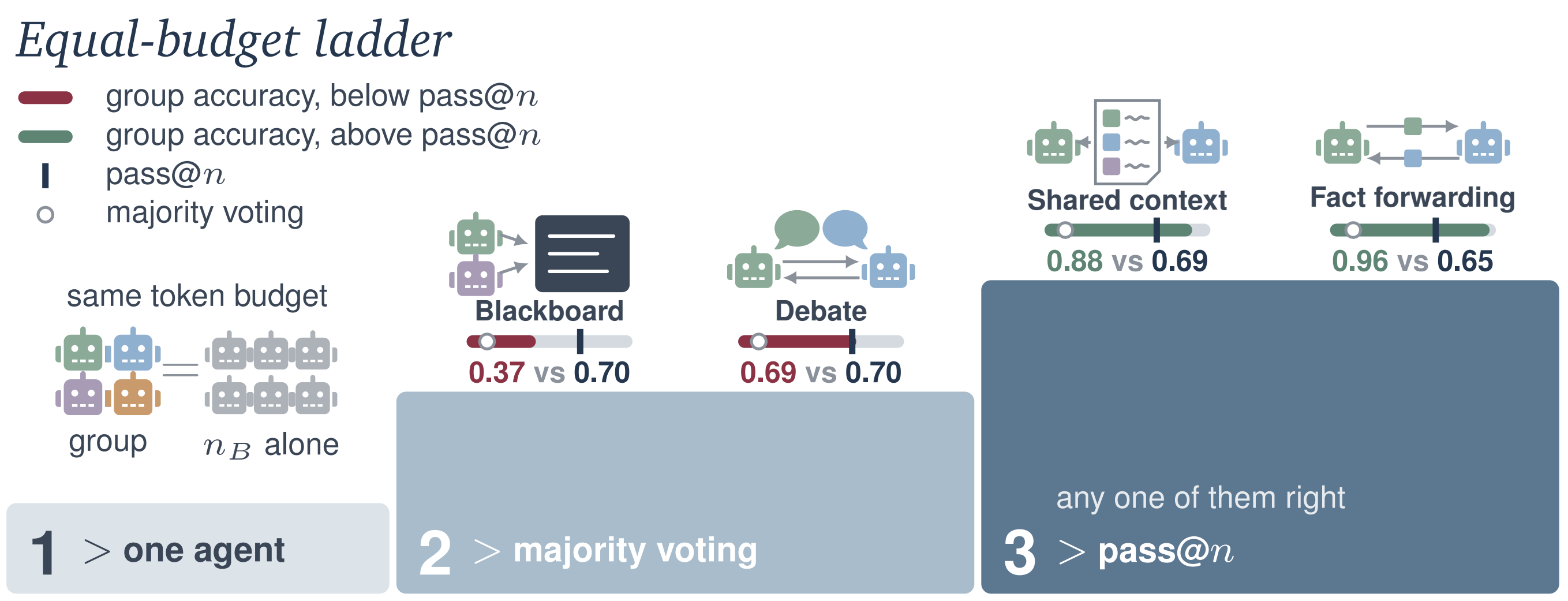

The systems reviewed in Section~\ref{sec:related} differ mainly in how information moves between agents. We use the following terms throughout the paper. A \emph{group} is a set of agents that run the same model with the same tools and return one joint answer. Its \emph{organization} is the rule that determines which messages each agent writes and reads, such as a majority vote over independent answers, a debate, a central coordinator or a shared blackboard. Its \emph{token budget} is the number of completion tokens that it spends on a task, expressed as the number $n_B$ of independent agents that the same tokens would pay for. Adding agents to any organization raises the token budget, and we think three constraints determine what this larger budget can buy.

\paragraph{Information structure.} When every agent sees the whole task, additional agents add samples from the same distribution, and their value is bounded by the best selection among independent answers, which is pass@$n$ (Proposition~\ref{prop:bound}). When the decisive facts are split across agents, solving the task requires combining them, which only interaction can do~\citep{stasser1985pooling,li2025hiddenbench}. We therefore report the two regimes separately.

\paragraph{Correlated errors.} Copies of one model make correlated errors. A majority vote over their answers saturates quickly~\citep{brown2024monkeys,chen2024morecalls}. Exchanging messages, moreover, tends to align agents with the majority even when a minority holds the right answer~\citep{lorenz2011social,bikhchandani1992cascades,choi2025debatevote}.

\paragraph{Coordination cost.} Every message that an agent writes or reads consumes tokens that independent agents could spend on their own attempts, and the failure of one agent can propagate through the messages of the others~\citep{cemri2025fail}. Interaction therefore pays only when the information it moves is worth more than its cost.

\paragraph{Collective intelligence.} Following studies of human groups, we measure collective intelligence against the same individuals working apart~\citep{woolley2010collective,steiner1972group,lorge1955twomodels,laughlin1986demonstrability}, and we hold the token budget fixed. A group is \emph{collectively intelligent} when it outperforms independent agents with the same token budget, and its strongest form exceeds their pass@$n$, which no selection among independent answers can reach. Its organization is \emph{emergent} when identical agents form it themselves under local rules, such as a limit on how many messages each agent may read. A group shows \emph{Recursive Collective Intelligence} when its own output, without a teacher or external labels, makes its next generation smarter at equal cost.

Each of these claims can also arise from the way groups are evaluated. A group that outperforms a single agent may simply have spent more tokens. Similarly, a generation that scores higher than its predecessor may only have been selected on the noise of a reused evaluation. Section~\ref{sec:method} therefore turns each claim into a test that a group can fail. The equal-budget ladder grades every group against compute-matched independent agents on the same pool and splits each margin into accuracy and cost, and gated recursion releases a new generation only when it outperforms its predecessor on tasks that no earlier round has used. Sections~\ref{sec:exp-when}, \ref{sec:exp-emergence} and~\ref{sec:exp-rsi} apply these tests to interaction, emergence and recursion.

%% file: figures/teaser_ladder.tex
\begingroup%
\definecolor{tzNavy}{HTML}{25374F}\definecolor{tzSky}{HTML}{40627F}\definecolor{tzSage}{HTML}{5E8574}%
\definecolor{tzRed}{HTML}{8B3344}\definecolor{tzAmber}{HTML}{B5651D}\definecolor{tzGray}{HTML}{8A919A}%
\definecolor{tzInk}{HTML}{3A4656}\definecolor{tzRule}{HTML}{D4DAE0}\definecolor{tzTint}{HTML}{EEF2F5}\definecolor{tzCard}{HTML}{C9D3DB}%
\tikzset{%
  ttl/.style={font=\fontsize{10.5}{12}\selectfont\itshape, text=tzNavy, anchor=west},%
  lab/.style={font=\sffamily\bfseries\fontsize{7.6}{9}\selectfont, text=tzInk},%
  tck/.style={font=\sffamily\fontsize{7}{8}\selectfont, text=tzGray},%
  val/.style={font=\sffamily\bfseries\fontsize{7.2}{8.4}\selectfont},%
  card/.style={fill=white, draw=tzCard, line width=0.7pt, rounded corners=6pt},%
}%
\definecolor{mA}{HTML}{8BAA98}\definecolor{mB}{HTML}{8FB0CF}\definecolor{mC}{HTML}{A79BB5}\definecolor{mD}{HTML}{C99A6B}%
\definecolor{stepA}{HTML}{DCE4EA}\definecolor{stepB}{HTML}{A9BCCB}\definecolor{stepC}{HTML}{5C7790}%
\newcommand{\tzrb}[3]{\node[inner sep=0pt, text=#3, font=\fontsize{7.6}{7.6}\selectfont] at (#1,#2) {\faIcon{robot}};}%
\newcommand{\tzchip}[3]{\fill[#3, rounded corners=0.6pt] (#1-0.055,#2-0.055) rectangle (#1+0.055,#2+0.055);}%
\tikzset{tzarr/.style={-{Latex[length=2.8pt,width=2.8pt]}, tzGray, line width=0.6pt}}%
\newcommand{\icoBoard}[2]{%
  \tzrb{#1-0.46}{#2+0.13}{mA}\tzrb{#1-0.46}{#2-0.13}{mC}%
  \draw[tzarr] (#1-0.31,#2+0.12) -- (#1-0.13,#2+0.06);%
  \draw[tzarr] (#1-0.31,#2-0.12) -- (#1-0.13,#2-0.06);%
  \fill[tzInk, rounded corners=1.5pt] (#1-0.09,#2-0.24) rectangle (#1+0.5,#2+0.24);%
  \foreach \dy/\w in {0.11/0.4, 0/0.3, -0.11/0.36} {\draw[white, line width=0.7pt, line cap=round] (#1+0.0,#2+\dy) -- (#1+\w,#2+\dy);}}%
\newcommand{\icoDebate}[2]{%
  \tzrb{#1-0.42}{#2-0.08}{mA}\tzrb{#1+0.42}{#2-0.08}{mB}%
  \node[inner sep=0pt, text=mA, font=\fontsize{8}{8}\selectfont] at (#1-0.15,#2+0.15) {\faIcon{comment}};%
  \node[inner sep=0pt, text=mB, font=\fontsize{8}{8}\selectfont] at (#1+0.15,#2+0.15) {\reflectbox{\faIcon{comment}}};%
  \draw[tzarr] (#1-0.24,#2-0.05) -- (#1+0.24,#2-0.05);%
  \draw[tzarr] (#1+0.24,#2-0.15) -- (#1-0.24,#2-0.15);}%
\newcommand{\icoShared}[2]{%
  \tzrb{#1-0.46}{#2}{mA}\tzrb{#1+0.46}{#2}{mB}%
  \draw[fill=white, draw=tzInk!65, line width=0.55pt] (#1-0.2,#2-0.26) -- (#1+0.12,#2-0.26) -- (#1+0.2,#2-0.18) -- (#1+0.2,#2+0.26) -- (#1-0.2,#2+0.26) -- cycle;%
  \foreach \dy/\c in {0.15/mA, 0/mB, -0.15/mC} {\tzchip{#1-0.1}{#2+\dy}{\c}\draw[tzGray, line width=0.6pt, line cap=round, decorate, decoration={snake, amplitude=0.3pt, segment length=2.4pt}] (#1-0.01,#2+\dy) -- (#1+0.13,#2+\dy);}%
  \draw[tzarr] (#1-0.23,#2) -- (#1-0.33,#2);%
  \draw[tzarr] (#1+0.23,#2) -- (#1+0.33,#2);}%
\newcommand{\icoForward}[2]{%
  \tzrb{#1-0.44}{#2}{mA}\tzrb{#1+0.44}{#2}{mB}%
  \draw[tzarr] (#1-0.29,#2+0.1) -- (#1+0.29,#2+0.1); \tzchip{#1}{#2+0.1}{mA}%
  \draw[tzarr] (#1+0.29,#2-0.1) -- (#1-0.29,#2-0.1); \tzchip{#1}{#2-0.1}{mB}}%
\begin{tikzpicture}[x=1cm, y=1cm]

\node[ttl] at (0.2,4.25) {Equal-budget ladder};
\newcommand{\bigico}[3]{\begin{scope}[shift={(#2,#3)}, scale=1.15, every node/.append style={scale=1.15}]#1{0}{0}\end{scope}}
\newcommand{\cmp}[6]{%
  \draw[tzRule, line width=2.6pt, line cap=round] (#1-0.55,#2) -- (#1+0.55,#2);
  \draw[#6, line width=2.6pt, line cap=round] (#1-0.55,#2) -- (#1-0.55+1.1*#3,#2);
  \draw[tzGray, fill=white, line width=0.6pt] (#1-0.55+1.1*#5,#2) circle (1.5pt);
  \draw[tzNavy, line width=1.3pt] (#1-0.55+1.1*#4,#2-0.09) -- (#1-0.55+1.1*#4,#2+0.09);}
\tikzset{pname/.style={font=\sffamily\bfseries\fontsize{6.3}{8}\selectfont, text=tzInk},
         num/.style={font=\sffamily\bfseries\fontsize{6.6}{8}\selectfont}}
\fill[stepA, rounded corners=2pt] (0.25,0.3) rectangle (3.0,0.95);
\fill[stepB, rounded corners=2pt] (3.05,0.3) rectangle (7.2,1.75);
\fill[stepC, rounded corners=2pt] (7.25,0.3) rectangle (11.4,2.55);
\foreach \x/\n/\t/\c in {0.4/1/{$>$ one agent}/tzInk, 3.2/2/{$>$ majority voting}/white, 7.4/3/{$>$ pass@$n$}/white} {
  \node[font=\sffamily\bfseries\fontsize{14}{14}\selectfont, text=\c, anchor=south west, inner sep=0pt] at (\x,0.43) {\n};
  \node[font=\sffamily\bfseries\fontsize{7.8}{9}\selectfont, text=\c, anchor=south west, inner sep=0pt] at (\x+0.38,0.47) {\t};}
\node[tck, text=white!85!stepC, anchor=south west, inner sep=0pt] at (7.78,0.86) {any one of them right};
\draw[tzRed, line width=2.6pt, line cap=round] (0.38,3.86) -- (0.68,3.86);
\node[tck, text=tzInk, anchor=west] at (0.84,3.86) {group accuracy, below pass@$n$};
\draw[tzSage, line width=2.6pt, line cap=round] (0.38,3.58) -- (0.68,3.58);
\node[tck, text=tzInk, anchor=west] at (0.84,3.58) {group accuracy, above pass@$n$};
\draw[tzNavy, line width=1.3pt] (0.53,3.21) -- (0.53,3.39);  \node[tck, text=tzInk, anchor=west] at (0.84,3.30) {pass@$n$};
\draw[tzGray, fill=white, line width=0.6pt] (0.53,3.02) circle (1.5pt); \node[tck, text=tzInk, anchor=west] at (0.84,3.02) {majority voting};
\node[tck, text=tzInk] at (1.62,2.42) {same token budget};
\foreach \dx/\dy/\c in {-0.18/0.16/mA,0.18/0.16/mB,-0.18/-0.16/mC,0.18/-0.16/mD} {\node[inner sep=0pt, text=\c, font=\fontsize{9}{9}\selectfont] at (0.98+\dx,1.9+\dy) {\faIcon{robot}};}
\node[font=\fontsize{11}{11}\selectfont, text=tzGray] at (1.5,1.88) {$=$};
\foreach \c in {0,1,2} \foreach \r in {0,1} {\node[inner sep=0pt, text=tzGray!70, font=\fontsize{8}{8}\selectfont] at (1.85+0.3*\c,2.05-0.32*\r) {\faIcon{robot}};}
\node[tck, text=tzInk] at (0.98,1.36) {group};
\node[tck, text=tzInk] at (2.15,1.36) {$n_B$ alone};
\bigico{\icoBoard}{4.15}{2.74}  \node[pname] at (4.15,2.33) {Blackboard};
\cmp{4.15}{2.11}{0.369}{0.700}{0.092}{tzRed}   \node[num] at (4.15,1.89) {\textcolor{tzRed}{0.37} \textcolor{tzGray}{vs} \textcolor{tzNavy}{0.70}};
\bigico{\icoDebate}{6.1}{2.74}  \node[pname] at (6.1,2.33) {Debate};
\cmp{6.1}{2.11}{0.688}{0.703}{0.093}{tzRed}    \node[num] at (6.1,1.89) {\textcolor{tzRed}{0.69} \textcolor{tzGray}{vs} \textcolor{tzNavy}{0.70}};
\bigico{\icoShared}{8.3}{3.54} \node[pname] at (8.3,3.13) {Shared context};
\cmp{8.3}{2.91}{0.881}{0.691}{0.094}{tzSage}  \node[num] at (8.3,2.69) {\textcolor{tzSage}{0.88} \textcolor{tzGray}{vs} \textcolor{tzNavy}{0.69}};
\bigico{\icoForward}{10.35}{3.54} \node[pname] at (10.35,3.13) {Fact forwarding};
\cmp{10.35}{2.91}{0.959}{0.649}{0.110}{tzSage}  \node[num] at (10.35,2.69) {\textcolor{tzSage}{0.96} \textcolor{tzGray}{vs} \textcolor{tzNavy}{0.65}};

\pgfresetboundingbox\path (0.15,0.25) rectangle (11.45,4.5);
\end{tikzpicture}
\endgroup

%% file: subtex/method.tex
\section{An Equal-Budget Ladder for Collective Intelligence}
\label{sec:method}

\subsection{Compute-matched ensembles and three levels}
\label{sec:nominal}
\label{sec:ladder}

A group consists of $N$ members that run the same model. The members interact through a protocol, and the group commits to one answer with accuracy $Y$. As a reference, we also run the same model as independent agents. Each independent agent receives the information of one member, and we draw ensembles of these agents seat by seat. If a group run spends $k$ completion tokens, its budget is $n_B=k/c_1$ agents, where $c_1$ is the mean cost of one independent agent. The ladder needs only each agent's final answer and token count, so it applies to agents in any modality, including agents that read page images (Section~\ref{sec:exp-symmetric}). \looseness=-1

For an ensemble of $n$ independent agents, $V(n)$ is the accuracy of their majority vote, and $T(n)$ is their pass@$n$ accuracy, the probability that at least one of them is correct. We define three levels of collective intelligence.
\begin{itemize}[leftmargin=*,itemsep=1pt,topsep=2pt]
  \item \textbf{Level 1}, $Y>V(1)$: the group outperforms a single agent, so its extra tokens buy a better answer.
  \item \textbf{Level 2}, $Y>V(n_B)$: the group outperforms the majority vote of $n_B$ independent agents.
  \item \textbf{Level 3}, $Y>T(n_B)$: the group exceeds pass@$n_B$, so it produces correct answers that no selection among independent answers can produce.
\end{itemize}
Levels are tested in order. A group reaches level $k$ when one-sided paired tests at 0.025, clustered by question, confirm every level from 1 to $k$. A group below a single agent sits at level 0. We say that a group \emph{beats the vote} at level 2 and \emph{beats pass@$n$} at level 3, and we write pass@$n$ for pass@$n_B$ when the budget is clear. Three properties of the references explain the levels (proofs in Appendix~\ref*{app:proofs}).

\begin{proposition}[Structure of the references]
\label{prop:bound}
Let each independent agent answer question $x$ correctly with probability $p_x$, independently of the others.
\textbf{(i)} $T(n)=1-\mathbb{E}_x[(1-p_x)^n]$ is nondecreasing and concave: each additional agent adds $\mathbb{E}_x[p_x(1-p_x)^n]$.
\textbf{(ii)} Any rule that commits to one of the answers of $n$ independent agents is correct at most $T(n)$ of the time, even if it also uses their texts, confidences or a judge's opinion. Hence $T(n)\ge\max\{V(1),V(n)\}$, and level 3 implies levels 1 and 2.
\textbf{(iii)} $T(n)\le\min\{1,\,n\,V(1)\}$, so adding $n-m$ agents raises pass@$n$ by at most $(n-m)\,V(1)$.
\end{proposition}

Part (ii) makes the ladder a test of interaction. Every selection rule, from a majority vote to a learned judge, picks one of the independent answers, so its accuracy stays at or below pass@$n$ at equal cost. A group above pass@$n$ must therefore have produced correct answers through what its members exchanged. Part (iii) shows when such answers are possible. With split facts, a single agent is rarely correct, so pass@$n$ grows at most linearly with the budget. A group that pools facts can then outperform every selection rule, so its gain comes mainly from accuracy. With shared information, $V(1)$ is large and pass@$n$ saturates within a few agents, so a group must outperform a nearly perfect selector. \looseness=-1

\paragraph{Accuracy and cost.} Consider a protocol X and a reference R graded against the same pool. The change in margin decomposes into a change in accuracy and a change in pass@$n$ caused by the different budgets,
\begin{equation}
\label{eq:channels}
\begin{aligned}
&\big[Y_X-T(n_X)\big]-\big[Y_R-T(n_R)\big]\\
&\quad=\underbrace{\big(Y_X-Y_R\big)}_{\text{accuracy}}+\underbrace{\big[T(n_R)-T(n_X)\big]}_{\text{cost}}.
\end{aligned}
\end{equation}
Proposition~\ref{prop:bound} turns this bookkeeping identity into a statement about cost. The cost term is the accuracy that $n_R-n_X$ additional independent agents would add. By Proposition~\ref{prop:bound}(iii), it is at most $(n_R-n_X)\,V(1)$, so a cheaper protocol gains much only when additional agents would still improve accuracy. X reaches level 3 \emph{through accuracy} when its counterfactual margin $Y_X-T(n_R)$, its margin at R's budget, is confirmed above zero. Otherwise, X needs both higher accuracy and lower cost.

\subsection{Organizations on the ladder}
\label{sec:orgs}

We evaluate several common group organizations on the ladder, and Figure~\ref{fig:ladder}a groups them into seven families by what members exchange (Appendix Table~\ref*{tab:methods}). In a \emph{debate}, members exchange answers and reasons in free text for two or three rounds~\citep{du2024debate}. In a \emph{blackboard}, members publish and read messages on a common board~\citep{han2025blackboard,mao2026delm}. Evidence protocols exchange facts instead. In \emph{fact forwarding}, members forward only their own private facts. \emph{Shared context} attaches every member's facts to the shared record. A \emph{facts-only blackboard} accepts only forwarded facts, and a \emph{gated blackboard} lets a member post an analysis only after it has read every fact. To study self-organization, we give each member a reading budget of one message, which it may transfer to another member. A member that collects enough budget can read every fact and becomes a hub. A cap on each member's budget prevents a single hub, so the group must build relays instead. An optional planning round lets members agree on roles before reading. Designed organizations assign these roles by hand. A \emph{designed hub} gives one member the whole budget, and a \emph{designed tree} assigns hub and relay roles. On a shared corpus, a \emph{designed aggregator} lets three readers split the documents and a fourth member combine their findings. With a \emph{self-written prompt}, the group follows organization instructions that the model writes itself. Finally, a token accounting labels every model call as waiting, negotiation or work (Appendix~\ref*{app:ledger}). \looseness=-1

\subsection{Self-improvement and Shapley credit}
\label{sec:recursion-setup}
\label{sec:credit}

This section defines recursive collective intelligence. A group can improve itself in two ways: by writing a prompt at test time or by training its weights. We also introduce the credit rule for training on the group's outcome.

\input{figures/overview}

\paragraph{Recursion.} A group is recursively collectively intelligent when its own output, without a teacher or external labels, improves its next generation on the same ladder (Figure~\ref{fig:overview}). At test time, the model writes a prompt for its own group. In the weights, each generation trains on turns that its parent group's protocol has verified. These are mainly first turns, which members write before reading each other and which must cover the member's private facts. The second generation samples four turns per seat and keeps the shortest one that passes. At 27B, first turns are trained in the standard-debate context, and the 4B and 9B standard-debate runs also train on commit turns (Appendix~\ref*{app:recursion}). Groups play one of two debates. The \emph{standard debate} lets members discuss freely for three rounds with the hint that facts are split, and the \emph{3-step debate} limits each member to three steps. In both, the protocol's check on each turn replaces an answer label. We call a result recursion when three conditions hold: the first generation (G1) improves on the base model (G0), the second generation (G2) improves on G1, and G2 beats a \emph{control} trained on G0's turns.

\paragraph{Shapley credit.} When no protocol checks individual turns, members that share one policy must train on the group's outcome, and the credit rule determines what the policy learns. On a set-valued task, the group delivers the union $P_S=\bigcup_{i\in S}P_i$ of its members' claims. A coalition $S$ is worth $v(S)=s(P_S)$ under the task score $s$, with $v(\emptyset)=0$, and the group's reward is $R=v(\{1,\dots,N\})$. Suppose the group holds $r$ distinct claims, $h$ of them true, out of $t$ true findings. A member then adds a claim with calibrated confidence $q$, the probability that it is true.

\begin{proposition}[Reporting thresholds]
\label{prop:threshold}
Under recall $h/t$, the claim raises the expected score for every $q>0$. Under the F1 score $2h/(r+t)$, it does so exactly when $q>\mathrm{F1}/2$. Under a score that adds $1/t$ for a true claim and subtracts $c/t$ for a false one, it does so exactly when $q>c/(1+c)$.
\end{proposition}

\emph{Shapley credit} pays each member its Shapley share of the group's score~\citep{shapley1953value},
\begin{equation}
\phi_i=\sum_{S\subseteq\{1,\dots,N\}\setminus\{i\}}\frac{|S|!\,(N-|S|-1)!}{N!}\,\big(v(S\cup\{i\})-v(S)\big),
\label{eq:share}
\end{equation}
so a finding reported by several members is split among them, and a wrong claim is charged to its author. A fraction $\gamma=0.5$ of each positive share is then passed along citations to the members whose messages it builds on, which gives $f_i$. Member $i$ receives the credit $\cred_i=\lambda R+(1-\lambda)N f_i$\phantomsection\label{eq:credit} with $\lambda=0.25$, and training uses a clipped group-relative policy gradient~\citep{schulman2017ppo,yu2025dapo} (Appendix~\ref*{app:credit-results}).

\subsection{Gated recursion over generations}
\label{sec:chain-protocol}

Two label-free generations show that a group can improve itself, which raises the question of whether the improvement continues. Over longer sequences we see two risks. Selecting on a noisy evaluation can release a generation that merely matches its predecessor, and each generation may also learn less than the one before. We therefore run recursion on PlanBench, HiddenBench and Hanabi as a gated sequence of generations (Figure~\ref{fig:chain-protocol}). On these three benchmarks, we call a run recursive when every released generation outperforms its predecessor on unseen tasks, while the three conditions above define recursion on SplitBench.

\paragraph{One round.} The current generation plays a fresh set of training tasks with its own collective protocol, and an automatic check accepts or rejects each output without human-written answer labels: the plan simulator in PlanBench, agreement with the answer that at least three of four base-model members give with every fact in HiddenBench and the game score in Hanabi. Candidates continue from the current generation's weights, train mainly on its latest accepted outputs, and are ranked on a selection set. The best candidate that passes the health gates then meets the current generation on confirmation tasks that no earlier round has used, paired by task. It becomes the next generation only when its paired gain $d$ exceeds a threshold, for example $d \ge 1.645\,\mathrm{SE}$ in PlanBench; otherwise the round is flat. Fresh confirmation tasks in every round prevent a candidate from being selected on the noise of a reused set, and health gates anchored at the base model reject candidates whose token budget, answer format or accuracy on control tasks degrade.

\paragraph{Single-agent self-training.} To test whether a gain belongs to the group, on HiddenBench one run, T3, adds a matched run that trains each agent on its own solo outputs under the same caps on tasks and training tokens, and Hanabi adds a non-recursive control run that trains on games of a fixed generation. Collective training and single-agent self-training then differ only in whether the training data comes from the group's interaction. Appendix~\ref*{app:chains} lists every run, its task sets and its release rule.

\input{figures/chain_protocol}

\input{figures/ladder}

%% file: figures/overview.tex
\begin{figure*}[t]
\centering
\includegraphics[width=\textwidth]{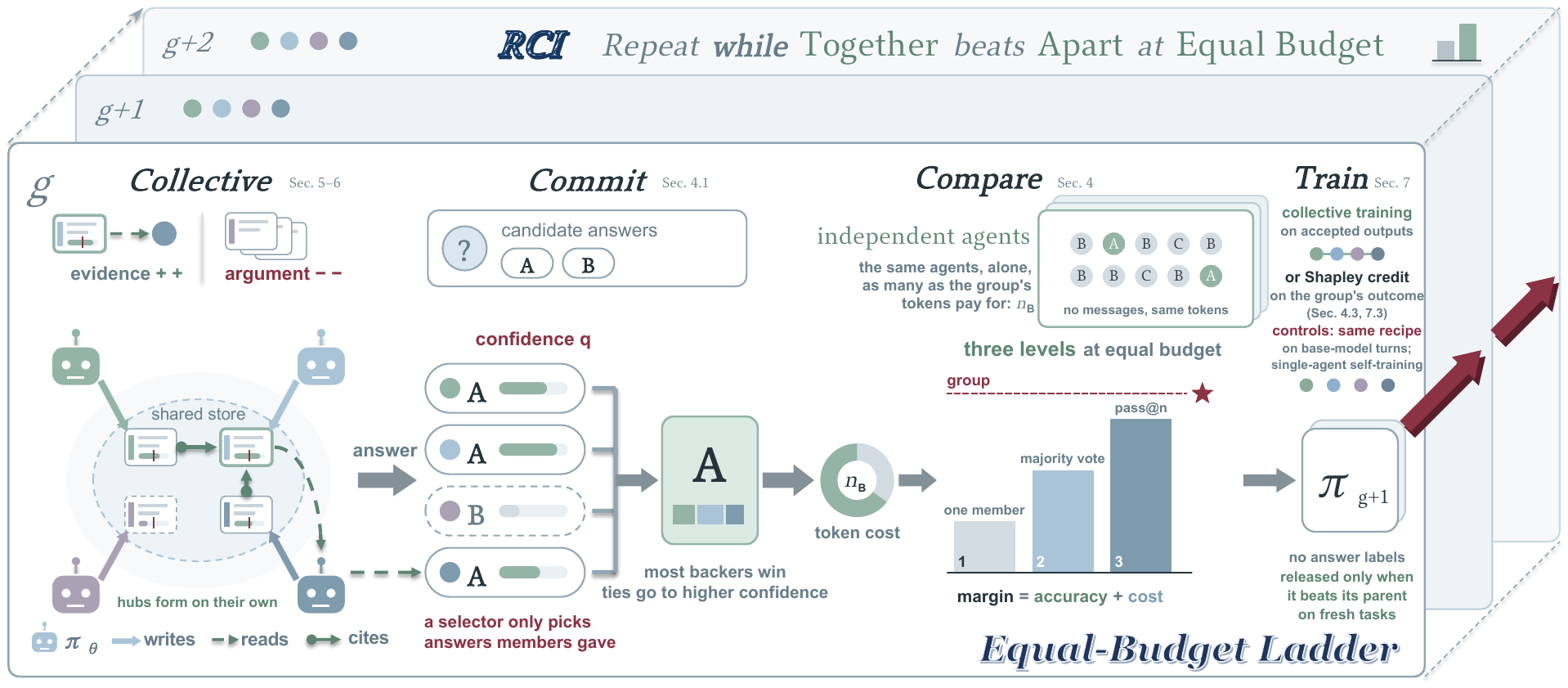}
\caption{\textbf{Overview of one generation $g$.} \textit{Collective:} copies of the policy $\pi_\theta$ write messages to a shared store, read and cite each other's messages, and can hand their reading budget to another member, so hubs form on their own; evidence helps ($++$) and arguments mislead ($--$). \textit{Commit:} every member answers with a confidence $q$, and the group commits the answer with the most backers, ties going to higher confidence, so the selector only picks among answers that members gave. \textit{Compare:} the record's completion tokens set its cost $n_B$, the number of independent agents they pay for; the same agents run alone with the same information, and the group is graded against $n_B$ of them on three levels, one member, their majority vote and pass@$n$, with a margin that splits into accuracy and cost (Equation~\ref{eq:channels}). \textit{Train:} the next generation $\pi_{g+1}$ trains on the outputs that the group's own protocol accepts, with no teacher and no answer label (collective training), or, when no protocol checks single turns, on the group's outcome with each member paid its Shapley credit (Section~\ref{sec:credit}), and controls test the gain: a control trained with the same recipe on base-model turns and, on HiddenBench, single-agent self-training, which trains each agent on its own outputs; a candidate is released only when it outperforms its parent on fresh tasks (Figure~\ref{fig:chain-protocol}), and the loop repeats. Panel titles give the sections that develop each step.}
\label{fig:overview}
\end{figure*}

%% file: figures/chain_protocol.tex
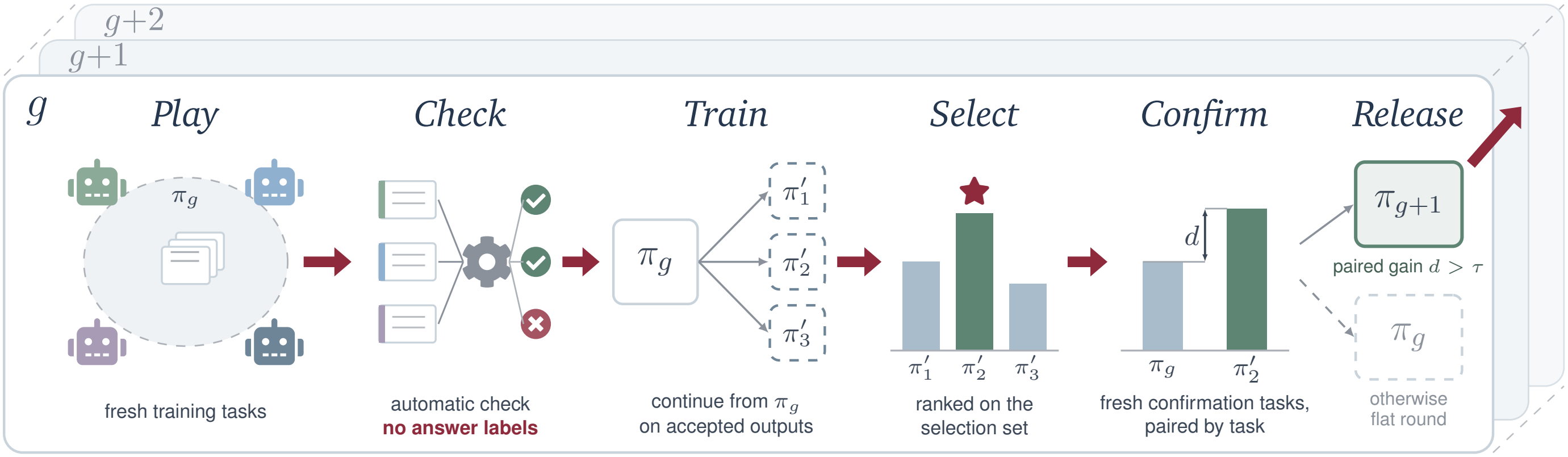
\begin{figure}[!t]
\centering
\begingroup
\definecolor{tzNavy}{HTML}{25374F}\definecolor{tzSage}{HTML}{5E8574}\definecolor{tzCrim}{HTML}{8E2A3B}
\definecolor{tzGray}{HTML}{8A919A}\definecolor{tzInk}{HTML}{3A4656}\definecolor{tzTint}{HTML}{EEF2F5}
\definecolor{tzCard}{HTML}{C9D3DB}\definecolor{tzBar}{HTML}{A9BCCB}\definecolor{tzBarD}{HTML}{6E8597}
\definecolor{mA}{HTML}{8BAA98}\definecolor{mB}{HTML}{8FB0CF}\definecolor{mC}{HTML}{A79BB5}\definecolor{mD}{HTML}{6E8597}
\resizebox{\linewidth}{!}{%
\begin{tikzpicture}[x=1cm, y=1cm,
  stage/.style={font=\rmfamily\itshape\fontsize{12}{13}\selectfont, text=tzNavy, anchor=base},
  lbl/.style={font=\sffamily\fontsize{6.9}{8}\selectfont, text=tzInk, align=center},
  lblr/.style={font=\sffamily\bfseries\fontsize{6.9}{8}\selectfont, text=tzCrim, align=center},
  red/.style={-{Triangle[length=5.5pt,width=9pt]}, line width=3.4pt, tzCrim},
  thin/.style={-{Latex[length=3.2pt,width=3pt]}, line width=0.65pt, tzGray},
  thind/.style={thin, dashed},
  pcard/.style={draw=tzCard, fill=white, line width=0.8pt, rounded corners=3pt, minimum width=0.95cm, minimum height=0.95cm, inner sep=0pt, font=\rmfamily\fontsize{13}{13}\selectfont, text=tzInk}]
\fill[tzTint!55, draw=tzCard!70, line width=0.5pt, rounded corners=6pt] (0.8,1.15) rectangle (17.6,5.45);
\fill[tzTint!80, draw=tzCard!80, line width=0.5pt, rounded corners=6pt] (0.4,0.77) rectangle (17.2,5.05);
\fill[white, draw=tzCard, line width=0.8pt, rounded corners=6pt] (0,0.4) rectangle (16.8,4.65);
\foreach \a/\b in {{(0,4.65)}/{(0.8,5.45)}, {(16.8,4.65)}/{(17.6,5.45)}, {(16.8,0.4)}/{(17.6,1.15)}} {\draw[tzGray!60, dashed, line width=0.45pt] \a -- \b;}
\node[font=\rmfamily\itshape\fontsize{11}{12}\selectfont, text=tzGray, anchor=west] at (1.0,5.25) {$g{+}2$};
\node[font=\rmfamily\itshape\fontsize{11}{12}\selectfont, text=tzGray, anchor=west] at (0.6,4.85) {$g{+}1$};
\node[font=\rmfamily\itshape\fontsize{15}{16}\selectfont, text=tzNavy, anchor=west] at (0.12,4.25) {$g$};
\foreach \x/\t in {2.05/Play, 5.15/Check, 8.15/Train, 10.95/Select, 13.55/Confirm, 15.85/Release} {\node[stage] at (\x,4.08) {\t};}
\fill[tzTint] (2.05,2.55) ellipse (1.15cm and 0.95cm);
\draw[tzGray!70, dashed, line width=0.5pt] (2.05,2.55) ellipse (1.15cm and 0.95cm);
\foreach \x/\y/\c in {1.05/3.45/mA, 3.05/3.45/mB, 1.05/1.65/mC, 3.05/1.65/mD} {\node[inner sep=0pt, text=\c, font=\fontsize{15}{15}\selectfont] at (\x,\y) {\faIcon{robot}};}
\foreach \dy in {0.16,0.08,0} {\draw[tzCard, fill=white, line width=0.6pt, rounded corners=1pt] (1.78+\dy,2.3+\dy) rectangle (2.32+\dy,2.72+\dy);}
\foreach \dy in {0.04,-0.06} {\draw[tzGray!70, line width=0.6pt] (1.88,2.5+\dy+0.12) -- (2.2,2.5+\dy+0.12);}
\node[font=\rmfamily\itshape\fontsize{8.5}{9}\selectfont, text=tzInk] at (2.05,3.25) {$\pi_g$};
\node[lbl] at (2.05,0.85) {fresh training tasks};
\newcommand{\outdoc}[3]{\draw[tzCard, fill=white, line width=0.6pt, rounded corners=1pt] (#1-0.32,#2-0.21) rectangle (#1+0.32,#2+0.21);
  \fill[#3] (#1-0.32,#2-0.21) rectangle (#1-0.25,#2+0.21);
  \foreach \dy in {0.08,-0.04} {\draw[tzGray!60, line width=0.55pt] (#1-0.17,#2+\dy) -- (#1+0.2,#2+\dy);}}
\outdoc{4.55}{3.25}{mA}\outdoc{4.55}{2.55}{mB}\outdoc{4.55}{1.85}{mC}
\node[inner sep=0pt, text=tzGray, font=\fontsize{17}{17}\selectfont] (gear) at (5.45,2.55) {\faIcon{cog}};
\foreach \y in {3.25,2.55,1.85} {\draw[tzGray, line width=0.55pt] (4.9,\y) -- (5.18,2.55);}
\node[inner sep=0pt, text=tzSage, font=\fontsize{10}{10}\selectfont] at (6.0,3.25) {\faIcon{check-circle}};
\node[inner sep=0pt, text=tzSage, font=\fontsize{10}{10}\selectfont] at (6.0,2.55) {\faIcon{check-circle}};
\node[inner sep=0pt, text=tzCrim!80, font=\fontsize{10}{10}\selectfont] at (6.0,1.85) {\faIcon{times-circle}};
\foreach \y in {3.25,2.55,1.85} {\draw[tzGray!70, line width=0.5pt] (5.7,2.55) -- (5.86,\y);}
\node[lbl] at (5.15,0.95) {automatic check};\node[lblr] at (5.15,0.68) {no answer labels};
\node[pcard] (pg) at (7.35,2.55) {$\pi_g$};
\foreach \y/\n in {3.35/1, 2.55/2, 1.75/3} {\node[pcard, minimum width=0.62cm, minimum height=0.62cm, draw=tzBarD, dashed, font=\rmfamily\fontsize{9}{9}\selectfont] (c\n) at (8.95,\y) {$\pi'_{\n}$};
  \draw[thin] (pg.east) -- (c\n.west);}
\node[lbl] at (8.15,0.95) {continue from $\pi_g$};\node[lbl] at (8.15,0.68) {on accepted outputs};
\foreach \x/\h/\c in {10.35/1.0/tzBar, 10.95/1.55/tzSage, 11.55/0.75/tzBar} {\fill[\c] (\x-0.21,1.55) rectangle (\x+0.21,1.55+\h);}
\draw[tzGray!70, line width=0.6pt] (10.0,1.55) -- (11.9,1.55);
\foreach \x/\n in {10.35/1, 10.95/2, 11.55/3} {\node[font=\rmfamily\fontsize{7.5}{8}\selectfont, text=tzInk] at (\x,1.36) {$\pi'_{\n}$};}
\node[inner sep=0pt, text=tzCrim, font=\fontsize{9}{9}\selectfont] at (10.95,3.35) {\faIcon{star}};
\node[lbl] at (10.95,0.95) {ranked on the};\node[lbl] at (10.95,0.68) {selection set};
\fill[tzBar] (12.85,1.55) rectangle (13.3,2.55);
\fill[tzSage] (13.8,1.55) rectangle (14.25,3.15);
\draw[tzGray!70, line width=0.6pt] (12.6,1.55) -- (14.5,1.55);
\node[font=\rmfamily\fontsize{8.5}{9}\selectfont, text=tzInk] at (13.075,1.34) {$\pi_g$};
\node[font=\rmfamily\fontsize{8.5}{9}\selectfont, text=tzInk] at (14.025,1.34) {$\pi'_2$};
\draw[tzGray!80, line width=0.5pt] (12.85,2.55) -- (13.55,2.55);
\draw[tzInk, line width=0.7pt, {Latex[length=2.4pt]}-{Latex[length=2.4pt]}] (13.55,2.55) -- (13.55,3.15) node[midway, left, font=\rmfamily\itshape\fontsize{9}{9}\selectfont, inner sep=1.5pt] {$d$};
\draw[tzGray!80, line width=0.5pt] (13.55,3.15) -- (13.8,3.15);
\node[lbl] at (13.55,0.95) {fresh confirmation tasks,};\node[lbl] at (13.55,0.68) {paired by task};
\node[pcard, draw=tzSage, line width=1.1pt, fill=tzSage!12, minimum width=1.2cm, font=\rmfamily\fontsize{11.5}{12}\selectfont] (nx) at (15.85,3.2) {$\pi_{g+1}$};
\node[pcard, draw=tzCard, dashed, text=tzGray, minimum width=1.2cm] (kp) at (15.85,1.7) {$\pi_g$};
\node[font=\sffamily\fontsize{6.5}{7.5}\selectfont, text=tzSage!70!black, anchor=north] at (nx.south) {paired gain $d>\tau$};
\node[font=\sffamily\fontsize{6.5}{7.5}\selectfont, text=tzGray, anchor=north, align=center] at (kp.south) {otherwise\\[-1pt]flat round};
\draw[thin] (14.62,2.75) -- (nx.west);
\draw[thind] (14.62,2.35) -- (kp.west);
\foreach \a/\b in {3.38/3.92, 6.3/6.72, 9.4/9.9, 12.0/12.45} {\draw[red] (\a,2.55) -- (\b,2.55);}
\draw[red] (16.55,3.65) -- (17.12,4.3);
\end{tikzpicture}}
\endgroup
\caption{\textbf{Gated recursion over generations.} In each round, generation $g$ plays fresh training tasks with its collective protocol, and an automatic check (plan simulator, protocol check or game score) accepts or rejects its outputs without human-written answer labels. Candidates continue from $\pi_g$ on the accepted outputs and are ranked on a selection set; the best one meets $\pi_g$ on fresh confirmation tasks, paired by task, and is released as generation $g{+}1$ only when its paired gain $d$ exceeds a threshold $\tau$. Otherwise the round is flat.}
\label{fig:chain-protocol}
\end{figure}

%% file: figures/ladder.tex
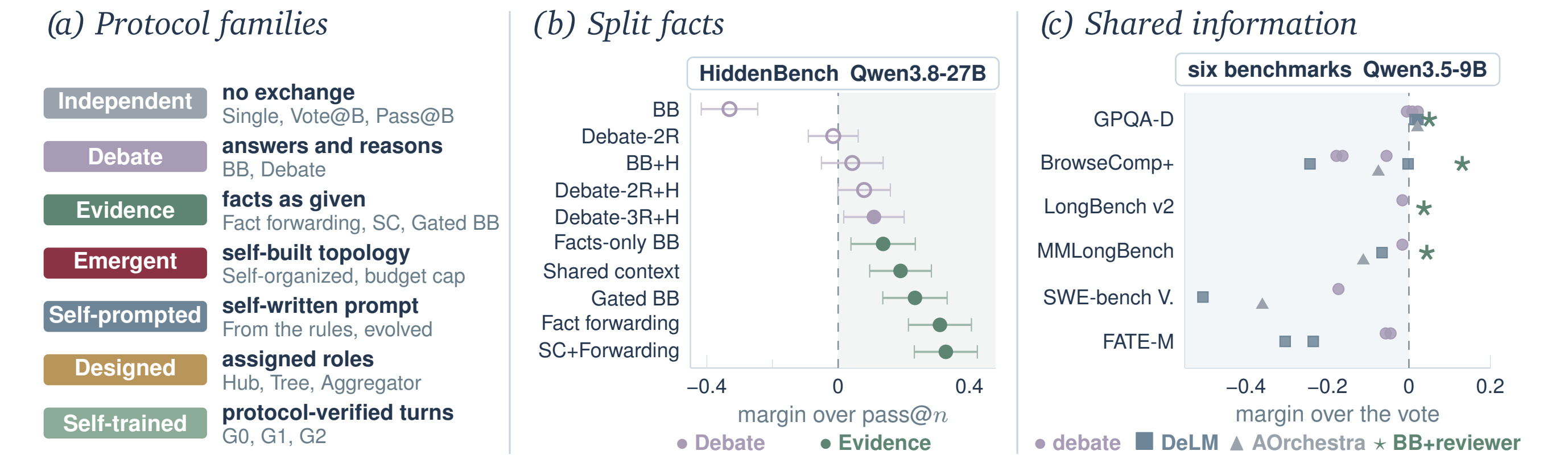
\begin{figure*}[t]
\centering
\begingroup
\definecolor{fgT}{HTML}{687D89}\definecolor{fgN}{HTML}{25374F}\definecolor{fgL}{HTML}{C9D6DF}
\definecolor{fgP}{HTML}{F3F6F8}\definecolor{fgY}{HTML}{9AA3AB}\definecolor{fgD}{HTML}{6E8597}
\tikzset{
  qt/.style={font=\fontsize{11}{12}\selectfont\itshape, text=fgN, anchor=base west, inner sep=0pt},
  qn/.style={font=\sffamily\fontsize{8}{9}\selectfont, text=fgT, anchor=base west, inner sep=0pt},
  qk/.style={font=\sffamily\bfseries\fontsize{7.8}{9}\selectfont, anchor=base, inner sep=0pt},
  key/.style={draw=fgL, line width=0.5pt, rounded corners=2pt, fill=white, inner xsep=4pt, inner ysep=2.5pt,
              font=\sffamily\bfseries\fontsize{8}{9}\selectfont, text=fgN},
  fam/.style={rounded corners=1.5pt, fill=#1, text=white, inner sep=0pt, minimum width=1.9cm, minimum height=0.36cm,
              font=\sffamily\bfseries\fontsize{8.3}{9}\selectfont},
}
\pgfplotsset{ladderax/.style={scale only axis, axis line style={fgL}, tick style={fgL}, axis x line*=bottom,
  axis y line*=left, ytick style={draw=none}, clip=false,
  tick label style={font=\sffamily\fontsize{7.5}{8.5}\selectfont, text=fgN},
  xlabel style={font=\sffamily\fontsize{8.3}{9.3}\selectfont, text=fgT, yshift=3pt}}}
\begin{tikzpicture}[x=1cm, y=1cm]
\draw[fgL, line width=0.6pt] (5.45,0.0) -- (5.45,5.15);
\draw[fgL, line width=0.6pt] (11.35,0.0) -- (11.35,5.15);

\node[qt] at (0.05,4.88) {(a) Protocol families};
\foreach \i/\c/\code/\name/\who in {
  0/famSolo/Solo/{no exchange}/{Single, Vote@B, Pass@B},
  1/famArgue/Argue/{answers and reasons}/{BB, Debate},
  2/famFact/Fact/{facts as given}/{Fact forwarding, SC, Gated BB},
  3/famSelf/Self/{self-built topology}/{Self-organized, budget cap},
  4/famNotes/Notes/{self-written prompt}/{From the rules, evolved},
  5/famDesign/Design/{assigned roles}/{Hub, Tree, Aggregator},
  6/famTrain/Train/{protocol-verified turns}/{G0, G1, G2}} {
  \node[fam=\c] at (0.98,4.08-0.62*\i) {\csname famname\code\endcsname};
  \node[qn, font=\sffamily\bfseries\fontsize{8}{9}\selectfont, text=fgN] at (2.1,4.12-0.62*\i) {\name};
  \node[qn, font=\sffamily\fontsize{7.5}{8.5}\selectfont] at (2.1,3.85-0.62*\i) {\who};}

\node[qt] at (5.7,4.88) {(b) Split facts};
\begin{axis}[ladderax, at={(7.55cm,1.0cm)}, anchor=south west, width=3.55cm, height=3.2cm,
  xmin=-0.45, xmax=0.48, ymin=0.4, ymax=10.6, xtick={-0.4,0,0.4},
  xticklabels={\textminus0.4,0,0.4}, minor xtick={-0.2,0.2,0.6},
  ytick={1,...,10}, yticklabels={SC+Forwarding, Fact forwarding, Gated BB, Shared context, Facts-only BB, Debate-3R+H, Debate-2R+H, BB+H, Debate-2R, BB},
  xlabel={margin over pass@$n$}]
\fill[famFact!8] (axis cs:0,0.4) rectangle (axis cs:0.48,10.6);
\draw[fgN!55, dashed, line width=0.5pt] (axis cs:0,0.4) -- (axis cs:0,10.6);
\addplot[only marks, mark=*, mark size=2.2pt, color=famFact, error bars/.cd, x dir=both, x explicit, error bar style={famFact!60, line width=0.7pt}]
  coordinates {(0.328,1) +- (0.096,0) (0.310,2) +- (0.096,0) (0.234,3) +- (0.098,0) (0.190,4) +- (0.094,0) (0.137,5) +- (0.098,0)};
\addplot[only marks, mark=*, mark size=2.2pt, color=famArgue, error bars/.cd, x dir=both, x explicit, error bar style={famArgue!60, line width=0.7pt}]
  coordinates {(0.109,6) +- (0.092,0)};
\addplot[only marks, mark=o, mark size=2.2pt, mark options={line width=0.9pt}, color=famArgue, error bars/.cd, x dir=both, x explicit, error bar style={famArgue!60, line width=0.7pt}]
  coordinates {(0.079,7) +- (0.080,0) (0.043,8) +- (0.094,0) (-0.015,9) +- (0.076,0) (-0.331,10) +- (0.086,0)};
\node[key, anchor=south] at (axis cs:0.015,10.7) {HiddenBench\enspace Qwen3.8-27B};
\end{axis}
\node[qk, text=famArgue] at (7.9,0.05) {$\bullet$ Debate};
\node[qk, text=famFact] at (9.7,0.05) {$\bullet$ Evidence};

\node[qt] at (11.6,4.88) {(c) Shared information};
\begin{axis}[ladderax, at={(13.3cm,1.0cm)}, anchor=south west, width=3.55cm, height=3.2cm,
  xmin=-0.55, xmax=0.2, ymin=0.4, ymax=6.6, xtick={-0.4,-0.2,0,0.2},
  xticklabels={\textminus0.4,\textminus0.2,0,0.2},
  ytick={1,...,6}, yticklabels={FATE-M, SWE-bench V., MMLongBench, LongBench v2, BrowseComp+, GPQA-D},
  xlabel={margin over the vote}]
\fill[fgP] (axis cs:-0.55,0.4) rectangle (axis cs:0,6.6);
\draw[fgN!55, dashed, line width=0.5pt] (axis cs:0,0.4) -- (axis cs:0,6.6);
\addplot[only marks, mark=*, mark size=1.8pt, color=famArgue, fill opacity=0.85]
  coordinates {(-0.005,6.18) (0.008,6.18) (0.022,6.18) (-0.055,5.18) (-0.178,5.18) (-0.163,5.18) (-0.016,4.18) (-0.016,3.18) (-0.173,2.18) (-0.057,1.18) (-0.045,1.18)};
\addplot[only marks, mark=square*, mark size=1.7pt, color=fgD, fill opacity=0.85]
  coordinates {(0.022,6) (0.015,6) (0.022,6) (-0.002,5) (-0.243,5) (-0.066,3) (-0.506,2) (-0.235,1) (-0.304,1)};
\addplot[only marks, mark=triangle*, mark size=2.1pt, color=fgY]
  coordinates {(0.021,5.82) (-0.075,4.82) (-0.112,2.82) (-0.360,1.82)};
\addplot[only marks, mark=star, mark size=2.6pt, mark options={line width=1pt}, color=famFact]
  coordinates {(0.050,6) (0.131,5) (0.037,4) (0.044,3)};
\node[key, anchor=south] at (axis cs:-0.175,6.75) {six benchmarks\enspace Qwen3.5-9B};
\end{axis}
\node[qk, text=famArgue] at (12.05,0.05) {$\bullet$ debate};
\node[qk, text=fgD] at (13.2,0.05) {$\blacksquare$ DeLM};
\node[qk, text=fgY] at (14.6,0.05) {$\blacktriangle$ AOrchestra};
\node[qk, text=famFact] at (16.35,0.05) {$\star$ BB+reviewer};
\end{tikzpicture}
\endgroup
\caption{\textbf{Protocol families and their positions on the ladder.} \textbf{(a)} We group protocols into seven families by what members exchange, and each family keeps its colour throughout. BB is a blackboard, a message board shared by all members. SC is shared context, which attaches every member's facts to the record. Vote@B and Pass@B are the majority vote and pass@$n$ of independent agents at the group's token budget. \textbf{(b)} With split facts, \famtag{Fact} protocols rise above pass@$n$. Filled markers reach level 3, bars show 95\% intervals, and +H marks the hint that facts are distributed. \textbf{(c)} With shared information, debate, DeLM and AOrchestra never outperform the vote, on text or on the page images of MMLongBench-Doc. Stars mark our blackboard with a 27B reviewer that selects among the 9B members' answers, which outperforms the vote on all four benchmarks where it was run.}
\label{fig:ladder}
\end{figure*}

%% file: subtex/experiment.tex
\section{When Is a Group Smarter Than Its Members Working Apart?}
\label{sec:exp-when}

\subsection{Experimental setup}
\label{sec:exp-design}

\textbf{Models.} We evaluate four Qwen models (Qwen3.5-4B, Qwen3.5-9B, Qwen3.6-35B-A3B and Qwen3.8-27B~\citep{yang2025qwen3,qwen2026qwen35}) and two models from other families (GLM-4.5-Air~\citep{glm2025glm45} and gpt-oss-120b~\citep{openai2025gptoss}). All models are served with vLLM~\citep{kwon2023vllm}, and each group uses one model. \textbf{Evaluation modes.} The three modes differ in what each member sees. In the \emph{shared-information} mode, every member sees the full question; we use GPQA-Diamond~\citep{rein2024gpqa}, LongBench v2~\citep{bai2025longbenchv2}, MMLongBench-Doc~\citep{ma2024mmlongbenchdoc}, BrowseComp-Plus~\citep{chen2025browsecompplus}, SWE-bench Verified~\citep{jimenez2024swebench} and FATE-M~\citep{jiang2025fate}. In the \emph{split-information} mode, the decisive facts are spread across members. HiddenBench poses 65 such decisions~\citep{li2025hiddenbench}. The ladder of Table~\ref{tab:ladder} and the free-budget organization experiments use all 65 tasks, while the prompt and budget-cap experiments, and the free-budget references they are compared with, use the 56 tasks with exactly four private facts (Appendix~\ref*{app:protocols}). We also build SplitBench in the same format. It keeps the 697 of 1{,}500 tasks written by Qwen3.8-27B that members can solve only by pooling facts, with 501 for training and 196 held out (Appendix~\ref*{app:audit}). In the \emph{shared-corpus} mode, a contradiction audit and DocSweep spread the answer over many documents that every member can open (Appendix~\ref*{app:audit}).

\input{tables/main_ladder}
\input{figures/qualitative}

\subsection{Prior multi-agent methods do not outperform majority voting under shared information}
\label{sec:exp-symmetric}
\label{sec:exp-docsweep}

Figure~\ref{fig:ladder}c compares three prior methods with the compute-matched majority vote on six shared-information benchmarks. With Qwen3.5-9B agents, we re-implement multi-agent debate~\citep{du2024debate}, DeLM, in which agents share a blackboard~\citep{mao2026delm}, and AOrchestra, in which a coordinator delegates subtasks~\citep{ruan2026aorchestra}. Table~\ref{tab:shared} further evaluates groups of four Qwen3.8-27B agents on GPQA-Diamond and DocSweep. \looseness=-1

All three methods stay at or below the compute-matched vote on every benchmark, so they remain below level 2. Since every agent already holds the evidence, a message mostly repeats what its reader knows. This also holds for page images: on MMLongBench-Doc, debate performs on par with the vote, and DeLM and AOrchestra fall below it (Appendix Table~\ref*{tab:prior}). On GPQA-Diamond, the 27B blackboard and debate remain at level 0 (Table~\ref{tab:shared}). As in Figure~\ref{fig:qualitative}a, confident agents outvote the one correct agent. \looseness=-1

DocSweep shows what changes this outcome. Its answer is spread over a shared corpus that no agent can read in full. Self-organized agents open overlapping documents, so they outperform the vote but not pass@$n$. A self-written prompt or the designed aggregator instead assigns documents to readers, and these groups exceed pass@$n$ (Table~\ref{tab:shared}). When every agent sees the whole question, prior methods therefore stay at the vote, and only a reviewer that selects among the members' answers rises above it (Appendix Table~\ref*{tab:prior}). On a shared corpus, dividing the reading helps, and organized groups exceed pass@$n$. \looseness=-1

\subsection{Forwarding split facts reaches level 3}
\label{sec:exp-split}

Table~\ref{tab:ladder} and Figure~\ref{fig:ladder}b report results on HiddenBench. Each of four Qwen3.8-27B members holds part of the decisive facts, and the shared facts point to a decoy. An agent working alone therefore follows the decoy, so pass@$n$ stays low and a group that pools the facts can exceed it. For the same reason, level 2 is an easy bar: agents who each hold only part of the facts mostly vote for the decoy. Every \famtag{Fact} protocol reaches level 3. The \famtag{Argue} protocols mostly outperform the vote and remain at level 2. Their messages carry conclusions drawn from partial facts, and readers adopt these conclusions instead of the facts they lack. A third round with the hint that facts are split lifts debate to level 3 at a higher budget. Fact forwarding reaches level 3 at 9B, 27B and 35B-A3B. At 27B, four agents who each hold part of the facts are even more accurate than an equal-budget full-information agent (0.959 vs.\ 0.942, Table~\ref{tab:ladder}). Forwarding facts also outperforms shared context, where the comments attached to the facts carry arguments that mislead even an agent holding every fact (Figure~\ref{fig:qualitative}b). \looseness=-1

\section{Emergent Organization and Its Coordination Cost}
\label{sec:exp-emergence}

\input{tables/main_emergence}

\subsection{Hubs and relays built by identical agents}
\label{sec:exp-hubs}
\label{sec:exp-relay}

\input{figures/qual_structure}
\input{figures/qual_family}
We call a structure emergent when agents build it themselves. In Table~\ref{tab:emergence}, four identical Qwen3.8-27B agents on HiddenBench may each read one message, and each agent may transfer this reading budget to another agent. Panel (a) evaluates three settings: a free budget, self-written prompts, and a cap that prevents any agent from reading every fact. Figure~\ref{fig:structure} follows individual records.

Under the free budget, most agents give their budget away. In about a third of the records, one agent collects enough budget to read every fact and becomes a hub. The hub answers like one agent given every fact, so accuracy rises by 18 percentage points over a group without transfers. A planning round strengthens this behaviour and brings the group close to a designed hub. Under the cap, agents build relays instead, mostly two-level trees (Figure~\ref{fig:structure}b), and these relays outperform a group that may forward only its own facts. All these structures outperform the vote and remain at level 2. Their budgets explain why they climb no higher. A self-organized group spends the tokens of about 21 independent agents, twice the budget of a designed hub, because its agents wait for the hub and negotiate transfers (Appendix Table~\ref*{tab:channels}b). \looseness=-1

\subsection{Accuracy, cost and coordination overhead}
\label{sec:exp-price}

\input{figures/channels}

Figure~\ref{fig:channels} applies Equation~(\ref{eq:channels}) to every protocol family and compares each protocol with a reference on the same pool (Appendix Table~\ref*{tab:channels}). Its vertical axis is the counterfactual margin $Y_X-T(n_R)$, the margin at the reference's budget. \famtag{Fact} protocols remain confirmed above pass@$n$ even at the budget of their reference protocols, so accuracy alone explains their gain. \famtag{Design} and \famtag{Notes} organizations also gain accuracy and exceed pass@$n$ only at their own lower budget, so they need both higher accuracy and lower cost. \famtag{Self} structures gain accuracy over their references and stay below a confirmed margin over pass@$n$ at either budget. A self-organizing group spends about two thirds of its tokens on coordination (Appendix Table~\ref*{tab:channels}b). In the free-budget group, most of it goes to negotiating transfers, and the rest to waiting for the hub. Without this waiting, the hub would exceed pass@$n$. \looseness=-1

\label{sec:exp-general}\textbf{Across model families.} Among Qwen models, only 27B forms hubs; GLM-4.5-Air and gpt-oss-120b agents also rarely form them, and allowing transfers does not change either family's accuracy detectably (Table~\ref{tab:emergence}b); Figure~\ref{fig:family} shows one record of each family with a planning round.

\section{Can a Group Improve Itself?}
\label{sec:exp-rsi}

\subsection{Self-written prompts at test time}
\label{sec:exp-notes}

At test time, the model writes a prompt for its own group. It either writes the prompt from the rules or evolves it over four generations on unlabeled records. The prompt lifts the 27B group from level 2 to level 3 (descriptively, Table~\ref{tab:emergence}a, Figure~\ref{fig:qualitative}d). Even the prompt written from the rules describes the same hub as the evolved prompts, and a group following it spends less than half the tokens per record of the self-organized group. Experience helps where the rules leave a choice. Under the budget cap, an evolved prompt raises the margin of the rules-only prompt by 0.16. A supervisor with the same weights can also rewrite the 27B team from unlabeled records, which adds 0.11 (Appendix~\ref*{app:recursion}). \looseness=-1

\subsection{Training on verified turns}
\label{sec:exp-weights}

\input{tables/recursion}

Table~\ref{tab:recursion} reports the label-free generations. Each \famtag{Train} generation trains on its parent group's protocol-verified turns on 501 SplitBench tasks and is graded on 196 held-out tasks.

The first generation improves the group at every scale. It raises the margin by 0.225 at 9B and 0.421 at 4B, and it lifts the 27B margin from $-0.009$ to $+0.201$. A second generation climbs further at all three scales. At 9B, the second generation keeps the shortest passing turn, which halves the first call. The group then rises above pass@$n$ and surpasses the control by 0.192 on the first training seed. Pooled over four further seeds, the gain over the control is 0.115 and positive in every pair, with accuracy carrying the gain (Appendix~\ref*{app:recursion}). All three conditions of recursion therefore hold at 9B. At 27B, the second step comes from cost: the budget falls from 9.6 to 6.4 agents while accuracy stays at its parent's level. Both steps also hold in the standard debate and on the official HiddenBench tasks (Appendix~\ref*{app:recursion}). \looseness=-1

The gain belongs to the group (Appendix Table~\ref*{tab:collective}). In the standard debate, the first step leaves trained 9B agents as accurate alone as before, while their group gains 0.177 on the tasks that members of both generations solve alone. In the 3-step debate, the second generation stays above pass@$n$ of its own members.

\subsection{Training on the group's outcome}
\label{sec:exp-credit}

\input{tables/credit}

When a task scores only the group's joint deliverable, the credit rule determines what the shared policy learns. Like the ladder's budget, it must charge each agent for the claims it adds. Table~\ref{tab:credit} tests this on the contradiction audit, which scores the group's claimed contradictory pairs by recall, precision and F1. Rewarding individual success or the team makes agents report more pairs and collapses the group's precision. Proposition~\ref{prop:threshold} explains why: an agent's own score rewards low-confidence claims, while the group pays for their errors. Shapley credit instead charges each wrong claim to its author. It alone raises held-out precision above the untrained group, and it is best on every metric. \looseness=-1

\subsection{Recursion on PlanBench, HiddenBench and Hanabi}
\label{sec:exp-chain}

Recursion should continue beyond two generations and hold across different tasks. We therefore apply the gated recursion of Section~\ref{sec:chain-protocol} to three benchmarks that test different abilities: plan generation in PlanBench~\citep{valmeekam2023planbench}, pooling of split facts in HiddenBench~\citep{li2025hiddenbench} and cooperation under hidden information in Hanabi~\citep{bard2020hanabi}. Every run starts from four Qwen3.5-9B agents, and each generation trains on its predecessor's outputs that a plan simulator, agreement with the base model's full-information answer or the game score accepts, without human-written answer labels. On all three benchmarks, every released generation outperforms its predecessor on unseen confirmation tasks, for four to six successive generations (Figure~\ref{fig:chains}). \looseness=-1

\input{figures/chains}

\paragraph{PlanBench.} All four rounds release a generation, and the group's accuracy on the selection set rises from 0.613 to 0.950 (Table~\ref{tab:bench-planbench}). The agents improve together with the group: a single agent's first draft and final plan both improve in every generation, and from G1 on, the group commits a valid plan on nearly every task where any of its four agents finds one. The confirmed gain narrows as accuracy approaches its ceiling. Because fewer calls reach the token cap in later generations, part of the rise may come from shorter thinking; the table therefore reports the tokens of every generation beside its accuracy.

\input{tables/bench_planbench}

\paragraph{HiddenBench.} Two runs, T3 and T3k, differ in how they build their training data and in the step sizes of their candidates (Appendix~\ref*{app:chains}). Collective training raises the group from 0.509 to 0.886 over five generations of T3k and to 0.836 over three generations of T3 (Figure~\ref{fig:bench-hiddenbench}a). Single-agent self-training under the same caps also improves, and the collective stays far above four agents trained this way in every generation. When facts are split across agents, the group thus learns what its agents cannot learn alone. For reference, Figure~\ref{fig:bench-hiddenbench}b shows the protocols that we evaluated with the same base model on the 65 official tasks, and Figure~\ref{fig:bench-hiddenbench}c the groups reported by the benchmark's authors. \looseness=-1

\input{figures/bench_hiddenbench}

\paragraph{Hanabi.} The base agents lose all their lives in every game and score zero. Six generations of recursive training each outperform their predecessor on fresh deals, and the challenge score rises to 4.5 out of 25 (Table~\ref{tab:bench-hanabi}). A control run that trains on games of a fixed generation instead of its own improves for two generations and stays near 2.9, well below recursive training. Early candidates first lost fewer lives by simply playing fewer cards, because the training filter had removed every uncertain play. \looseness=-1

\input{tables/bench_hanabi}

\subsection{HiddenBench generations on the equal-budget ladder}
\label{sec:exp-hb-ladder}

\input{figures/hb_ladder}

Accuracy alone shows that each generation improves on its predecessor. To see what this improvement means at equal cost, we place every HiddenBench generation on the ladder (Figure~\ref{fig:hb-ladder}). For each generation, 24 independent agents of that generation's own model answer every one of the 200 selection tasks, six seeds for each of the four seats, and each agent holds the shared facts and one seat's private facts, just as a member of the group does. The group is then graded against these agents at its own budget.

The base group beats one agent and the vote of its independent agents, so it starts at level 2, and it lies 0.15 below pass@$n_B$. From the first trained generation on, every generation exceeds pass@$n_B$ and reaches level 3. The margin grows from $+0.06$ at G1 to $+0.23$ at G5, while the budget falls from 13.9 to 7.9 independent agents. Pass@$n_B$ of the trained agents stays near 0.65 in every generation, and an agent working alone stays below 0.18, so the trained agents still need each other's facts. Recursive training therefore climbs the ladder through interaction: each generation learns to pool what its members hold, at a lower cost than the base group. \looseness=-1

Four properties of this reading frame it. Because each independent agent holds only a quarter of the private facts, level 3 here measures how well the group pools distributed facts. The reference answers are the base model's full-information consensus, as in training, so a full-information agent would grade itself against its own answers and cannot serve as a reference. The 200 tasks form the selection set used during training, so the group's accuracy is optimistic, whereas the independent agents run fresh. Later generations often answer alone in an unparseable form, which this reading counts as wrong; letting these answers abstain raises the vote at G3 to G5 and leaves every level unchanged. \looseness=-1

%% file: tables/main_ladder.tex
\begin{table}[t]
\centering
\tsz
\setlength{\tabcolsep}{24pt}
\renewcommand{\arraystretch}{1.08}
\caption{\textbf{Shared information and a shared corpus.} Four Qwen3.8-27B agents. $n_B$ counts the agents that a group's tokens pay for, and the margin is accuracy minus pass@$n_B$. Levels, tested in order, beat one agent, the majority vote and pass@$n$ by a confirmed margin; level 0 does not beat one agent.}
\label{tab:shared}
\centering
\begin{tabular}{@{}lcccc@{}}
\toprule
\textbf{Protocol} & Acc. & $n_B$ & Margin & Lvl. \\
\midrule
\tgroup{5}{GPQA-Diamond: the same question for all} \\
Blackboard & 0.909 & 4.22 & $-0.039$ & \Lv{0} \\
Debate, 2 rounds & 0.899 & 4.24 & $-0.049$ & \Lv{0} \\
\tgroup{5}{DocSweep: the answer is spread over a shared corpus} \\
Self-organized & 0.200 & 4.93 & $+0.026$ & \Lv{2} \\
Self-written prompt & 0.861 & 4.02 & $+0.706$ & \Lv{3} \\
Designed aggregator & \textbf{0.930} & 4.56 & $\mathbf{+0.759}$ & \Lv{3} \\
\bottomrule
\end{tabular}
\vspace{10pt}
\caption{\textbf{Results with split facts.} Four Qwen3.8-27B agents on HiddenBench, each holding part of the decisive facts; in rows with a hint, the agents are told that the facts are split. Passing on facts reaches level 3, and fact forwarding exceeds an equal-budget full-information agent.}
\label{tab:ladder}
\centering
\begin{tabular}{@{}lcccc@{}}
\toprule
\textbf{Protocol} & Acc. & $n_B$ & Margin & Lvl. \\
\midrule
\tgroup{5}{Debate: members exchange answers and reasons} \\
Blackboard & 0.369 & 8.6 & $-0.331$ & \Lv{2} \\
Debate, 2 rounds & 0.688 & 8.8 & $-0.015$ & \Lv{2} \\
Debate, 3 rounds + hint & 0.841 & 11.1 & $+0.109$ & \Lv{3} \\
\tgroup{5}{Evidence: members pass on their facts} \\
Shared context & 0.881 & 7.9 & $+0.190$ & \Lv{3} \\
\oursrow Fact forwarding & 0.959 & 6.2 & $+0.310$ & \Lv{3} \\
\oursrow Shared context + forwarding & \textbf{0.972} & 6.0 & $\mathbf{+0.328}$ & \Lv{3} \\
\midrule
\textit{Full-information agent, equal budget} & \textit{0.942} & & & \\
\bottomrule
\end{tabular}
\end{table}

%% file: figures/qualitative.tex
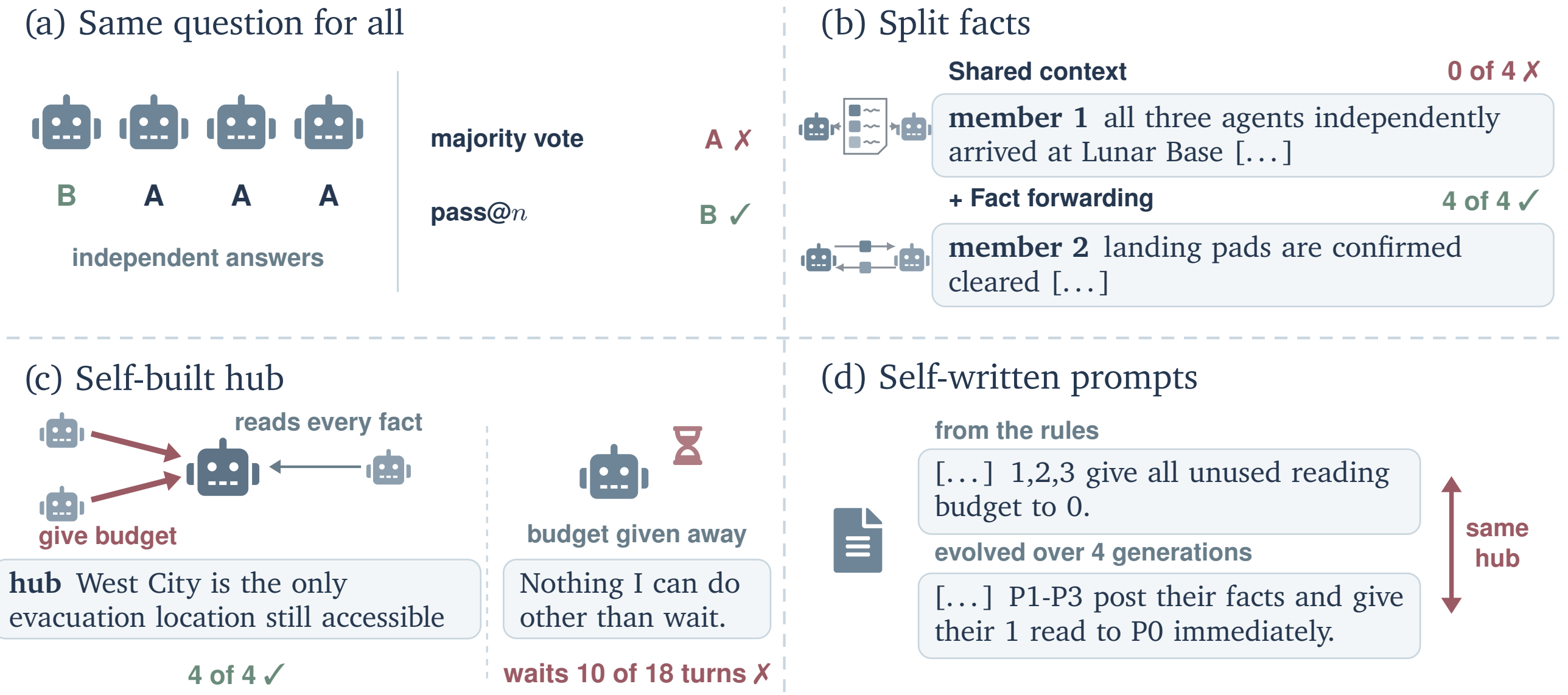
\begin{figure*}[!t]
\centering
\begingroup
\definecolor{fgA}{HTML}{6E8597}\definecolor{fgB}{HTML}{8D9FAE}\definecolor{fgC}{HTML}{AAB8C3}\definecolor{fgD}{HTML}{5E7385}
\definecolor{fgT}{HTML}{687D89}\definecolor{fgR}{HTML}{9A5A64}\definecolor{fgG}{HTML}{6A8B7B}\definecolor{fgN}{HTML}{25374F}
\definecolor{fgL}{HTML}{C9D6DF}\definecolor{fgP}{HTML}{F3F6F8}\definecolor{fgI}{HTML}{3A4656}\definecolor{fgY}{HTML}{8A919A}
\tikzset{
  qt/.style={font=\fontsize{11}{12}\selectfont, text=fgN, anchor=west},
  ql/.style={font=\sffamily\bfseries\fontsize{8.5}{9.5}\selectfont, text=fgT, align=center},
  qq/.style={font=\fontsize{9}{10.6}\selectfont, text=fgN, fill=fgP, draw=fgL, line width=0.5pt,
             rounded corners=4pt, inner xsep=5pt, inner ysep=3.5pt, align=left, text width=#1,
             execute at begin node={\hyphenpenalty=10000\exhyphenpenalty=10000}},
  qq/.default=4.4cm,
  card/.style={draw=none, fill=none},
  ico/.style={font=\fontsize{#1}{#1}\selectfont, inner sep=0pt},
  pill/.style={inner sep=0pt, font=\sffamily\bfseries\fontsize{9}{10}\selectfont, text=#1},
  ql/.append style={inner sep=0pt},
  tag/.style={rounded corners=2.5pt, inner xsep=4pt, inner ysep=2.2pt, fill=#1, text=white,
              font=\sffamily\bfseries\fontsize{8}{9}\selectfont, anchor=east},
  ga/.style={-{Triangle[length=5pt,width=6pt]}, line width=1.8pt, draw=fgR},
  ra/.style={-{Triangle[length=3.6pt,width=4.2pt]}, line width=0.8pt, draw=fgT},
  qarr/.style={-{Latex[length=2.8pt,width=2.8pt]}, fgY, line width=0.6pt},
}
\newcommand{\qbot}[3]{\node[ico=#3, text=#2] at (#1) {\faIcon{robot}};}
\newcommand{\qrb}[3]{\node[inner sep=0pt, text=#3, font=\fontsize{7.6}{7.6}\selectfont] at (#1,#2) {\faIcon{robot}};}
\newcommand{\qchip}[3]{\fill[#3, rounded corners=0.6pt] (#1-0.055,#2-0.055) rectangle (#1+0.055,#2+0.055);}
\newcommand{\qicoShared}[2]{%
  \qrb{#1-0.46}{#2}{fgA}\qrb{#1+0.46}{#2}{fgB}
  \draw[fill=white, draw=fgI!65, line width=0.55pt] (#1-0.2,#2-0.26) -- (#1+0.12,#2-0.26) -- (#1+0.2,#2-0.18) -- (#1+0.2,#2+0.26) -- (#1-0.2,#2+0.26) -- cycle;
  \foreach \dy/\c in {0.15/fgA, 0/fgB, -0.15/fgC} {\qchip{#1-0.1}{#2+\dy}{\c}\draw[fgY, line width=0.6pt, line cap=round, decorate, decoration={snake, amplitude=0.3pt, segment length=2.4pt}] (#1-0.01,#2+\dy) -- (#1+0.13,#2+\dy);}
  \draw[qarr] (#1-0.23,#2) -- (#1-0.33,#2);
  \draw[qarr] (#1+0.23,#2) -- (#1+0.33,#2);}
\newcommand{\qicoForward}[2]{%
  \qrb{#1-0.44}{#2}{fgA}\qrb{#1+0.44}{#2}{fgB}
  \draw[qarr] (#1-0.29,#2+0.1) -- (#1+0.29,#2+0.1); \qchip{#1}{#2+0.1}{fgA}
  \draw[qarr] (#1+0.29,#2-0.1) -- (#1-0.29,#2-0.1); \qchip{#1}{#2-0.1}{fgB}}
\newcommand{\qbig}[3]{\begin{scope}[shift={(#2,#3)}, scale=1.15, every node/.append style={scale=1.15}]#1{0}{0}\end{scope}}
\begin{tikzpicture}[x=1cm, y=1cm]

\begin{scope}[shift={(0,4.05)}]
\path[card] (0,0) rectangle (8.45,3.6);
\node[qt] at (0.15,3.3) {(a) Same question for all};
\foreach \x/\c/\a/\p in {0.75/fgA/B/fgG, 1.7/fgA/A/fgN, 2.65/fgA/A/fgN, 3.6/fgA/A/fgN} {
  \qbot{\x,2.25}{\c}{17}
  \node[pill=\p, font=\sffamily\bfseries\fontsize{10}{11}\selectfont] at (\x,1.45) {\a};}
\node[ql] at (2.18,0.75) {independent answers};
\draw[fgL, line width=0.6pt] (4.35,0.4) -- (4.35,2.8);
\node[ql, anchor=west, text=fgN] at (4.7,2.05) {majority vote};
\node[pill=fgR, anchor=east] at (8.2,2.05) {A \,\ding{55}};
\node[ql, anchor=west, text=fgN] at (4.7,1.25) {pass@$n$};
\node[pill=fgG, anchor=east] at (8.2,1.25) {B \,\ding{51}};
\end{scope}

\begin{scope}[shift={(8.65,4.05)}]
\path[card] (0,0) rectangle (8.45,3.6);
\node[qt] at (0.15,3.3) {(b) Split facts};
\qbig{\qicoShared}{0.78}{2.2}
\node[ql, anchor=west, text=fgN] at (1.68,2.8) {Shared context};
\node[qq=6.4cm, anchor=north west] at (1.5,2.56) {\textbf{\upshape member 1}\enspace all three agents independently arrived at Lunar Base [\ldots]};
\node[pill=fgR, anchor=east] at (8.12,2.8) {0 of 4 \ding{55}};
\qbig{\qicoForward}{0.78}{0.78}
\node[ql, anchor=west, text=fgN] at (1.68,1.4) {+ Fact forwarding};
\node[qq=6.4cm, anchor=north west] at (1.5,1.15) {\textbf{\upshape member 2}\enspace landing pads are confirmed cleared~[\ldots]};
\node[pill=fgG, anchor=east] at (8.12,1.4) {4 of 4 \ding{51}};
\end{scope}

\path[card] (0,0) rectangle (8.45,3.85);
\node[qt] at (0.15,3.5) {(c) Self-built hub};
\qbot{2.45,2.55}{fgD}{18}
\qbot{0.7,2.95}{fgB}{11}\qbot{0.7,2.15}{fgB}{11}\qbot{4.25,2.55}{fgB}{11}
\draw[ga] (1.02,2.91) -- (2.0,2.66); \draw[ga] (1.02,2.19) -- (2.0,2.44);
\node[ql, text=fgR] at (1.2,1.78) {give budget};
\draw[ra] (3.95,2.55) -- (2.95,2.55);
\node[ql] at (3.6,3.02) {reads every fact};
\node[qq=4.95cm, anchor=north] at (2.6,1.55) {\textbf{\upshape hub}\enspace West City is the only evacuation location still accessible};
\node[pill=fgG] at (2.6,0.3) {4 of 4 \ding{51}};
\draw[fgL, line width=0.6pt, dash pattern=on 2pt off 2pt] (5.32,0.25) -- (5.32,3.05);
\qbot{6.7,2.5}{fgA}{17}
\node[ico=12, text=fgR!80] at (7.5,2.78) {\faIcon{hourglass-half}};
\node[ql] at (6.95,1.8) {budget given away};
\node[qq=2.55cm, anchor=north] at (6.95,1.55) {Nothing I can do other than wait.};
\node[pill=fgR] at (6.95,0.3) {waits 10 of 18 turns \ding{55}};

\begin{scope}[shift={(8.65,0)}]
\path[card] (0,0) rectangle (8.45,3.85);
\node[qt] at (0.15,3.5) {(d) Self-written prompts};
\node[ico=20, text=fgA] at (0.7,1.75) {\faIcon{file-alt}};
\node[ql, anchor=west] at (1.53,2.95) {from the rules};
\node[qq=5.1cm, anchor=north west] at (1.35,2.75) {[\ldots] 1,2,3 give all unused reading budget to 0.};
\node[ql, anchor=west] at (1.53,1.6) {evolved over 4 generations};
\node[qq=5.1cm, anchor=north west] at (1.35,1.4) {[\ldots] P1-P3 post their facts and give their 1 read to P0 immediately.};
\draw[fgR, line width=1.4pt, {Triangle[length=5pt,width=6pt]}-{Triangle[length=5pt,width=6pt]}] (7.15,2.45) -- (7.15,0.95);
\node[ql, text=fgR, anchor=west] at (7.3,1.7) {same\\hub};
\end{scope}
\draw[fgL, dashed, line width=0.8pt, dash pattern=on 4pt off 3pt] (8.55,0.1) -- (8.55,7.55);
\draw[fgL, dashed, line width=0.8pt, dash pattern=on 4pt off 3pt] (0.1,3.95) -- (17.0,3.95);
\end{tikzpicture}
\endgroup
\caption{\textbf{Qualitative results}, Qwen3.8-27B records. \textbf{(a)} The majority vote overrules the one member that answers correctly. \textbf{(b)} Under shared context, an argument misleads even the member that holds every fact, whereas forwarded facts lead all four members to the answer. \textbf{(c)} Members hand their reading budget to a hub, and a member of another group then waits for its own hub. \textbf{(d)} A prompt written from the rules alone already describes the same hub. Boxed text is quoted verbatim.}
\label{fig:qualitative}
\end{figure*}

%% file: tables/main_emergence.tex
\begin{table}[t]
\centering
\tsz
\setlength{\tabcolsep}{4pt}
\renewcommand{\arraystretch}{1.08}
\caption{\textbf{Emergent organization.} Identical Qwen3.8-27B agents on HiddenBench, each reading one message. \textbf{(a)} Members may give their budget away (free budget, all 65 tasks), or each budget is capped so that no single hub can form (budget cap, the 56 tasks with four private facts). The prompt rows of the free budget run on the 56 tasks with a pool from another batch, so their levels are descriptive; the tree's level holds before multiple-testing correction. \textbf{(b)} Records with a hub or a transfer without a planning round, and the accuracy change from allowing transfers, on the 65 tasks.}
\label{tab:emergence}
\label{tab:general}
\begin{tabular}[t]{@{}llcccccccc@{}}
\toprule
&& \multicolumn{4}{c}{\textbf{Free budget}} & \multicolumn{4}{c}{\textbf{Budget cap}} \\
\cmidrule(lr){3-6}\cmidrule(l){7-10}
\multicolumn{2}{@{}l}{\textbf{(a) Organized by}} & Acc. & $n_B$ & Margin & Lvl. & Acc. & $n_B$ & Margin & Lvl. \\
\midrule
\multirow{3}{*}{Members} & No transfer or relay & 0.496 & 20.3 & $-0.300$ & \Lv{2} & 0.571 & 37.0 & $-0.267$ & \Lv{2} \\
 & Self-organized & 0.677 & 21.3 & $-0.124$ & \Lv{2} & 0.689 & 40.7 & $-0.159$ & \Lv{2} \\
 & \quad + planning round & 0.822 & 18.3 & $+0.037$ & \Lv{2} & 0.677 & 44.4 & $-0.180$ & \Lv{2} \\[2pt]
\multirow{2}{*}{Prompt} & From the rules & \textbf{0.879} & 9.2 & $\mathbf{+0.205}$ & \Lv{3} & 0.732 & 40.2 & $-0.115$ & \Lv{2} \\
 & Evolved & 0.862 & 9.7 & $+0.180$ & \Lv{3} & \textbf{0.840} & 25.0 & $+0.043$ & \Lv{2} \\[2pt]
Design & Hub / tree & 0.863 & 11.3 & $+0.130$ & \Lv{3} & 0.802 & 10.8 & $\mathbf{+0.107}$ & \Lv{3} \\
\bottomrule
\end{tabular}\hspace{2.2em}%
\begin{tabular}[t]{@{}lccc@{}}
\toprule
& & & \\
\cmidrule{2-4}
\textbf{(b) Model} & Hubs & Transfers & Gain \\
\midrule
Qwen3.5-4B & 0.000 & 0.024 & $+0.027$ \\
Qwen3.5-9B & 0.000 & 0.008 & $-0.001$ \\
Qwen3.6-35B-A3B & 0.000 & 0.004 & $+0.025$ \\
\textbf{Qwen3.8-27B} & \textbf{0.345} & \textbf{0.611} & $\mathbf{+0.181}$ \\
GLM-4.5-Air & 0.000 & 0.017 & $+0.025$ \\
gpt-oss-120b & 0.045 & 0.372 & $+0.036$ \\
\bottomrule
\end{tabular}
\end{table}

%% file: figures/qual_structure.tex
\begin{figure*}[t]
\centering
\begingroup
\definecolor{qfA}{HTML}{8BAA98}\definecolor{qfB}{HTML}{A9C4DC}\definecolor{qfC}{HTML}{A79BB5}\definecolor{qfD}{HTML}{6E8597}
\definecolor{qbg}{HTML}{FFFFFF}\definecolor{qln}{HTML}{C9D6DF}\definecolor{qan}{HTML}{25374F}\definecolor{qik}{HTML}{25374F}\definecolor{qik2}{HTML}{687D89}
\definecolor{qrd}{HTML}{8E9FAE}\definecolor{qgv}{HTML}{8B3344}\definecolor{qwt}{HTML}{9AA3AB}
\tikzset{
  qh/.style={font=\fontsize{11}{12}\selectfont, text=qik, align=left},
  qs/.style={font=\sffamily\bfseries\fontsize{8.8}{9.8}\selectfont, text=qik2},
  qlane/.style={draw=qln, line width=0.5pt},
  qread/.style={draw=qrd, line width=0.6pt, -{Stealth[length=2.4pt]}},
  qgive/.style={draw=qgv, line width=1.5pt, -{Triangle[length=3.6pt,width=4.6pt]}},
  qbc/.style={draw=qan!40, line width=0.5pt, -{Stealth[length=2.2pt]}},
  qrel/.style={draw=qan},
  qeye/.style={circle, fill=qrd, inner sep=0pt, minimum size=3.2pt},
  qcoin/.style={circle, fill=qgv, inner sep=0pt, minimum size=4.2pt},
  qwait/.style={circle, draw=qwt, line width=0.6pt, inner sep=0pt, minimum size=4.2pt},
}
\newcommand{\qok}{\textcolor{sage!80!black}{\ding{51}}}
\newcommand{\qno}{\textcolor{BrickRed!70!black}{\ding{55}}}
\begin{tikzpicture}[x=1cm, y=1cm]
\fill[qbg, draw=qln, line width=0.7pt, rounded corners=6pt] (0,3.75) rectangle (8.45,7.35);
\node[qh, anchor=north west, text width=8.15cm] at (0.15,7.25) {(a) Hub};
\draw[qlane] (1.750,6.500) -- (8.250,6.500);
\node[qs, anchor=east] at (1.530,6.500) {member 0};
\draw[qlane] (1.750,5.950) -- (8.250,5.950);
\node[qs, anchor=east] at (1.530,5.950) {member 1};
\draw[qlane] (1.750,5.400) -- (8.250,5.400);
\node[qs, anchor=east] at (1.530,5.400) {member 2};
\draw[qlane] (1.750,4.850) -- (8.250,4.850);
\node[qs, anchor=east] at (1.530,4.850) {member 3};
\node[qs, text=qik2] at (2.000,4.430) {0};
\node[qs, text=qik2] at (3.300,4.430) {1};
\node[qs, text=qik2] at (4.600,4.430) {2};
\node[qs, text=qik2] at (5.900,4.430) {3};
\node[qs, text=qik2] at (7.200,4.430) {4};
\node[qs, text=qik2, anchor=east] at (1.530,4.430) {step};
\fill[qfA, rounded corners=0.6pt] (1.945,6.445) rectangle ++(0.11,0.11);
\draw[qread] (2.100,4.850) to[out=0,in=180] (4.480,6.500);
\draw[qread] (2.100,5.400) to[out=0,in=180] (4.480,6.500);
\draw[qread] (2.100,5.950) to[out=0,in=180] (4.480,6.500);
\node[qeye] at (4.600,6.500) {};
\fill[qan, rounded corners=1.2pt] (5.540,6.380) rectangle (6.260,6.620);
\fill[qfA, rounded corners=0.6pt] (5.633,6.450) rectangle ++(0.1,0.1);
\fill[qfB, rounded corners=0.6pt] (5.777,6.450) rectangle ++(0.1,0.1);
\fill[qfC, rounded corners=0.6pt] (5.923,6.450) rectangle ++(0.1,0.1);
\fill[qfD, rounded corners=0.6pt] (6.068,6.450) rectangle ++(0.1,0.1);
\node[font=\fontsize{10}{10}\selectfont] at (7.220,6.500) {\qok};
\fill[qfB, rounded corners=0.6pt] (1.945,5.895) rectangle ++(0.11,0.11);
\draw[qgive] (3.300,5.950) -- (3.300,6.400);
\node[qcoin] at (3.300,5.950) {};
\node[qwait] at (4.600,5.950) {};
\draw[qbc] (7.150,6.500) -- (7.150,5.950);
\node[font=\fontsize{10}{10}\selectfont] at (7.220,5.950) {\qok};
\fill[qfC, rounded corners=0.6pt] (1.945,5.345) rectangle ++(0.11,0.11);
\draw[qgive] (3.300,5.400) -- (3.300,6.400);
\node[qcoin] at (3.300,5.400) {};
\node[qwait] at (4.600,5.400) {};
\draw[qbc] (7.150,6.500) -- (7.150,5.400);
\node[font=\fontsize{10}{10}\selectfont] at (7.220,5.400) {\qok};
\fill[qfD, rounded corners=0.6pt] (1.945,4.795) rectangle ++(0.11,0.11);
\draw[qread] (2.100,6.500) to[out=0,in=180] (3.180,4.850);
\node[qeye] at (3.300,4.850) {};
\node[qwait] at (5.900,4.850) {};
\draw[qbc] (7.150,6.500) -- (7.150,4.850);
\node[font=\fontsize{10}{10}\selectfont] at (7.220,4.850) {\qok};
\fill[qbg, draw=qln, line width=0.7pt, rounded corners=6pt] (8.65,3.75) rectangle (17.1,7.35);
\node[qh, anchor=north west, text width=8.15cm] at (8.8,7.25) {(b) Relay};
\draw[qlane] (10.400,6.500) -- (16.900,6.500);
\node[qs, anchor=east] at (10.180,6.500) {member 0};
\draw[qlane] (10.400,5.950) -- (16.900,5.950);
\node[qs, anchor=east] at (10.180,5.950) {member 1};
\draw[qlane] (10.400,5.400) -- (16.900,5.400);
\node[qs, anchor=east] at (10.180,5.400) {member 2};
\draw[qlane] (10.400,4.850) -- (16.900,4.850);
\node[qs, anchor=east] at (10.180,4.850) {member 3};
\node[qs, text=qik2] at (10.650,4.430) {0};
\node[qs, text=qik2] at (11.630,4.430) {1};
\node[qs, text=qik2] at (12.610,4.430) {2};
\node[qs, text=qik2] at (13.590,4.430) {3};
\node[qs, text=qik2] at (14.570,4.430) {4};
\node[qs, text=qik2] at (15.550,4.430) {5};
\node[qs, text=qik2] at (16.530,4.430) {6};
\node[qs, text=qik2, anchor=east] at (10.180,4.430) {step};
\fill[qfA, rounded corners=0.6pt] (10.595,6.445) rectangle ++(0.11,0.11);
\node[qwait] at (12.610,6.500) {};
\draw[qread] (10.750,5.950) to[out=0,in=180] (14.450,6.500);
\node[qeye] at (14.570,6.500) {};
\node[qwait] at (15.550,6.500) {};
\draw[qbc] (16.480,4.850) -- (16.480,6.500);
\node[font=\fontsize{10}{10}\selectfont] at (16.550,6.500) {\qok};
\fill[qfB, rounded corners=0.6pt] (10.595,5.895) rectangle ++(0.11,0.11);
\draw[qgive] (11.630,5.950) -- (11.630,5.500);
\node[qcoin] at (11.630,5.950) {};
\node[qwait] at (13.590,5.950) {};
\node[qwait] at (14.570,5.950) {};
\node[qwait] at (15.550,5.950) {};
\draw[qbc] (16.480,4.850) -- (16.480,5.950);
\node[font=\fontsize{10}{10}\selectfont] at (16.550,5.950) {\qok};
\fill[qfC, rounded corners=0.6pt] (10.595,5.345) rectangle ++(0.11,0.11);
\draw[qread] (10.750,6.500) to[out=0,in=180] (12.490,5.400);
\draw[qread] (10.750,5.950) to[out=0,in=180] (12.490,5.400);
\node[qeye] at (12.610,5.400) {};
\draw[qrel, line width=0.5pt, rounded corners=1pt] (13.365,5.305) rectangle (13.815,5.495);
\fill[qfA, rounded corners=0.6pt] (13.405,5.345) rectangle ++(0.11,0.11);
\fill[qfB, rounded corners=0.6pt] (13.535,5.345) rectangle ++(0.11,0.11);
\fill[qfC, rounded corners=0.6pt] (13.665,5.345) rectangle ++(0.11,0.11);
\node[qwait] at (14.570,5.400) {};
\node[qwait] at (15.550,5.400) {};
\draw[qbc] (16.480,4.850) -- (16.480,5.400);
\node[font=\fontsize{10}{10}\selectfont] at (16.550,5.400) {\qok};
\fill[qfD, rounded corners=0.6pt] (10.595,4.795) rectangle ++(0.11,0.11);
\node[qwait] at (11.630,4.850) {};
\draw[qread] (13.690,5.400) to[out=0,in=180] (14.450,4.850);
\node[qeye] at (14.570,4.850) {};
\fill[qan, rounded corners=1.2pt] (15.190,4.730) rectangle (15.910,4.970);
\fill[qfA, rounded corners=0.6pt] (15.283,4.800) rectangle ++(0.1,0.1);
\fill[qfB, rounded corners=0.6pt] (15.428,4.800) rectangle ++(0.1,0.1);
\fill[qfC, rounded corners=0.6pt] (15.573,4.800) rectangle ++(0.1,0.1);
\fill[qfD, rounded corners=0.6pt] (15.718,4.800) rectangle ++(0.1,0.1);
\node[font=\fontsize{10}{10}\selectfont] at (16.550,4.850) {\qok};
\fill[qbg, draw=qln, line width=0.7pt, rounded corners=6pt] (0,0) rectangle (11.65,3.6);
\node[qh, anchor=north west, text width=11.35cm] at (0.15,3.5) {(c) Failure under the cap};
\draw[qlane] (1.750,2.750) -- (11.450,2.750);
\node[qs, anchor=east] at (1.530,2.750) {member 0};
\draw[qlane] (1.750,2.200) -- (11.450,2.200);
\node[qs, anchor=east] at (1.530,2.200) {member 1};
\draw[qlane] (1.750,1.650) -- (11.450,1.650);
\node[qs, anchor=east] at (1.530,1.650) {member 2};
\draw[qlane] (1.750,1.100) -- (11.450,1.100);
\node[qs, anchor=east] at (1.530,1.100) {member 3};
\node[qs, align=center, text=qik2] at (9.125,1.945) {waits through\\steps 7--15};
\node[qs, text=qik2] at (2.000,0.680) {0};
\node[qs, text=qik2] at (2.950,0.680) {1};
\node[qs, text=qik2] at (3.900,0.680) {2};
\node[qs, text=qik2] at (4.850,0.680) {3};
\node[qs, text=qik2] at (5.800,0.680) {4};
\node[qs, text=qik2] at (6.750,0.680) {5};
\node[qs, text=qik2] at (7.700,0.680) {6};
\node[qs, text=qik2] at (10.550,0.680) {16--19};
\node[qs, text=qik2, anchor=east] at (1.530,0.680) {step};
\fill[qfA, rounded corners=0.6pt] (1.945,2.695) rectangle ++(0.11,0.11);
\node[qwait] at (2.950,2.750) {};
\draw[qgive] (4.850,2.750) -- (4.850,2.300);
\node[qcoin] at (4.850,2.750) {};
\node[qwait] at (5.800,2.750) {};
\node[qwait] at (6.750,2.750) {};
\node[qwait] at (7.700,2.750) {};
\node[font=\fontsize{10}{10}\selectfont] at (10.570,2.750) {\qno};
\fill[qfB, rounded corners=0.6pt] (3.845,2.145) rectangle ++(0.11,0.11);
\node[qwait] at (4.850,2.200) {};
\draw[qread] (2.100,2.750) to[out=0,in=180] (5.680,2.200);
\draw[qread] (2.100,1.650) to[out=0,in=180] (5.680,2.200);
\node[qeye] at (5.800,2.200) {};
\draw[qrel, line width=0.5pt, rounded corners=1pt] (6.525,2.105) rectangle (6.975,2.295);
\fill[qfA, rounded corners=0.6pt] (6.565,2.145) rectangle ++(0.11,0.11);
\fill[qfB, rounded corners=0.6pt] (6.695,2.145) rectangle ++(0.11,0.11);
\fill[qfC, rounded corners=0.6pt] (6.825,2.145) rectangle ++(0.11,0.11);
\node[qwait] at (7.700,2.200) {};
\node[font=\fontsize{10}{10}\selectfont] at (10.570,2.200) {\qno};
\fill[qfC, rounded corners=0.6pt] (1.945,1.595) rectangle ++(0.11,0.11);
\node[qwait] at (2.950,1.650) {};
\draw[qread] (2.100,2.750) to[out=0,in=180] (5.680,1.650);
\node[qeye] at (5.800,1.650) {};
\node[qwait] at (6.750,1.650) {};
\node[font=\fontsize{10}{10}\selectfont] at (10.570,1.650) {\qno};
\fill[qfD, rounded corners=0.6pt] (1.945,1.045) rectangle ++(0.11,0.11);
\node[qwait] at (3.900,1.100) {};
\draw[qread] (2.100,2.750) to[out=0,in=180] (5.680,1.100);
\node[qeye] at (5.800,1.100) {};
\draw[qrel, line width=0.5pt, rounded corners=1pt] (6.655,1.005) rectangle (6.845,1.195);
\fill[qfA, rounded corners=0.6pt] (6.695,1.045) rectangle ++(0.11,0.11);
\draw[qik2, line width=0.6pt] (7.700,1.020) -- (7.700,1.180);
\node[font=\fontsize{10}{10}\selectfont] at (10.570,1.100) {\qno};
\fill[qbg, draw=qln, line width=0.7pt, rounded corners=6pt] (11.85,0) rectangle (17.1,3.6);
\node[qh, anchor=north west, text width=4.95cm] at (12.0,3.5) {(d) Token use};
\node[qs, anchor=east] at (13.420,2.450) {relay (b)};
\fill[qgv!65] (13.500,2.300) rectangle ++(0.516,0.3);
\fill[qfC!75] (14.016,2.300) rectangle ++(0.495,0.3);
\fill[qfA] (14.511,2.300) rectangle ++(1.020,0.3);
\node[qs, anchor=north west, text=qik2, inner xsep=0pt] at (13.500,2.230) {65k \textperiodcentered{} 50\,\% coordination};
\node[qs, anchor=east] at (13.420,1.500) {failure (c)};
\fill[qgv!65] (13.500,1.350) rectangle ++(1.370,0.3);
\fill[qfC!75] (14.870,1.350) rectangle ++(0.794,0.3);
\fill[qfA] (15.664,1.350) rectangle ++(1.086,0.3);
\node[qs, anchor=north west, text=qik2, inner xsep=0pt] at (13.500,1.280) {103k \textperiodcentered{} 67\,\% coordination};
\fill[qgv!65] (12.050,0.330) rectangle ++(0.2,0.14);
\node[qs, anchor=west] at (12.290,0.400) {waiting};
\fill[qfC!75] (13.550,0.330) rectangle ++(0.2,0.14);
\node[qs, anchor=west] at (13.790,0.400) {negotiation};
\fill[qfA] (15.700,0.330) rectangle ++(0.2,0.14);
\node[qs, anchor=west] at (15.940,0.400) {work};
\fill[qfC, rounded corners=0.7pt] (0.360,-0.445) rectangle ++(0.13,0.13);
\node[qs, anchor=west, text=qik2] at (0.750,-0.380) {private fact};
\draw[qgive] (0.250,-0.830) -- ++(0.35,0);
\node[qs, anchor=west, text=qik2] at (0.750,-0.830) {budget transfer};
\draw[qrel, line width=0.5pt, rounded corners=1pt] (4.550,-0.475) rectangle (4.870,-0.285);
\fill[qfA, rounded corners=0.6pt] (4.590,-0.435) rectangle ++(0.11,0.11);
\fill[qfB, rounded corners=0.6pt] (4.720,-0.435) rectangle ++(0.11,0.11);
\node[qs, anchor=west, text=qik2] at (5.050,-0.380) {relay};
\node[qwait] at (4.710,-0.830) {};
\node[qs, anchor=west, text=qik2] at (5.050,-0.830) {waiting};
\fill[qan, rounded corners=1.2pt] (8.850,-0.480) rectangle ++(0.4,0.2);
\node[qs, anchor=west, text=qik2] at (9.430,-0.380) {analysis with every fact};
\draw[qik2, line width=0.6pt] (9.050,-0.910) -- ++(0,0.16);
\node[qs, anchor=west, text=qik2] at (9.430,-0.830) {held submission};
\draw[qread] (13.150,-0.380) -- ++(0.35,0);
\node[qs, anchor=west, text=qik2] at (13.700,-0.380) {read};
\node[font=\fontsize{9}{9}\selectfont] at (13.325,-0.830) {\qok\,\qno};
\node[qs, anchor=west, text=qik2] at (13.700,-0.830) {final answer};
\end{tikzpicture}
\endgroup
\caption{\textbf{Self-built structures, traced read by read.} Each panel follows one Qwen3.8-27B record on HiddenBench. Panel (a) traces the hub of Figure~\ref{fig:qualitative}c, (b) a relay under the budget cap and (c) a failure under the cap, while (d) compares the token use of (b) and (c).}
\label{fig:structure}
\end{figure*}
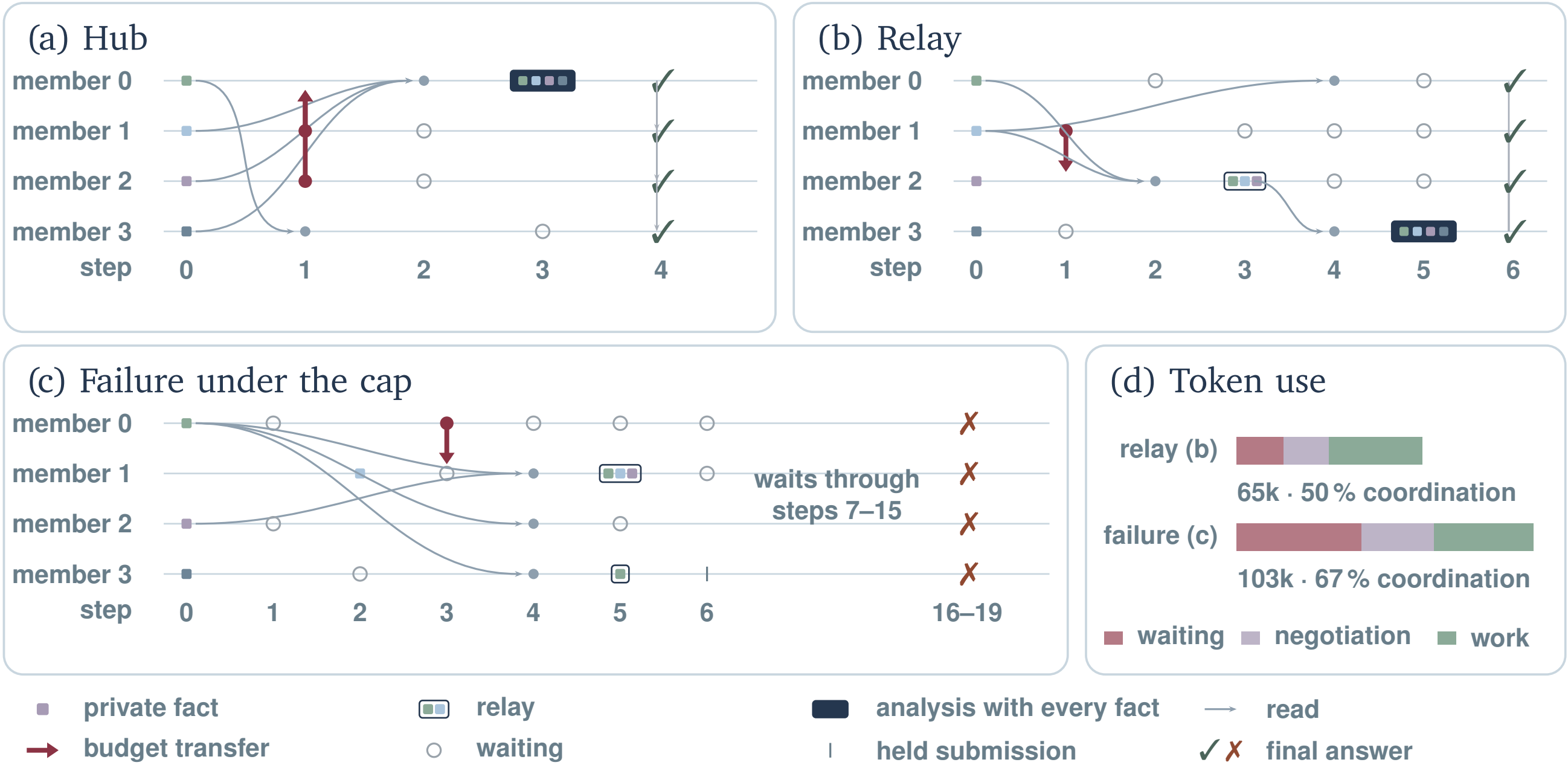

%% file: figures/qual_family.tex
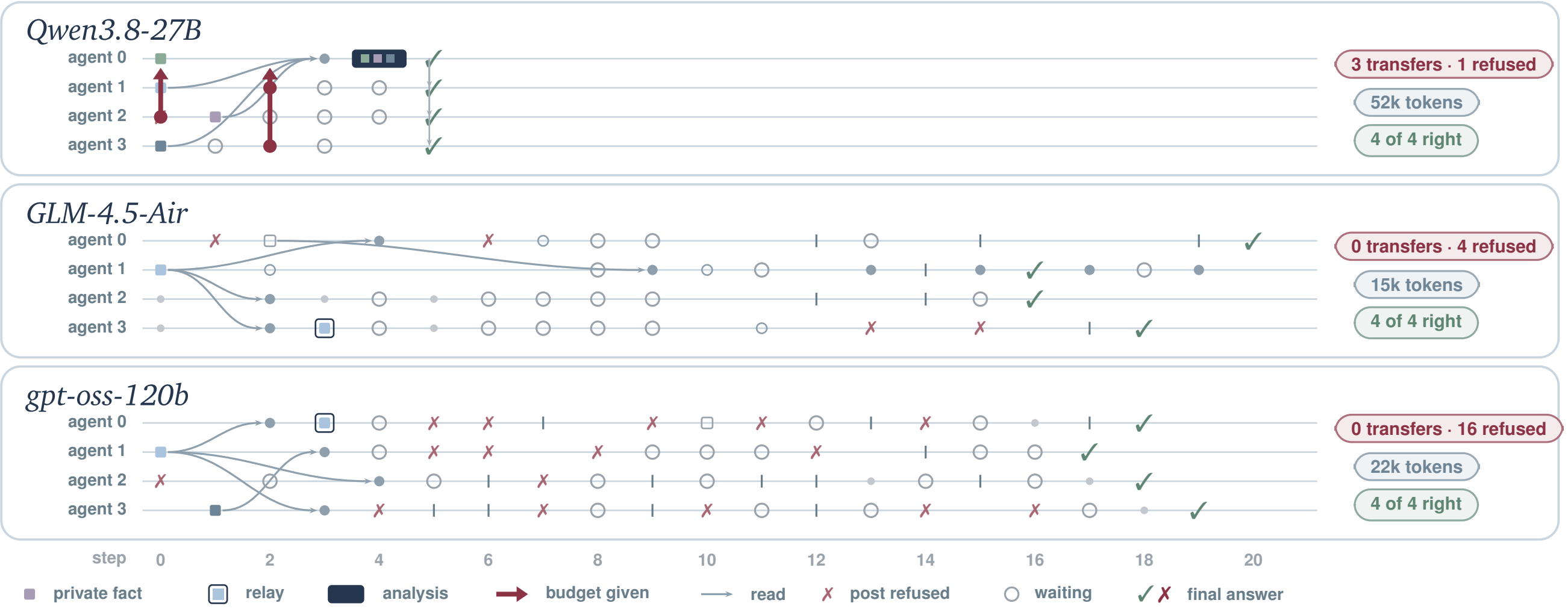
\begin{figure*}[t]
\centering
\begingroup
\definecolor{fgA}{HTML}{8BAA98}\definecolor{fgB}{HTML}{A9C4DC}\definecolor{fgC}{HTML}{A79BB5}\definecolor{fgD}{HTML}{6E8597}
\definecolor{fgT}{HTML}{687D89}\definecolor{fgR}{HTML}{8B3344}\definecolor{fgG}{HTML}{5E8574}\definecolor{fgN}{HTML}{25374F}
\definecolor{fgL}{HTML}{C9D6DF}\definecolor{fgP}{HTML}{F3F6F8}\definecolor{fgY}{HTML}{9AA3AB}\definecolor{fgQ}{HTML}{8E9FAE}
\tikzset{
  qh/.style={font=\fontsize{9.5}{10.5}\selectfont\itshape, text=fgN, anchor=north west},
  qs/.style={font=\sffamily\bfseries\fontsize{5.8}{6.8}\selectfont, text=fgT},
  qq/.style={font=\fontsize{6.8}{8}\selectfont\itshape, text=fgN, fill=fgP, draw=fgL, line width=0.5pt,
             rounded corners=4pt, inner xsep=5pt, inner ysep=3pt, align=left, text width=#1},
  card/.style={fill=white, draw=fgL, line width=0.7pt, rounded corners=6pt},
  pill/.style={rounded corners=5pt, inner xsep=5pt, inner ysep=2.2pt, line width=0.6pt,
               font=\sffamily\bfseries\fontsize{6.2}{7}\selectfont, draw=#1!70, fill=#1!10, text=#1},
  ico/.style={font=\fontsize{#1}{#1}\selectfont, inner sep=0pt},
  qlane/.style={draw=fgL, line width=0.5pt},
  qread/.style={draw=fgQ, line width=0.6pt, -{Stealth[length=2.4pt]}},
  qgive/.style={draw=fgR, line width=1.5pt, -{Triangle[length=3.6pt,width=4.6pt]}},
  qbc/.style={draw=fgN!40, line width=0.5pt, -{Stealth[length=2.2pt]}},
  qeye/.style={circle, fill=fgQ, inner sep=0pt, minimum size=3.2pt},
  qeyeo/.style={circle, draw=fgQ, line width=0.5pt, inner sep=0pt, minimum size=3.2pt},
  qcoin/.style={circle, fill=fgR, inner sep=0pt, minimum size=4.2pt},
  qwait/.style={circle, draw=fgY, line width=0.6pt, inner sep=0pt, minimum size=4.2pt},
  qnoop/.style={circle, fill=fgY!60, inner sep=0pt, minimum size=2.4pt},
}
\newcommand{\qok}{\textcolor{fgG}{\ding{51}}}
\newcommand{\qno}{\textcolor{fgR}{\ding{55}}}
\begin{tikzpicture}[x=1cm, y=1cm]
\path[card] (0,4.200) rectangle (17.1,6.100);
\node[qh] at (0.15,6.040) {Qwen3.8-27B};
\draw[qlane] (1.550,5.480) -- (14.450,5.480);
\node[qs, anchor=east] at (1.500,5.480) {agent 0};
\draw[qlane] (1.550,5.160) -- (14.450,5.160);
\node[qs, anchor=east] at (1.500,5.160) {agent 1};
\draw[qlane] (1.550,4.840) -- (14.450,4.840);
\node[qs, anchor=east] at (1.500,4.840) {agent 2};
\draw[qlane] (1.550,4.520) -- (14.450,4.520);
\node[qs, anchor=east] at (1.500,4.520) {agent 3};
\fill[fgA, rounded corners=0.6pt] (1.690,5.420) rectangle ++(0.12,0.12);
\draw[qread] (1.830,5.160) to[out=0,in=180] (3.470,5.480);
\draw[qread] (1.830,4.520) to[out=0,in=180] (3.470,5.480);
\draw[qread] (2.430,4.840) to[out=0,in=180] (3.470,5.480);
\node[qeye] at (3.550,5.480) {};
\fill[fgN, rounded corners=1.2pt] (3.850,5.380) rectangle (4.450,5.580);
\fill[fgA] (3.950,5.435) rectangle ++(0.09,0.09);
\fill[fgC] (4.090,5.435) rectangle ++(0.09,0.09);
\fill[fgD] (4.230,5.435) rectangle ++(0.09,0.09);
\node[font=\fontsize{7.5}{7.5}\selectfont] at (4.750,5.480) {\qok};
\fill[fgB, rounded corners=0.6pt] (1.690,5.100) rectangle ++(0.12,0.12);
\draw[qgive] (2.950,5.160) -- (2.950,5.380);
\node[qcoin] at (2.950,5.160) {};
\node[qwait] at (3.550,5.160) {};
\node[qwait] at (4.150,5.160) {};
\draw[qbc] (4.700,5.480) -- (4.700,5.160);
\node[font=\fontsize{7.5}{7.5}\selectfont] at (4.750,5.160) {\qok};
\node[font=\fontsize{6}{6}\selectfont, text=fgR!75] at (1.750,4.840) {\ding{55}};
\draw[qgive] (1.750,4.840) -- (1.750,5.380);
\node[qcoin] at (1.750,4.840) {};
\fill[fgC, rounded corners=0.6pt] (2.290,4.780) rectangle ++(0.12,0.12);
\node[qwait] at (2.950,4.840) {};
\node[qwait] at (3.550,4.840) {};
\node[qwait] at (4.150,4.840) {};
\draw[qbc] (4.700,5.480) -- (4.700,4.840);
\node[font=\fontsize{7.5}{7.5}\selectfont] at (4.750,4.840) {\qok};
\fill[fgD, rounded corners=0.6pt] (1.690,4.460) rectangle ++(0.12,0.12);
\node[qwait] at (2.350,4.520) {};
\draw[qgive] (2.950,4.520) -- (2.950,5.380);
\node[qcoin] at (2.950,4.520) {};
\node[qwait] at (3.550,4.520) {};
\draw[qbc] (4.700,5.480) -- (4.700,4.520);
\node[font=\fontsize{7.5}{7.5}\selectfont] at (4.750,4.520) {\qok};
\node[pill=fgR, anchor=west] at (14.640,5.420) {3 transfers \textperiodcentered{} 1 refused};
\node[pill=fgD, anchor=west] at (14.850,5.000) {52k tokens};
\node[pill=fgG, anchor=west] at (14.850,4.580) {4 of 4 right};
\path[card] (0,2.200) rectangle (17.1,4.100);
\node[qh] at (0.15,4.040) {GLM-4.5-Air};
\draw[qlane] (1.550,3.480) -- (14.450,3.480);
\node[qs, anchor=east] at (1.500,3.480) {agent 0};
\draw[qlane] (1.550,3.160) -- (14.450,3.160);
\node[qs, anchor=east] at (1.500,3.160) {agent 1};
\draw[qlane] (1.550,2.840) -- (14.450,2.840);
\node[qs, anchor=east] at (1.500,2.840) {agent 2};
\draw[qlane] (1.550,2.520) -- (14.450,2.520);
\node[qs, anchor=east] at (1.500,2.520) {agent 3};
\node[font=\fontsize{6}{6}\selectfont, text=fgR!75] at (2.350,3.480) {\ding{55}};
\draw[fgY, line width=0.5pt, rounded corners=0.6pt] (2.890,3.420) rectangle ++(0.12,0.12);
\draw[qread] (1.830,3.160) to[out=0,in=180] (4.070,3.480);
\node[qeye] at (4.150,3.480) {};
\node[font=\fontsize{6}{6}\selectfont, text=fgR!75] at (5.350,3.480) {\ding{55}};
\node[qeyeo] at (5.950,3.480) {};
\node[qwait] at (6.550,3.480) {};
\node[qwait] at (7.150,3.480) {};
\draw[fgT, line width=0.6pt] (8.950,3.410) -- (8.950,3.550);
\node[qwait] at (9.550,3.480) {};
\draw[fgT, line width=0.6pt] (10.750,3.410) -- (10.750,3.550);
\draw[fgT, line width=0.6pt] (13.150,3.410) -- (13.150,3.550);
\node[font=\fontsize{7.5}{7.5}\selectfont] at (13.750,3.480) {\qok};
\fill[fgB, rounded corners=0.6pt] (1.690,3.100) rectangle ++(0.12,0.12);
\node[qeyeo] at (2.950,3.160) {};
\node[qwait] at (6.550,3.160) {};
\draw[qread] (3.030,3.480) to[out=0,in=180] (7.070,3.160);
\node[qeye] at (7.150,3.160) {};
\node[qeyeo] at (7.750,3.160) {};
\node[qwait] at (8.350,3.160) {};
\node[qeye] at (9.550,3.160) {};
\draw[fgT, line width=0.6pt] (10.150,3.090) -- (10.150,3.230);
\node[qeye] at (10.750,3.160) {};
\node[font=\fontsize{7.5}{7.5}\selectfont] at (11.350,3.160) {\qok};
\node[qeye] at (11.950,3.160) {};
\node[qwait] at (12.550,3.160) {};
\node[qeye] at (13.150,3.160) {};
\node[qnoop] at (1.750,2.840) {};
\draw[qread] (1.830,3.160) to[out=0,in=180] (2.870,2.840);
\node[qeye] at (2.950,2.840) {};
\node[qnoop] at (3.550,2.840) {};
\node[qwait] at (4.150,2.840) {};
\node[qnoop] at (4.750,2.840) {};
\node[qwait] at (5.350,2.840) {};
\node[qwait] at (5.950,2.840) {};
\node[qwait] at (6.550,2.840) {};
\node[qwait] at (7.150,2.840) {};
\draw[fgT, line width=0.6pt] (8.950,2.770) -- (8.950,2.910);
\draw[fgT, line width=0.6pt] (10.150,2.770) -- (10.150,2.910);
\node[qwait] at (10.750,2.840) {};
\node[font=\fontsize{7.5}{7.5}\selectfont] at (11.350,2.840) {\qok};
\node[qnoop] at (1.750,2.520) {};
\draw[qread] (1.830,3.160) to[out=0,in=180] (2.870,2.520);
\node[qeye] at (2.950,2.520) {};
\fill[fgB, rounded corners=0.6pt] (3.490,2.460) rectangle ++(0.12,0.12);
\draw[fgN, line width=0.5pt, rounded corners=1pt] (3.450,2.420) rectangle ++(0.200,0.2);
\node[qwait] at (4.150,2.520) {};
\node[qnoop] at (4.750,2.520) {};
\node[qwait] at (5.350,2.520) {};
\node[qwait] at (5.950,2.520) {};
\node[qwait] at (6.550,2.520) {};
\node[qwait] at (7.150,2.520) {};
\node[qeyeo] at (8.350,2.520) {};
\node[font=\fontsize{6}{6}\selectfont, text=fgR!75] at (9.550,2.520) {\ding{55}};
\node[font=\fontsize{6}{6}\selectfont, text=fgR!75] at (10.750,2.520) {\ding{55}};
\draw[fgT, line width=0.6pt] (11.950,2.450) -- (11.950,2.590);
\node[font=\fontsize{7.5}{7.5}\selectfont] at (12.550,2.520) {\qok};
\node[pill=fgR, anchor=west] at (14.640,3.420) {0 transfers \textperiodcentered{} 4 refused};
\node[pill=fgD, anchor=west] at (14.850,3.000) {15k tokens};
\node[pill=fgG, anchor=west] at (14.850,2.580) {4 of 4 right};
\path[card] (0,0.200) rectangle (17.1,2.100);
\node[qh] at (0.15,2.040) {gpt-oss-120b};
\draw[qlane] (1.550,1.480) -- (14.450,1.480);
\node[qs, anchor=east] at (1.500,1.480) {agent 0};
\draw[qlane] (1.550,1.160) -- (14.450,1.160);
\node[qs, anchor=east] at (1.500,1.160) {agent 1};
\draw[qlane] (1.550,0.840) -- (14.450,0.840);
\node[qs, anchor=east] at (1.500,0.840) {agent 2};
\draw[qlane] (1.550,0.520) -- (14.450,0.520);
\node[qs, anchor=east] at (1.500,0.520) {agent 3};
\draw[qread] (1.830,1.160) to[out=0,in=180] (2.870,1.480);
\node[qeye] at (2.950,1.480) {};
\fill[fgB, rounded corners=0.6pt] (3.490,1.420) rectangle ++(0.12,0.12);
\draw[fgN, line width=0.5pt, rounded corners=1pt] (3.450,1.380) rectangle ++(0.200,0.2);
\node[qwait] at (4.150,1.480) {};
\node[font=\fontsize{6}{6}\selectfont, text=fgR!75] at (4.750,1.480) {\ding{55}};
\node[font=\fontsize{6}{6}\selectfont, text=fgR!75] at (5.350,1.480) {\ding{55}};
\draw[fgT, line width=0.6pt] (5.950,1.410) -- (5.950,1.550);
\node[font=\fontsize{6}{6}\selectfont, text=fgR!75] at (7.150,1.480) {\ding{55}};
\draw[fgY, line width=0.5pt, rounded corners=0.6pt] (7.690,1.420) rectangle ++(0.12,0.12);
\node[font=\fontsize{6}{6}\selectfont, text=fgR!75] at (8.350,1.480) {\ding{55}};
\node[qwait] at (8.950,1.480) {};
\draw[fgT, line width=0.6pt] (9.550,1.410) -- (9.550,1.550);
\node[font=\fontsize{6}{6}\selectfont, text=fgR!75] at (10.150,1.480) {\ding{55}};
\node[qwait] at (10.750,1.480) {};
\node[qnoop] at (11.350,1.480) {};
\draw[fgT, line width=0.6pt] (11.950,1.410) -- (11.950,1.550);
\node[font=\fontsize{7.5}{7.5}\selectfont] at (12.550,1.480) {\qok};
\fill[fgB, rounded corners=0.6pt] (1.690,1.100) rectangle ++(0.12,0.12);
\draw[qread] (2.430,0.520) to[out=0,in=180] (3.470,1.160);
\node[qeye] at (3.550,1.160) {};
\node[qwait] at (4.150,1.160) {};
\node[font=\fontsize{6}{6}\selectfont, text=fgR!75] at (4.750,1.160) {\ding{55}};
\node[font=\fontsize{6}{6}\selectfont, text=fgR!75] at (5.350,1.160) {\ding{55}};
\node[font=\fontsize{6}{6}\selectfont, text=fgR!75] at (6.550,1.160) {\ding{55}};
\node[qwait] at (7.150,1.160) {};
\node[qwait] at (7.750,1.160) {};
\node[qwait] at (8.350,1.160) {};
\node[font=\fontsize{6}{6}\selectfont, text=fgR!75] at (8.950,1.160) {\ding{55}};
\draw[fgT, line width=0.6pt] (10.150,1.090) -- (10.150,1.230);
\node[qwait] at (10.750,1.160) {};
\node[qwait] at (11.350,1.160) {};
\node[font=\fontsize{7.5}{7.5}\selectfont] at (11.950,1.160) {\qok};
\node[font=\fontsize{6}{6}\selectfont, text=fgR!75] at (1.750,0.840) {\ding{55}};
\node[qwait] at (2.950,0.840) {};
\draw[qread] (1.830,1.160) to[out=0,in=180] (4.070,0.840);
\node[qeye] at (4.150,0.840) {};
\node[qwait] at (4.750,0.840) {};
\draw[fgT, line width=0.6pt] (5.350,0.770) -- (5.350,0.910);
\node[font=\fontsize{6}{6}\selectfont, text=fgR!75] at (5.950,0.840) {\ding{55}};
\node[qwait] at (6.550,0.840) {};
\draw[fgT, line width=0.6pt] (7.150,0.770) -- (7.150,0.910);
\node[qwait] at (7.750,0.840) {};
\draw[fgT, line width=0.6pt] (8.350,0.770) -- (8.350,0.910);
\draw[fgT, line width=0.6pt] (8.950,0.770) -- (8.950,0.910);
\node[qnoop] at (9.550,0.840) {};
\node[qwait] at (10.150,0.840) {};
\draw[fgT, line width=0.6pt] (10.750,0.770) -- (10.750,0.910);
\node[qwait] at (11.350,0.840) {};
\node[qnoop] at (11.950,0.840) {};
\node[font=\fontsize{7.5}{7.5}\selectfont] at (12.550,0.840) {\qok};
\fill[fgD, rounded corners=0.6pt] (2.290,0.460) rectangle ++(0.12,0.12);
\draw[qread] (1.830,1.160) to[out=0,in=180] (3.470,0.520);
\node[qeye] at (3.550,0.520) {};
\node[font=\fontsize{6}{6}\selectfont, text=fgR!75] at (4.150,0.520) {\ding{55}};
\draw[fgT, line width=0.6pt] (4.750,0.450) -- (4.750,0.590);
\draw[fgT, line width=0.6pt] (5.350,0.450) -- (5.350,0.590);
\node[font=\fontsize{6}{6}\selectfont, text=fgR!75] at (5.950,0.520) {\ding{55}};
\node[qwait] at (6.550,0.520) {};
\draw[fgT, line width=0.6pt] (7.150,0.450) -- (7.150,0.590);
\node[font=\fontsize{6}{6}\selectfont, text=fgR!75] at (7.750,0.520) {\ding{55}};
\node[qwait] at (8.350,0.520) {};
\draw[fgT, line width=0.6pt] (8.950,0.450) -- (8.950,0.590);
\node[qwait] at (9.550,0.520) {};
\node[font=\fontsize{6}{6}\selectfont, text=fgR!75] at (10.150,0.520) {\ding{55}};
\node[font=\fontsize{6}{6}\selectfont, text=fgR!75] at (11.350,0.520) {\ding{55}};
\node[qwait] at (11.950,0.520) {};
\node[qnoop] at (12.550,0.520) {};
\node[font=\fontsize{7.5}{7.5}\selectfont] at (13.150,0.520) {\qok};
\node[pill=fgR, anchor=west] at (14.640,1.420) {0 transfers \textperiodcentered{} 16 refused};
\node[pill=fgD, anchor=west] at (14.850,1.000) {22k tokens};
\node[pill=fgG, anchor=west] at (14.850,0.580) {4 of 4 right};
\node[qs, text=fgY] at (1.750,-0.020) {0};
\node[qs, text=fgY] at (2.950,-0.020) {2};
\node[qs, text=fgY] at (4.150,-0.020) {4};
\node[qs, text=fgY] at (5.350,-0.020) {6};
\node[qs, text=fgY] at (6.550,-0.020) {8};
\node[qs, text=fgY] at (7.750,-0.020) {10};
\node[qs, text=fgY] at (8.950,-0.020) {12};
\node[qs, text=fgY] at (10.150,-0.020) {14};
\node[qs, text=fgY] at (11.350,-0.020) {16};
\node[qs, text=fgY] at (12.550,-0.020) {18};
\node[qs, text=fgY] at (13.750,-0.020) {20};
\node[qs, text=fgY, anchor=east] at (1.500,-0.020) {step};
\fill[fgC, rounded corners=0.6pt] (0.250,-0.460) rectangle ++(0.12,0.12);
\node[qs, anchor=west] at (0.450,-0.400) {private fact};
\fill[fgB] (2.320,-0.460) rectangle ++(0.12,0.12);\draw[fgN, line width=0.5pt, rounded corners=1pt] (2.280,-0.500) rectangle ++(0.2,0.2);
\node[qs, anchor=west] at (2.560,-0.400) {relay};
\fill[fgN, rounded corners=1.2pt] (3.585,-0.500) rectangle ++(0.4,0.2);
\node[qs, anchor=west] at (4.065,-0.400) {analysis};
\draw[qgive] (5.435,-0.400) -- ++(0.35,0);
\node[qs, anchor=west] at (5.855,-0.400) {budget given};
\draw[qread] (7.685,-0.400) -- ++(0.35,0);
\node[qs, anchor=west] at (8.105,-0.400) {read};
\node[font=\fontsize{6}{6}\selectfont, text=fgR!75] at (9.075,-0.400) {\ding{55}};
\node[qs, anchor=west] at (9.195,-0.400) {post refused};
\node[qwait] at (11.095,-0.400) {};
\node[qs, anchor=west] at (11.225,-0.400) {waiting};
\node[font=\fontsize{7}{7}\selectfont] at (12.660,-0.400) {\qok\,\qno};
\node[qs, anchor=west] at (12.900,-0.400) {final answer};
\end{tikzpicture}
\endgroup
\caption{\textbf{One task, three model families.} One HiddenBench task, self-organized with a planning round (lanes: agents; chips: private facts). Qwen3.8-27B agents give their reading budget to one agent, which posts an analysis within six steps, while GLM-4.5-Air and gpt-oss-120b agents forward single facts.}
\label{fig:family}
\end{figure*}

%% file: figures/channels.tex
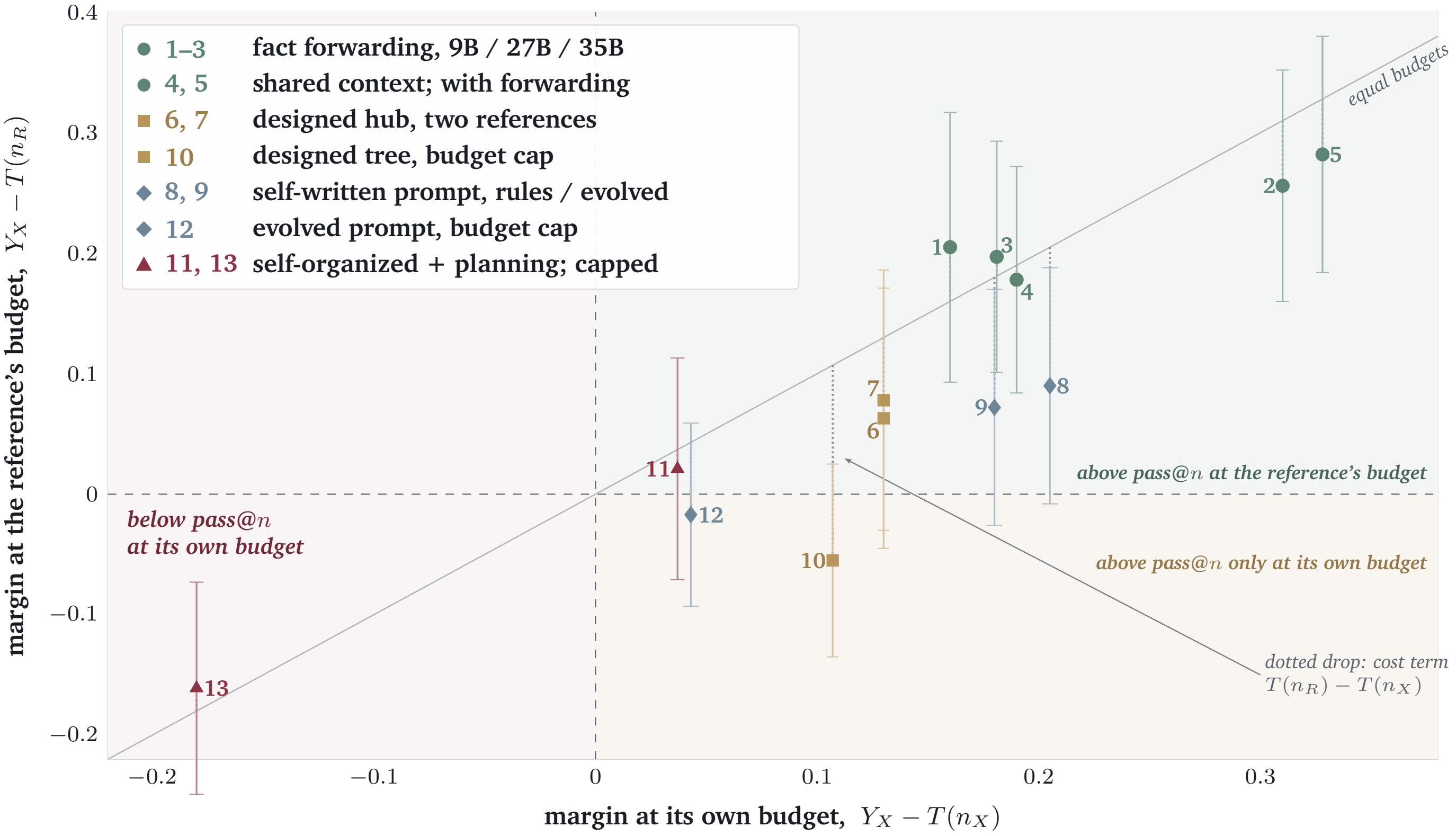
\begin{figure}[!htb]
\centering
\begin{tikzpicture}
\begin{axis}[
  width=\linewidth, height=0.6\linewidth,
  xmin=-0.22, xmax=0.38, ymin=-0.22, ymax=0.40,
  xlabel={margin at its own budget, \;$Y_X-T(n_X)$},
  ylabel={margin at the reference's budget, \;$Y_X-T(n_R)$},
  xlabel style={font=\rmfamily\bfseries\small, text=ink}, ylabel style={font=\rmfamily\bfseries\small, text=ink},
  tick label style={font=\rmfamily\footnotesize, text=ink},
  xtick={-0.2,-0.1,0,0.1,0.2,0.3}, ytick={-0.2,-0.1,0,0.1,0.2,0.3,0.4},
  axis line style={ivoryline}, tick style={ivoryline},
  clip=false, axis on top=false, name=ax,
]
\fill[famFact!7] (axis cs:0,0) rectangle (axis cs:0.38,0.40);
\fill[famDesign!9] (axis cs:0,-0.22) rectangle (axis cs:0.38,0);
\fill[famSelf!5] (axis cs:-0.22,-0.22) rectangle (axis cs:0,0.40);
\node[font=\rmfamily\bfseries\itshape\scriptsize, text=famFact!75!black, anchor=south east, align=right, inner sep=2pt] at (axis cs:0.378,0.004) {above pass@$n$ at the reference's budget};
\node[font=\rmfamily\bfseries\itshape\scriptsize, text=famDesign!80!black, anchor=north east, align=right, inner sep=2pt] at (axis cs:0.378,-0.045) {above pass@$n$ only at its own budget};
\node[font=\rmfamily\bfseries\itshape\footnotesize, text=famSelf!80!black, anchor=north west, align=left, inner sep=2pt] at (axis cs:-0.214,-0.008) {below pass@$n$\\at its own budget};
\draw[ink2, dashed, line width=0.4pt] (axis cs:-0.22,0) -- (axis cs:0.38,0);
\draw[ink2, dashed, line width=0.4pt] (axis cs:0,-0.22) -- (axis cs:0,0.40);
\draw[ink2!45, line width=0.5pt] (axis cs:-0.22,-0.22) -- (axis cs:0.38,0.38) node[pos=0.965, sloped, below, font=\rmfamily\itshape\scriptsize, text=ink2, inner sep=2pt] {equal budgets};
\foreach \x/\y in {0.160/0.205, 0.310/0.256, 0.181/0.197, 0.190/0.178, 0.328/0.282, 0.130/0.063, 0.130/0.078, 0.205/0.090, 0.180/0.072, 0.107/-0.055, 0.037/0.021, 0.043/-0.017, -0.180/-0.161}{\edef\temp{\noexpand\draw[ink2!70, densely dotted, line width=0.7pt] (axis cs:\x,\x) -- (axis cs:\x,\y);}\temp}
\draw[ink2!80, line width=0.45pt, -{Latex[length=2.6pt]}] (axis cs:0.30,-0.15) node[anchor=west, font=\rmfamily\itshape\scriptsize, text=ink2, inner sep=1.5pt, align=left] {dotted drop: cost term\\$T(n_R)-T(n_X)$} -- (axis cs:0.112,0.03);
\addplot[only marks, mark=*, mark size=2.3pt, color=famFact, error bars/.cd, y dir=both, y explicit, error bar style={famFact!60, line width=0.6pt}]
  coordinates {(0.160,0.205) +- (0,0.112) (0.310,0.256) +- (0,0.096) (0.181,0.197) +- (0,0.096) (0.190,0.178) +- (0,0.094) (0.328,0.282) +- (0,0.098)};
\addplot[only marks, mark=square*, mark size=2.0pt, color=famDesign, error bars/.cd, y dir=both, y explicit, error bar style={famDesign!55, line width=0.6pt}]
  coordinates {(0.130,0.063) +- (0,0.108) (0.130,0.078) +- (0,0.108) (0.107,-0.055) +- (0,0.080)};
\addplot[only marks, mark=diamond*, mark size=2.8pt, color=famNotes, error bars/.cd, y dir=both, y explicit, error bar style={famNotes!55, line width=0.6pt}]
  coordinates {(0.205,0.090) +- (0,0.098) (0.180,0.072) +- (0,0.098) (0.043,-0.017) +- (0,0.076)};
\addplot[only marks, mark=triangle*, mark size=2.5pt, color=famSelf, error bars/.cd, y dir=both, y explicit, error bar style={famSelf!60, line width=0.6pt}]
  coordinates {(0.037,0.021) +- (0,0.092) (-0.180,-0.161) +- (0,0.088)};
\tikzset{pl/.style={font=\rmfamily\bfseries\footnotesize, inner sep=2.2pt}}
\node[pl, text=famFact, anchor=east] at (axis cs:0.160,0.205) {1};
\node[pl, text=famFact, anchor=east] at (axis cs:0.310,0.256) {2};
\node[pl, text=famFact, anchor=south west, inner sep=1.2pt] at (axis cs:0.181,0.197) {3};
\node[pl, text=famFact, anchor=north west, inner sep=1.2pt] at (axis cs:0.190,0.178) {4};
\node[pl, text=famFact, anchor=west] at (axis cs:0.328,0.282) {5};
\node[pl, text=famDesign!85!black, anchor=north east, inner sep=1.2pt] at (axis cs:0.130,0.063) {6};
\node[pl, text=famDesign!85!black, anchor=south east, inner sep=1.2pt] at (axis cs:0.130,0.078) {7};
\node[pl, text=famNotes, anchor=west] at (axis cs:0.205,0.090) {8};
\node[pl, text=famNotes, anchor=east] at (axis cs:0.180,0.072) {9};
\node[pl, text=famDesign!85!black, anchor=east] at (axis cs:0.107,-0.055) {10};
\node[pl, text=famSelf, anchor=east] at (axis cs:0.037,0.021) {11};
\node[pl, text=famNotes, anchor=west] at (axis cs:0.043,-0.017) {12};
\node[pl, text=famSelf, anchor=west] at (axis cs:-0.180,-0.161) {13};
\end{axis}
\begin{scope}[shift={(ax.north west)}, xshift=0.18cm, yshift=-0.16cm, font=\rmfamily\bfseries\small]
\fill[white, opacity=0.93, draw=ivoryline, line width=0.4pt, rounded corners=1.5pt] (0,0) rectangle (8.45,-3.3);
\foreach \i/\c/\n/\t in {0/famFact/{1--3}/{fact forwarding, 9B / 27B / 35B}, 1/famFact/{4, 5}/{shared context; with forwarding}, 2/famDesign!85!black/{6, 7}/{designed hub, two references}, 3/famDesign!85!black/{10}/{designed tree, budget cap}, 4/famNotes/{8, 9}/{self-written prompt, rules / evolved}, 5/famNotes/{12}/{evolved prompt, budget cap}, 6/famSelf/{11, 13}/{self-organized + planning; capped}} {
  \node[anchor=west, text=\c, inner sep=0pt] at (0.52,-0.3-0.45*\i) {\n};
  \node[anchor=west, text=ink, inner sep=0pt] at (1.62,-0.3-0.45*\i) {\t};}
\fill[famFact] (0.27,-0.30) circle (2.3pt); \fill[famFact] (0.27,-0.75) circle (2.3pt);
\fill[famDesign] (0.27-0.075,-1.2-0.075) rectangle (0.27+0.075,-1.2+0.075); \fill[famDesign] (0.27-0.075,-1.65-0.075) rectangle (0.27+0.075,-1.65+0.075);
\foreach \y in {-2.1,-2.55} {\fill[famNotes] (0.27,\y-0.1) -- (0.37,\y) -- (0.27,\y+0.1) -- (0.17,\y) -- cycle;}
\fill[famSelf] (0.17,-3.07) -- (0.37,-3.07) -- (0.27,-2.89) -- cycle;
\end{scope}
\end{tikzpicture}
\caption{\textbf{Accuracy and cost to level 3.} Each point is a protocol X against its reference R on the same pool (Appendix Table~\ref*{tab:channels}; bars: 95\%). The diagonal marks equal budgets, and the dotted drop from it to each point is the cost term of Equation~\ref{eq:channels}, the change in pass@$n$ between the two budgets; a point lies below the diagonal when X spends less than R. Regions are drawn at the point estimates. \mbox{\mkFact{}~\famtag{Fact}} protocols are confirmed above pass@$n$ even at the reference's budget; \mbox{\mkDesign{}~\famtag{Design}} and \mbox{\mkNotes{}~\famtag{Notes}} organizations 6 to 10 are confirmed above it only at their own, lower budget; the capped prompt 12 and the \mbox{\mkSelf{}~\famtag{Self}} structures are not confirmed above it at either budget.}
\label{fig:channels}
\end{figure}

%% file: tables/recursion.tex
\begin{table}[t]
\centering
\tsz
\setlength{\tabcolsep}{4pt}
\renewcommand{\arraystretch}{1.08}
\caption{\textbf{Label-free generations.} Margins over pass@$n$ of base-model agents on 196 held-out SplitBench tasks. The control trains like G2 on G0's turns, and $\Delta$ is the change between generations. The 9B standard-debate steps are read on evaluation seeds 14 to 17, where they replicate the readings on seeds 10 to 13 (Appendix~\ref*{app:recursion}).}
\label{tab:recursion}
\begin{tabular}{@{}lccccc@{}}
\toprule
& \multicolumn{3}{c}{\textbf{Margin over pass@$n$}} & \multicolumn{2}{c}{\textbf{Step $\Delta$}} \\
\cmidrule(lr){2-4}\cmidrule(l){5-6}
\textbf{Model} & G0 & G1 & G2 & G0$\to$G1 & G1$\to$G2 \\
\midrule
\tgroup{6}{3-step debate} \\
Qwen3.5-9B & $-0.253$ & $-0.028$ & $\mathbf{\color{methodblue}+0.159}$ & $+0.225$ & $+0.187$ \\
Qwen3.8-27B & $-0.009$ & $+0.201$ & $\mathbf{+0.284}$ & $+0.210$ & $+0.083$ \\
\tgroup{6}{Standard debate} \\
Qwen3.5-9B & $-0.228$ & $-0.090$ & $+0.098$ & $+0.138$ & $+0.188$ \\
Qwen3.5-4B & $-0.498$ & $-0.077$ & $+0.117$ & $+0.421$ & $+0.194$ \\
\midrule
\textbf{Control} & & \multicolumn{2}{c}{Margin} & \multicolumn{2}{c}{G2 minus control} \\
\midrule
Qwen3.5-9B, 3-step & & \multicolumn{2}{c}{$-0.033$} & \multicolumn{2}{c}{$\mathbf{+0.192}$} \\
Qwen3.8-27B, 3-step & & \multicolumn{2}{c}{$+0.016$} & \multicolumn{2}{c}{$+0.268$} \\
\bottomrule
\end{tabular}
\end{table}

%% file: tables/credit.tex
\begin{table}[t]
\centering
\tsz
\setlength{\tabcolsep}{4.5pt}
\renewcommand{\arraystretch}{1.08}
\caption{\textbf{Credit decides whether training helps the group.} Held-out recall, F1 and precision (\%) of $N$ Qwen3.5-4B agents on the shared-corpus contradiction audit after 40 training steps on the group's exact-pair F1; baselines: own success (GRPO), imitating wins (RAFT), one trainee and team reward (best underlined).}
\label{tab:credit}
\begin{adjustbox}{max width=\linewidth}
\begin{tabular}{@{}lcccccc@{}}
\toprule
& \multicolumn{3}{c}{\textbf{$N=4$}} & \multicolumn{3}{c}{\textbf{$N=8$}} \\
\cmidrule(lr){2-4}\cmidrule(lr){5-7}
\textbf{Method} & Recall & F1 & Precision & Recall & F1 & Precision \\
\midrule
\refrow Untrained & 6.2 & 10.5 & 33.3 & 13.0 & 19.5 & 33.9 \\
\tgroup{7}{Baselines} \\
\quad Own success (GRPO) & 5.5 & 7.7 & 12.2 & 9.5 & 10.2 & 9.4 \\
\quad Imitate wins (RAFT) & 7.8 & \underline{12.3} & \underline{30.9} & 12.2 & \underline{16.3} & 22.7 \\
\quad One trainee & 5.2 & 8.4 & 25.6 & 10.9 & \underline{16.3} & \underline{30.2} \\
\quad Team reward & \underline{8.3} & 10.9 & 15.3 & \underline{13.7} & 15.8 & 16.0 \\
\midrule
\oursrow Shapley credit (ours) & \textbf{9.2} & \textbf{15.9} & \textbf{58.4} & \textbf{14.2} & \textbf{20.1} & \textbf{49.7} \\
\bottomrule
\end{tabular}
\end{adjustbox}

\end{table}

%% file: figures/chains.tex
\begin{figure}[t]
\centering
\pgfplotsset{chainax/.style={scale only axis, width=0.245\textwidth, height=0.20\textwidth, font=\scriptsize, tick label style={font=\scriptsize}, label style={font=\scriptsize}, title style={font=\scriptsize\bfseries, yshift=-2pt},
  legend style={font=\tiny, draw=none, fill=none, row sep=-1.5pt, column sep=3pt, at={(0.5,-0.24)}, anchor=north}, legend columns=2, legend cell align=left,
  axis line style={ink2}, tick style={ink2}, grid=major, grid style={gray!15}, clip=false}}
\tikzset{dlab/.style={font=\tiny, text=ink2, inner sep=0.5pt}, glab/.style={font=\tiny\bfseries, inner sep=0.8pt}}
\begin{tikzpicture}[baseline=(ax.north)]
\begin{axis}[chainax, name=ax, title={(a) PlanBench, selection set}, ylabel={fraction}, ymin=0.0, ymax=1.12, ytick={0,0.2,0.4,0.6,0.8,1}, xmin=-0.3, xmax=4.3, xtick={0,1,2,3,4}, xticklabels={G0,G1,G2,G3,G4}]
\addplot[methodblue, thick, mark=*, mark size=1.4pt] coordinates {(0,0.6133) (1,0.8567) (2,0.9367) (3,0.9400) (4,0.9500)};
\addlegendentry{group}
\addplot[methodblue!45, thick, densely dotted] coordinates {(0,0.6833) (1,0.8600) (2,0.9367) (3,0.9467) (4,0.9500)};
\addlegendentry{any of four agents}
\addplot[famDesign!80!black, thick, mark=square*, mark size=1.2pt] coordinates {(0,0.4225) (1,0.6100) (2,0.7067) (3,0.7517) (4,0.7725)};
\addlegendentry{single agent, final plan}
\addplot[famDesign!60, thick, densely dashed, mark=square, mark size=1.2pt] coordinates {(0,0.3708) (1,0.4800) (2,0.5908) (3,0.6300) (4,0.6717)};
\addlegendentry{single agent, draft}
\addplot[ink2, dashed, mark=o, mark size=1.2pt] coordinates {(0,0.5979) (1,0.4467) (2,0.3433) (3,0.2971) (4,0.2646)};
\addlegendentry{calls truncated at the cap}
\node[dlab, anchor=east] at (axis cs:-0.3,1.06) {$d$};
\foreach \x/\d in {0.5/+0.295, 1.5/+0.080, 2.5/+0.024, 3.5/+0.021} {\edef\temp{\noexpand\node[dlab] at (axis cs:\x,1.06) {$\d$};}\temp}
\end{axis}
\end{tikzpicture}\hfill
\begin{tikzpicture}[baseline=(ax.north)]
\begin{axis}[chainax, name=ax, title={(b) HiddenBench (T3k), every round}, ylabel={confirmation accuracy}, ymin=0.45, ymax=1.0, ytick={0.5,0.6,0.7,0.8,0.9}, xmin=0.4, xmax=8.6, xtick={1,...,8}, xlabel={round}]
\addplot[pend, mark=o, mark size=1.3pt, only marks] coordinates {(1,0.4913) (2,0.7344) (4,0.7800) (5,0.7900) (7,0.8237)};
\addlegendentry{predecessor}
\addplot[famSelf, mark=*, mark size=1.4pt, only marks] coordinates {(1,0.6946) (2,0.7788) (4,0.8238) (5,0.8550) (7,0.8700)};
\addlegendentry{released ($d>0.026$)}
\addplot[pend!70, mark=diamond, mark size=1.5pt, only marks] coordinates {(3,0.8075) (6,0.8263) (8,0.9183)};
\addlegendentry{flat round, candidate}
\addlegendimage{empty legend}\addlegendentry{}
\foreach \x/\a/\b in {1/0.4913/0.6946, 2/0.7344/0.7788, 4/0.7800/0.8238, 5/0.7900/0.8550, 7/0.8237/0.8700}{\edef\temp{\noexpand\draw[famSelf!60, thick, -{Stealth[length=2.5pt]}] (axis cs:\x,\a) -- (axis cs:\x,\b);}\temp}
\foreach \x/\a in {3/0.8069, 6/0.8250, 8/0.9038}{\edef\temp{\noexpand\draw[pend!70, fill=white] (axis cs:\x,\a) circle (1.3pt);}\temp}
\foreach \x/\y/\g in {1/0.6946/G1, 2/0.7788/G2, 4/0.8238/G3, 5/0.8550/G4, 7/0.8700/G5}{\edef\temp{\noexpand\node[glab, text=famSelf, anchor=south] at (axis cs:\x,\y+0.012) {\g};}\temp}
\node[dlab, anchor=east] at (axis cs:0.45,0.475) {$d$};
\foreach \x/\d in {1/+.20, 2/+.04, 3/+.00, 4/+.04, 5/+.07, 6/+.00, 7/+.05, 8/+.01} {\edef\temp{\noexpand\node[dlab] at (axis cs:\x,0.475) {$\d$};}\temp}
\end{axis}
\end{tikzpicture}\hfill
\begin{tikzpicture}[baseline=(ax.north)]
\begin{axis}[chainax, name=ax, title={(c) Hanabi, each release}, ylabel={challenge score}, ymin=-0.9, ymax=5.2, ytick={0,1,2,3,4,5}, xmin=0.5, xmax=6.5, xtick={1,...,6}, xticklabels={G1,G2,G3,G4,G5,G6}, xlabel={released generation}]
\addplot[pend, mark=o, mark size=1.3pt, only marks] coordinates {(1,0) (2,0.590) (3,3.008) (4,4.212) (5,4.437) (6,4.357)};
\addlegendentry{predecessor}
\addplot[famDesign!80!black, mark=*, mark size=1.4pt, only marks] coordinates {(1,0.678) (2,2.953) (3,4.110) (4,4.420) (5,4.523) (6,4.482)};
\addlegendentry{recursive training}
\addplot[ink2!70, mark=square*, mark size=1.1pt, only marks] coordinates {(2.28,2.000) (3.28,2.907)};
\addlegendentry{non-recursive control}
\addlegendimage{empty legend}\addlegendentry{}
\foreach \x/\a/\b in {1/0/0.678, 2/0.590/2.953, 3/3.008/4.110, 4/4.212/4.420, 5/4.437/4.523, 6/4.357/4.482}{\edef\temp{\noexpand\draw[famDesign!60, thick, -{Stealth[length=2.5pt]}] (axis cs:\x,\a) -- (axis cs:\x,\b);}\temp}
\foreach \x/\a/\b in {2.28/0.635/2.000, 3.28/1.925/2.907}{\edef\temp{\noexpand\draw[ink2!50, -{Stealth[length=2.2pt]}] (axis cs:\x,\a) -- (axis cs:\x,\b); \noexpand\draw[ink2!60, fill=white] (axis cs:\x,\a) circle (1.1pt);}\temp}
\node[dlab, anchor=east] at (axis cs:0.5,-0.55) {$d$};
\foreach \x/\d in {1/+0.68, 2/+2.36, 3/+1.10, 4/+0.21, 5/+0.09, 6/+0.13} {\edef\temp{\noexpand\node[dlab] at (axis cs:\x,-0.55) {$\d$};}\temp}
\end{axis}
\end{tikzpicture}
\caption{\textbf{Recursion on PlanBench, HiddenBench and Hanabi.} \textbf{(a)}~PlanBench, generations 0 to 4 on the selection set: the group, the oracle that counts a problem when any of the four agents is right, the single agent's final plan and first draft, and the fraction of calls that reach the token cap. The top row gives each generation's paired gain $d$ over its predecessor on a fresh 800-problem confirmation slice. \textbf{(b)}~HiddenBench (T3k), all eight rounds: each candidate against the current generation on the fresh confirmation seeds of its round. Five rounds clear the threshold and release G1 to G5; in the three flat rounds the candidate does not, and the current generation is kept. \textbf{(c)}~Hanabi: each released generation against its predecessor on 600 fresh deals, and the two generations of the non-recursive control run. Pairs are measured on different confirmation sets and are not connected across rounds.}
\label{fig:chains}
\end{figure}
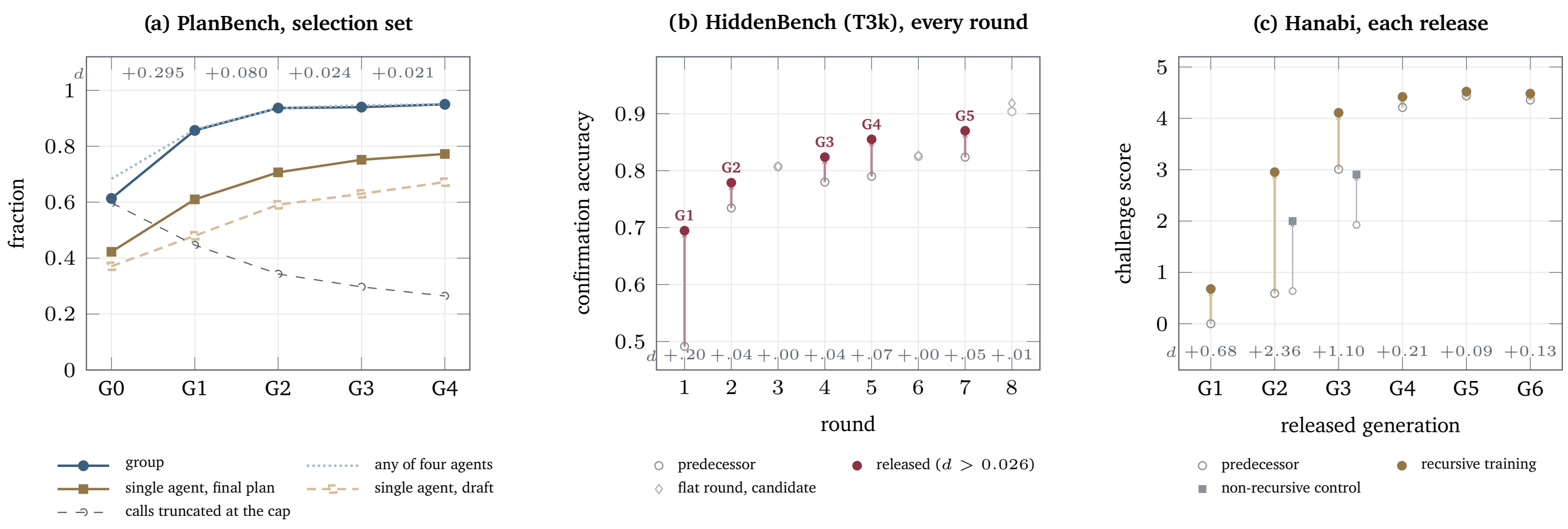

%% file: tables/bench_planbench.tex
\begin{table}[!htbp]
\centering
\tsz
\setlength{\tabcolsep}{4.5pt}
\renewcommand{\arraystretch}{1.08}
\caption{\textbf{PlanBench (Blocksworld plan generation).} Four Qwen3.5-9B agents on the selection set of 300 Blocksworld problems; every generation is trained from its predecessor on the group's own plans accepted by a plan simulator. The group submits the final plan that its agents' anonymous reviews rate highest. \emph{Single agent} scores one agent's first draft and its final plan, and \emph{Any of four} counts a problem when any agent's final plan is valid. \emph{Truncated} is the fraction of calls that reach the 8{,}192-token cap. $d$ is the paired gain over the predecessor on a fresh 800-problem confirmation slice, with its standard error.}
\label{tab:bench-planbench}
\begin{tabular}{@{}lcccccccc@{}}
\toprule
& & \multicolumn{2}{c}{\textbf{Single agent}} & & & & & \\
\cmidrule(lr){3-4}
\textbf{Generation} & \textbf{Group} & \textbf{Draft} & \textbf{Final} & \textbf{Any of four} & \textbf{Truncated} & \textbf{Steps / optimal} & \textbf{Tokens per call} & \textbf{$d$ (SE)} \\
\midrule
G0 (base) & 0.613 & 0.371 & 0.423 & 0.683 & 0.598 & 1.017 & 7{,}416 & \blank \\
G1 & 0.857 & 0.480 & 0.610 & 0.860 & 0.447 & 1.018 & 6{,}802 & $+0.295$ (0.020) \\
G2 & 0.937 & 0.591 & 0.707 & 0.937 & 0.343 & 1.019 & 6{,}651 & $+0.080$ (0.014) \\
G3 & 0.940 & 0.630 & 0.752 & 0.947 & 0.297 & 1.018 & 6{,}523 & $+0.024$ (0.012) \\
G4 & \textbf{0.950} & \textbf{0.672} & \textbf{0.773} & \textbf{0.950} & 0.265 & 1.022 & 6{,}541 & $+0.021$ (0.010) \\
\bottomrule
\end{tabular}
\end{table}

%% file: figures/bench_hiddenbench.tex
\begin{figure}[!htbp]
\centering
\pgfplotsset{hbax/.style={scale only axis, height=0.215\textwidth, font=\scriptsize, tick label style={font=\scriptsize}, label style={font=\scriptsize}, title style={font=\scriptsize\bfseries, yshift=-2pt, align=center},
  axis line style={ink2}, tick style={ink2}, clip=false,
  legend style={font=\scriptsize, draw=none, fill=none, row sep=-1pt, column sep=3pt, at={(0.5,-0.40)}, anchor=north}, legend cell align=left}}
\tikzset{gap/.style={font=\tiny\bfseries, inner sep=1pt}}
\begin{tikzpicture}[baseline=(ax.north)]
\begin{axis}[hbax, name=ax, width=0.25\textwidth, title={(a) Collective training vs.\ single-agent\\self-training (selection set)}, ylabel={accuracy}, ymin=0, ymax=1.0, ytick={0,0.2,0.4,0.6,0.8,1}, xmin=-0.3, xmax=5.3, xtick={0,...,5},
  xticklabels={G0\\[-1pt]{\tiny 19.2},G1\\[-1pt]{\tiny 22.6},G2\\[-1pt]{\tiny 22.4},G3\\[-1pt]{\tiny 22.1},G4\\[-1pt]{\tiny 21.2},G5\\[-1pt]{\tiny 20.3}}, xticklabel style={align=center},
  xlabel={generation \;(T3k tokens per game, k)}, grid=major, grid style={gray!15}, legend columns=1]
\addplot[famSelf, thick, mark=*, mark size=1.4pt] coordinates {(0,0.509) (1,0.713) (2,0.791) (3,0.811) (4,0.845) (5,0.886)};
\addlegendentry{collective T3k, group}
\addplot[famSelf!55, thin, densely dotted, mark=*, mark size=0.8pt] coordinates {(0,0.501) (1,0.710) (2,0.774) (3,0.780) (4,0.828) (5,0.878)};
\addlegendentry{collective T3k, one agent}
\addplot[heather, thick, mark=square*, mark size=1.2pt] coordinates {(0,0.509) (1,0.703) (2,0.755) (3,0.836)};
\addlegendentry{collective T3, group}
\addplot[pend, thick, dashed, mark=o, mark size=1.3pt] coordinates {(0,0.070) (1,0.396) (2,0.563) (3,0.658) (4,0.672)};
\addlegendentry{self-training, four agents}
\addplot[pend!70, thick, densely dotted, mark=triangle, mark size=1.4pt] coordinates {(0,0.148) (1,0.408) (2,0.475) (3,0.558) (4,0.564)};
\addlegendentry{self-training, one agent}
\foreach \x/\a/\b/\t in {1/0.396/0.703/+0.31, 2/0.563/0.755/+0.19, 3/0.658/0.836/+0.18}{\edef\temp{\noexpand\draw[heather!80, line width=0.6pt, {Latex[length=2pt]}-{Latex[length=2pt]}] (axis cs:\x+0.2,\a) -- (axis cs:\x+0.2,\b) node[midway, right, gap, text=heather, fill=white, fill opacity=0.9, text opacity=1] {\t};}\temp}
\end{axis}
\end{tikzpicture}\hfill
\begin{tikzpicture}[baseline=(ax.north)]
\begin{axis}[hbax, name=ax, width=0.19\textwidth, title={(b) Protocols with Qwen3.5-9B\\(65 official tasks)}, xbar, bar width=4.6pt, bar shift=0pt, xmin=0, xmax=1.0, xtick={0,0.5,1}, xlabel={accuracy},
  ymin=0.4, ymax=9.6, ytick={1,...,9}, y dir=reverse, axis y line*=left, axis x line*=bottom,
  yticklabels={\textbf{Fact forwarding (ours)}, Shared context, pass@8, Debate 3r + hint, Blackboard + hint, Debate 2r + hint, Single agent, Debate 2r, Majority vote of 8},
  yticklabel style={font=\tiny}, xmajorgrids, grid style={gray!15}, legend columns=1,
  nodes near coords, nodes near coords style={font=\tiny, anchor=west, inner sep=1.5pt, /pgf/number format/fixed, /pgf/number format/precision=3, /pgf/number format/fixed zerofill}]
\addplot[draw=none, fill=famFact, forget plot] coordinates {(0.834,1)};
\addplot[draw=none, fill=famFact!45, forget plot] coordinates {(0.748,2)};
\addplot[draw=none, fill=famSolo, forget plot] coordinates {(0.624,3)};
\addplot[draw=none, fill=famArgue, forget plot] coordinates {(0.569,4) (0.407,5) (0.285,6)};
\addplot[draw=none, fill=famSolo!60, forget plot] coordinates {(0.176,7)};
\addplot[draw=none, fill=famArgue, forget plot] coordinates {(0.132,8)};
\addplot[draw=none, fill=famSolo!60, forget plot] coordinates {(0.099,9)};
\addlegendimage{area legend, draw=none, fill=famFact}\addlegendentry{facts passed on}
\addlegendimage{area legend, draw=none, fill=famArgue}\addlegendentry{debate, blackboard}
\addlegendimage{area legend, draw=none, fill=famSolo!60}\addlegendentry{independent agents}
\addlegendimage{legend image code/.code={\draw[famSolo!80!black, dashed, line width=0.6pt] (0cm,0cm) -- (0.5cm,0cm);}}\addlegendentry{pass@8 level}
\draw[famSolo!80!black, dashed, line width=0.5pt] (axis cs:0.624,0.4) -- (axis cs:0.624,9.6);
\draw[famFact, line width=0.6pt, -{Latex[length=2pt]}] (axis cs:0.624,0.62) -- (axis cs:0.834,0.62) node[midway, above, gap, text=famFact!80!black] {+0.21};
\end{axis}
\end{tikzpicture}\hfill
\begin{tikzpicture}[baseline=(ax.north)]
\begin{axis}[hbax, name=ax, width=0.19\textwidth, title={(c) Published: group vs.\ one agent\\holding all facts}, xmin=0, xmax=1.0, xtick={0,0.5,1}, xlabel={accuracy},
  ymin=0.4, ymax=5.6, ytick={1,...,5}, y dir=reverse, yticklabels={Qwen3-8B, Qwen3-32B, GPT-4.1, Gemini-2.5-Flash, Gemini-2.5-Pro}, yticklabel style={font=\tiny},
  axis y line*=left, axis x line*=bottom, xmajorgrids, grid style={gray!15}, legend columns=1]
\foreach \y/\a/\b/\t in {1/0.196/0.587/$-$0.39, 2/0.191/0.824/$-$0.63, 3/0.342/0.969/$-$0.63, 4/0.550/0.972/$-$0.42, 5/0.671/0.981/$-$0.31}{\edef\temp{\noexpand\draw[gray!45, line width=1.6pt] (axis cs:\a,\y) -- (axis cs:\b,\y) node[midway, above, gap, text=ink2, yshift=0.5pt] {\t};}\temp}
\addplot[famArgue!80!black, mark=*, mark size=1.7pt, only marks] coordinates {(0.196,1) (0.191,2) (0.342,3) (0.550,4) (0.671,5)};
\addlegendentry{group of agents}
\addplot[ink2, mark=o, mark size=1.7pt, only marks, thick] coordinates {(0.587,1) (0.824,2) (0.969,3) (0.972,4) (0.981,5)};
\addlegendentry{one agent holding all facts}
\addlegendimage{famFact, thick, dashed}\addlegendentry{our fact forwarding, Qwen3.5-9B}
\draw[famFact, thick, dashed] (axis cs:0.834,0.4) -- (axis cs:0.834,5.6);
\end{axis}
\end{tikzpicture}
\caption{\textbf{HiddenBench.} \textbf{(a)}~Four Qwen3.5-9B agents in a three-round debate on the selection set of 200 generated hidden-profile tasks, run with two seeds. Runs T3 and T3k train every generation from its predecessor on the group's verified turns; one agent is an agent's own answer after the debate, and the numbers under the generations are the tokens per T3k game. Single-agent self-training trains each agent on its own solo turns under the same caps and is scored as four independent agents and as one agent; the arrows give the gap between T3 and its matched self-training run. \textbf{(b)}~Protocols on the 65 official tasks with the same base model, from the ladder experiments of Section~\ref{sec:exp-when}; the dashed line marks pass@8 of eight independent agents, which fact forwarding exceeds by 0.21; Appendix Table~\ref*{tab:channels} gives its equal-budget margin. \textbf{(c)}~Results reported by the benchmark's authors on the 65 official tasks~\citep{li2025hiddenbench}, read from their Figure~3: each group against one agent that holds all the facts, with the accuracy the group loses.}
\label{fig:bench-hiddenbench}
\end{figure}
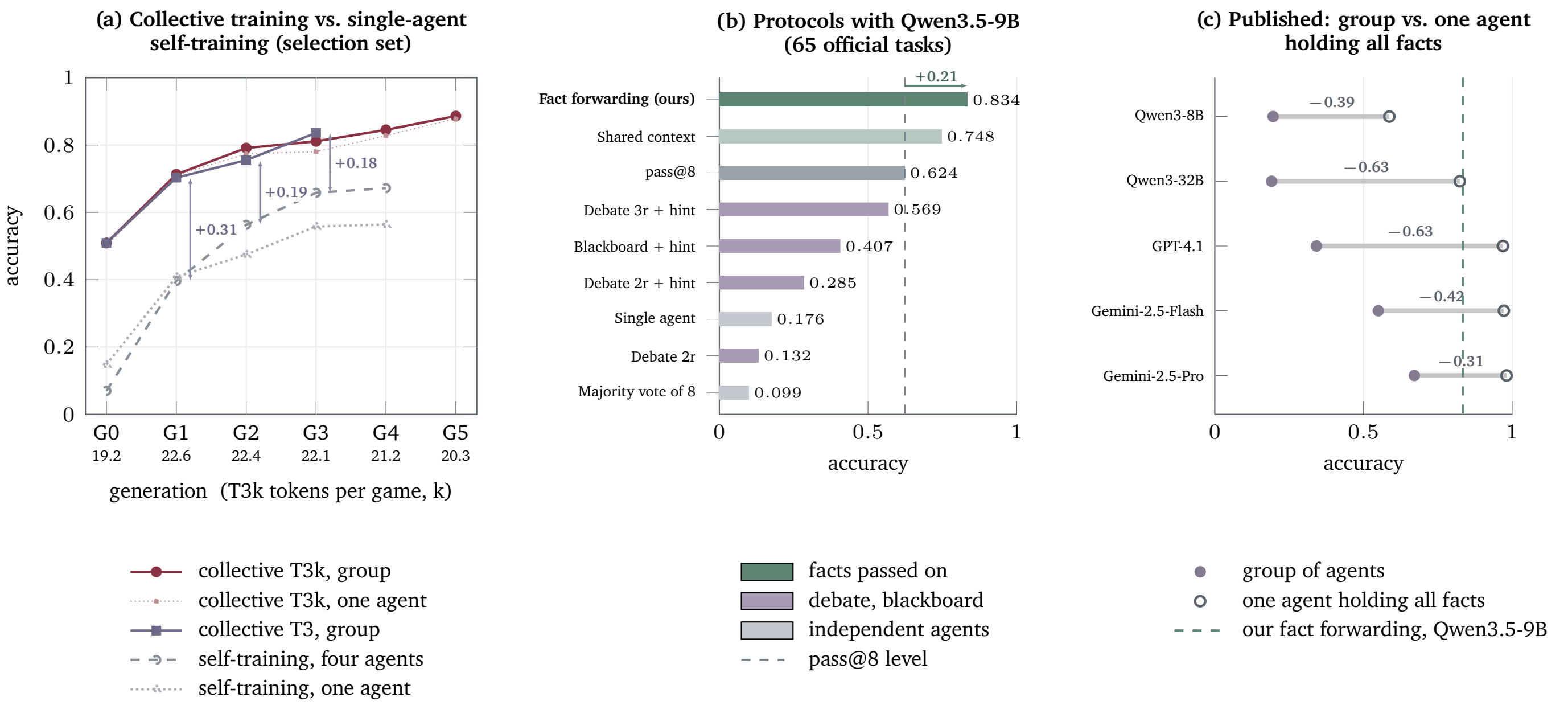

%% file: tables/bench_hanabi.tex
\begin{table}[!htbp]
\centering
\tsz
\setlength{\tabcolsep}{5pt}
\renewcommand{\arraystretch}{1.08}
\caption{\textbf{Hanabi.} Four Qwen3.5-9B agents play the cooperative card game; the challenge score counts the cards played and is zero when the team loses all its lives (maximum 25). Each round compares the candidate with its predecessor on 600 fresh deals, so each row pairs a generation with its predecessor on the same deals. Recursive training trains every generation on the elite games of its predecessor. The control run branches from G1 and trains on games of a fixed generation instead of its own.}
\label{tab:bench-hanabi}
\begin{tabular}{@{}llccccc@{}}
\toprule
& & \multicolumn{2}{c}{\textbf{Challenge score}} & & & \\
\cmidrule(lr){3-4}
\textbf{Run} & \textbf{Generation} & \textbf{Generation} & \textbf{Predecessor} & \textbf{Gain (SE)} & \textbf{Cards played} & \textbf{Lost all lives} \\
\midrule
\tgroup{7}{Recursive training} \\
& G0 (base) & 0.000 & \blank & \blank & 1.362 & 1.000 \\
& G1 & 0.678 & 0.000 & $+0.678$ (0.045) & 1.213 & 0.558 \\
& G2 & 2.953 & 0.590 & $+2.363$ (0.075) & 2.953 & 0.000 \\
& G3 & 4.110 & 3.008 & $+1.102$ (0.087) & 4.110 & 0.000 \\
& G4 & 4.420 & 4.212 & $+0.208$ (0.077) & 4.420 & 0.000 \\
& G5 & \textbf{4.523} & 4.437 & $+0.087$ (0.074) & 4.523 & 0.000 \\
& G6 & 4.482 & 4.357 & $+0.125$ (0.075) & 4.482 & 0.000 \\
\tgroup{7}{Non-recursive control} \\
& G2 & 2.000 & 0.635 & $+1.365$ (0.069) & 2.003 & 0.002 \\
& G3 & 2.907 & 1.925 & $+0.982$ (0.085) & 2.917 & 0.005 \\
\bottomrule
\end{tabular}
\end{table}

%% file: figures/hb_ladder.tex
\begin{figure}[t]
\centering
\begingroup
\definecolor{lzTwo}{HTML}{A9BCCB}\definecolor{lzThree}{HTML}{5C7790}\definecolor{lzNavy}{HTML}{25374F}
\pgfplotsset{hblad/.style={scale only axis, width=0.62\textwidth, height=0.25\textwidth, font=\rmfamily\footnotesize,
  tick label style={font=\rmfamily\footnotesize, text=ink}, label style={font=\rmfamily\bfseries\footnotesize, text=ink}, axis line style={ink2}, tick style={ink2}, clip=false,
  ymin=0, ymax=1.0, ytick={0,0.2,0.4,0.6,0.8,1}, ylabel={accuracy}, xmin=-0.55, xmax=5.55, xtick={0,...,5},
  xticklabels={\textbf{G0}\\[-1pt]{\scriptsize $n_B$ 13.9},\textbf{G1}\\[-1pt]{\scriptsize 9.5},\textbf{G2}\\[-1pt]{\scriptsize 9.6},\textbf{G3}\\[-1pt]{\scriptsize 9.6},\textbf{G4}\\[-1pt]{\scriptsize 8.5},\textbf{G5}\\[-1pt]{\scriptsize 7.9}},
  xticklabel style={align=center}, xlabel={generation on HiddenBench (budget $n_B$ in independent agents)},
  ymajorgrids, grid style={gray!12}, axis on top=false,
  legend style={font=\rmfamily\bfseries\small, text=ink, draw=none, fill=none, column sep=8pt, at={(0.5,1.03)}, anchor=south}, legend columns=4, legend cell align=left}}
\tikzset{lvl/.style={circle, inner sep=0pt, minimum size=11pt, font=\rmfamily\bfseries\scriptsize, text=white}}
\begin{tikzpicture}
\begin{axis}[hblad]
\foreach \g/\vone/\tn in {0/0.1350/0.6575, 1/0.1465/0.6525, 2/0.1644/0.6833, 3/0.1458/0.6479, 4/0.1542/0.6441, 5/0.1781/0.6564}{
  \edef\temp{\noexpand\fill[sage!20] (axis cs:\g-0.42,\tn) rectangle (axis cs:\g+0.42,1.0);
             \noexpand\fill[lzTwo!28] (axis cs:\g-0.42,\vone) rectangle (axis cs:\g+0.42,\tn);
             \noexpand\draw[sage, line width=1.3pt] (axis cs:\g-0.42,\tn) -- (axis cs:\g+0.42,\tn);
             \noexpand\draw[famSolo, line width=0.6pt] (axis cs:\g-0.42,\vone) -- (axis cs:\g+0.42,\vone);}\temp}
\node[font=\rmfamily\bfseries\small, text=sage!80!black, anchor=west] at (axis cs:5.62,0.83) {level 3};
\node[font=\rmfamily\bfseries\small, text=lzThree, anchor=west] at (axis cs:5.62,0.42) {level 2};
\addlegendimage{sage, line width=1.3pt}\addlegendentry{pass@$n_B$}
\addplot[famArgue, only marks, mark=diamond*, mark size=2pt] coordinates {(0,0.0437) (1,0.0559) (2,0.0863) (3,0.0831) (4,0.0910) (5,0.1131)};
\addlegendentry{vote of $n_B$ agents}
\addlegendimage{famSolo, line width=0.6pt}\addlegendentry{one agent}
\addplot[lzNavy, thick, mark=*, mark size=1.8pt, error bars/.cd, y dir=both, y explicit, error bar style={lzNavy, line width=0.6pt}, error mark options={rotate=90, mark size=1.6pt, lzNavy}]
  coordinates {(0,0.5088) +- (0,0.029) (1,0.7133) +- (0,0.026) (2,0.7913) +- (0,0.024) (3,0.8113) +- (0,0.022) (4,0.8450) +- (0,0.022) (5,0.8862) +- (0,0.018)};
\addlegendentry{group}
\node[lvl, fill=lzThree] at (axis cs:0.3,0.43) {2};
\foreach \g/\y in {1/0.7133, 2/0.7913, 3/0.8113, 4/0.8450, 5/0.8862}{
  \edef\temp{\noexpand\node[lvl, fill=sage] at (axis cs:\g+0.27,\y+0.045) {3};}\temp}
\draw[lzNavy!70, line width=0.5pt, {Latex[length=2pt]}-{Latex[length=2pt]}] (axis cs:-0.25,0.6575) -- (axis cs:-0.25,0.5088) node[midway, right, font=\rmfamily\bfseries\scriptsize, inner sep=1.5pt] {$-0.15$};
\draw[lzNavy!70, line width=0.5pt, {Latex[length=2pt]}-{Latex[length=2pt]}] (axis cs:4.75,0.6564) -- (axis cs:4.75,0.8862) node[midway, right, font=\rmfamily\bfseries\scriptsize, inner sep=1.5pt] {$+0.23$};
\end{axis}
\end{tikzpicture}
\endgroup
\caption{\textbf{HiddenBench generations climb the equal-budget ladder.} Each generation is graded on the 200 selection tasks against 24 independent agents of its own model, each holding the shared facts and one seat's private facts. Shaded steps mark the zone of each level: above one agent and the vote for level 2 (blue), and above pass@$n_B$ for level 3 (green). The base group starts at level 2, and every trained generation reaches level 3 at a smaller budget. The independent agents form a fresh pool, so one agent's accuracy differs slightly from Figure~\ref{fig:bench-hiddenbench}a.}
\label{fig:hb-ladder}
\end{figure}
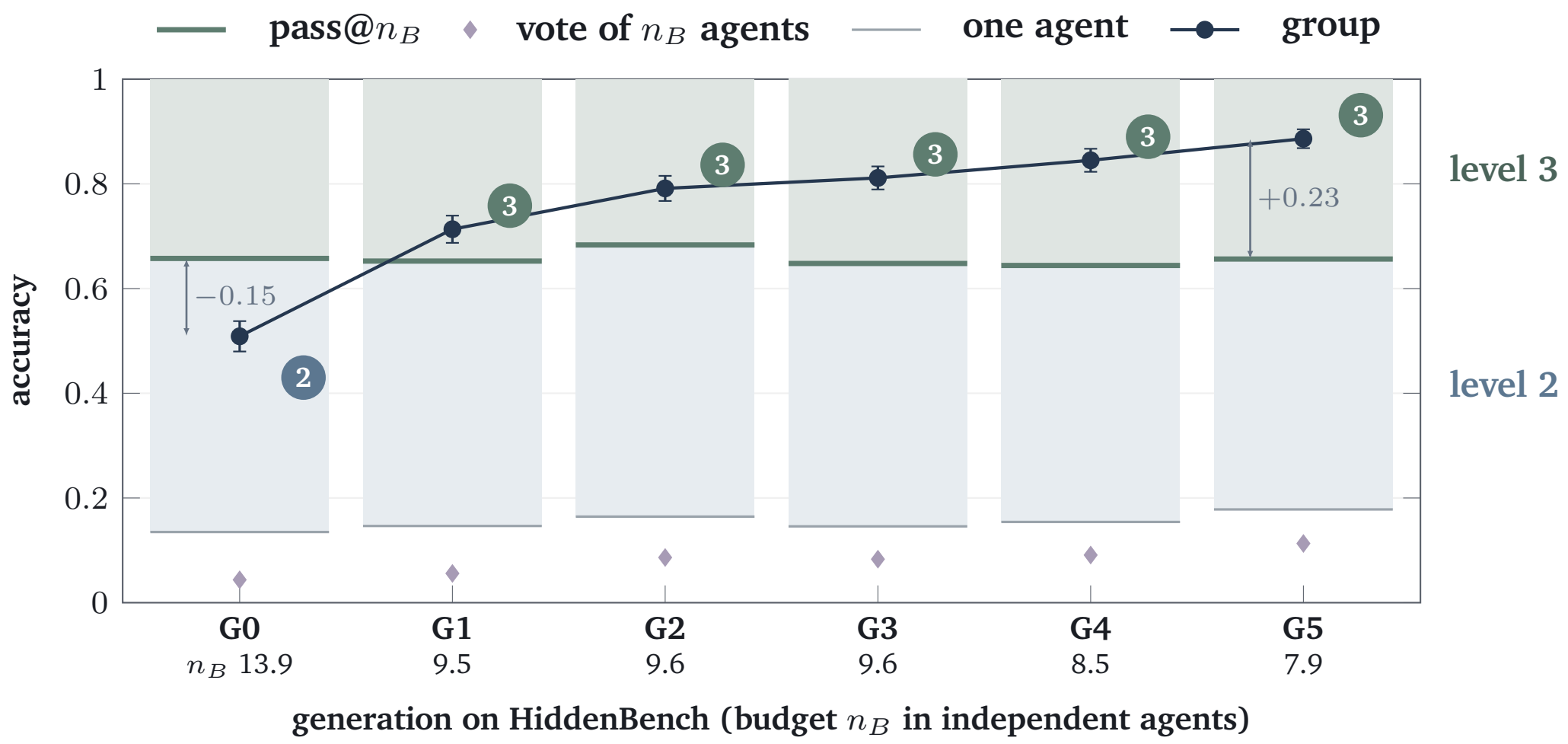

%% file: subtex/discussion.tex
\section{Discussion and Conclusion}
\label{sec:discussion}

The experiments above connect three questions that are usually studied apart: when a group outperforms its agents, how a group organizes itself, and whether a group can make its next generation better. Four lessons follow, and each one turns a qualitative argument about multi-agent systems into a measured condition.

\paragraph{Information structure governs both inference and training.} At test time, interaction exceeds pass@$n$ only when it moves facts that no single agent holds (Section~\ref{sec:exp-when}). The same condition reappears in training. On HiddenBench, where facts are split across agents, collective training stays far above single-agent self-training in every generation, although both use the same caps (Section~\ref{sec:exp-chain}). A collective learns what individuals cannot learn alone when its work contains information that no individual can gather alone, and adding agents to a task without this structure adds samples that the ladder bounds.

\paragraph{Recursion is limited by headroom at the frontier.} A group learns from problems that it solves some of the time. As it improves, more problems become easy and fewer remain at its frontier. On PlanBench the confirmed gain narrows from the first generation to the fourth as the group approaches the top of the selection set, and on Hanabi the later steps are small compared with the first ones. A self-improving collective therefore needs a source of new problems at its frontier as much as it needs a training rule.

\paragraph{Small steps need fresh evidence.} Continuing from the current generation sustains releases on three benchmarks, and its later steps are small. Each decision to release therefore rests on a slice that no earlier round has seen. A confirmation set reused over many rounds inflates the score of the generation it once selected, and true improvements then rarely clear the gate.

\paragraph{A gain can come from a cheaper behavior.} The ladder splits every margin into an accuracy channel and a cost channel (Equation~\ref{eq:channels}). Across generations the same split is needed. Hanabi candidates first lost fewer lives by playing fewer cards, and PlanBench generations truncate fewer calls as they spend fewer tokens per call than the base. Each of these changes raises a score without showing that the group plans better. Reporting the cost of every generation beside its accuracy, and filtering training data for the behaviors that a score can reward without the skill, separates recursive improvement from cheaper behavior.

%% file: subtex/conclusion.tex
\paragraph{Conclusion.}
\label{sec:conclusion}

The equal-budget ladder shows when a group of agents is smarter than its members working apart: interaction pays only when it gives agents information that they cannot gather alone. With shared information, prior multi-agent methods stay at the compute-matched vote, whereas forwarding split facts exceeds pass@$n$ at three model scales. A group therefore outperforms its members only when the information it exchanges is worth more than the tokens it spends on coordination. A group can also learn this from its own turns: two label-free 9B generations improve in turn and beat a control, establishing recursive collective intelligence. Multi-agent systems should thus be evaluated at equal cost and trained to move information efficiently. \looseness=-1

Recursion over more generations follows the same principle. On PlanBench, HiddenBench and Hanabi, each generation outperforms its predecessor for four to six successive generations, and collective training outperforms single-agent self-training where information is split across agents. Sustained recursion then depends on a supply of problems at the group's frontier and on fresh evidence for every release. We therefore think a self-improving collective should draw on a source of new problems. Each release should pass a gate on unseen problems, and every comparison should hold the cost equal. \looseness=-1

%% file: subtex/appendix.tex
\section{Overview of the Appendix}
\label{app:overview}

Appendix~\ref{app:proofs} proves the propositions of Section~\ref*{sec:method}. Appendix~\ref{app:ladder} specifies the equal-budget ladder, and Appendix~\ref{app:protocols} the hidden-profile protocols, scarce attention and the structure reader. Appendix~\ref{app:ledger} describes the token accounting, Appendix~\ref{app:recursion} the recursion experiments and Appendix~\ref{app:prereg} the rule that names a result recursion. Appendix~\ref{app:org} describes the blackboard and the agent harness, Appendix~\ref{app:audit} the constructed tasks and Appendix~\ref{app:eval} the evaluation protocols of the six shared-information benchmarks. Appendix~\ref{app:results} reports additional results, including the setup of the credit experiment, Appendix~\ref{app:chains} recursive training on PlanBench, HiddenBench and Hanabi, and Appendix~\ref{app:repro} compute and reproduction.

\section{Proofs}
\label{app:proofs}

\paragraph{Proposition~\ref*{prop:bound}.}
\emph{(i)} A draw of $n$ independent members fails on question $x$ only if all $n$ fail, which happens with probability $(1-p_x)^n$, so $T(n)=1-\mathbb{E}_x[(1-p_x)^n]$. The increment $T(n+1)-T(n)=\mathbb{E}_x[p_x(1-p_x)^n]$ is nonnegative and nonincreasing in $n$, so $T$ is nondecreasing and concave. Our references draw members without replacement from a pool of $m$ members, $k_x$ of them right on question $x$. The failure probability is then $q_n=\prod_{j<n}(m-k_x-j)/(m-j)$, the increment $q_n-q_{n+1}=q_n k_x/(m-n)$ is nonnegative, and the ratio of consecutive increments is $(m-k_x-n)/(m-n-1)\le1$ whenever $k_x\ge1$, so the same conclusions hold; linear interpolation between grid sizes preserves them.
\emph{(ii)} Let a rule observe answers $a_1,\dots,a_n$ and commit to some $a_j$. On any question the committed answer can be right only if some $a_i$ is right, so the rule's accuracy on that question is at most the probability that some member is right, and averaging over questions and draws gives at most $T(n)$. The argument holds for any way of choosing $j$, so it covers voting, confidence weighting, verifiers and judges. The majority vote over $n$ members is such a rule, so $V(n)\le T(n)$, and $V(1)=T(1)\le T(n)$ by part (i). A group above $T(n_B)$ is therefore above $V(1)$ and $V(n_B)$ as well.
\emph{(iii)} By part (i) the increments of $T$ never exceed the first one, $T(1)=V(1)$, so $T(n)\le n\,V(1)$, and $T(n)\le1$ trivially; for draws without replacement the first increment is $k_x/m$, again the accuracy of one member. The same bound gives $T(n)-T(m)\le(n-m)\,V(1)$ for $n>m$. \hfill$\square$

\paragraph{Proposition~\ref*{prop:threshold}.} Let the group have reported $r$ distinct claims, $h$ of them true, with $t$ true findings, and add one claim with calibrated confidence $q$, the probability that it is true. For a score that is linear in the numbers of true and false claims, the expected effect of the claim is $q$ times the effect of a true claim plus $(1-q)$ times the effect of a false one.
\emph{Recall} $h/t$ rises by $1/t$ for a true claim and stays the same for a false one, so the expected change $q/t$ is positive for every $q>0$.
\emph{F1} equals $2h/(r+t)$. After the claim it equals $2(h+1)/(r+1+t)$ if the claim is true and $2h/(r+1+t)$ otherwise, so its expectation is $2(h+q)/(r+t+1)$. This exceeds $2h/(r+t)$ exactly when $(h+q)(r+t)>h(r+t+1)$, that is when $q>h/(r+t)=\mathrm{F1}/2$.
\emph{Confidence-gated claims} score $(h-c(r-h))/t$. A true claim adds $1/t$ and a false one $-c/t$, so the expected change $(q-c(1-q))/t$ is positive exactly when $q>c/(1+c)$. \hfill$\square$

\paragraph{Shapley shares of confidence-gated claims.} Write $P_S=\bigcup_{i\in S}P_i$ and let $w(p)=+1$ for a true claim and $-c$ for a false one. Then $v(S)=\frac{1}{|T|}\sum_{p\in P_S}w(p)=\frac{1}{|T|}\sum_p w(p)\,u_p(S)$ with $u_p(S)=\mathbb{1}[\exists i\in S: p\in P_i]$. Each $u_p$ is a game in which a coalition wins as soon as it contains one of the $k_p$ members that reported $p$. By symmetry and the null-player property, its Shapley share is $1/k_p$ for each reporter and $0$ for everyone else. By additivity $\phi_i=\frac{1}{|T|}\sum_{p\in P_i}w(p)/k_p$, so every claim's payoff or cost is split equally among its reporters, and the shares sum to $R$. \hfill$\square$

\paragraph{Mean credit.} Summing the credit of Section~\ref*{sec:credit} over members gives $N\lambda R+(1-\lambda)N\sum_i f_i$. The flow moves positive shares between members and keeps every unit it receives, so $\sum_i f_i=\sum_i\phi_i=R$, and the mean credit is $R$. \hfill$\square$

\section{Grading on the Equal-Budget Ladder}
\label{app:ladder}

\paragraph{Unit cost.} $c_1$ is the total number of completion tokens of the independent pool of the same model, protocol family and platform, divided by the number of pool members that answer. Members that never answer are therefore paid for but not counted. DocSweep, whose questions differ in cost, uses a per-question $c_1(x)$. A group record's budget is its completion tokens divided by $c_1$, and a cell's budget $n_B$ is the mean over its records, so $n_B$ falls below one when a protocol spends less than an average independent member. Completion tokens are the unit of every level we report.

\paragraph{References.} $V(n)$ is the expected score of the group's own commitment rule applied to $n$ members drawn without replacement from the pool. The rule picks the answer with the most backers, breaks ties by the larger total stated confidence, and splits any remaining ties evenly. $T(n)$ is the probability that at least one of the $n$ members is right. Both curves are computed from the counts of right and wrong answers per question on the grid $1, 2, 3, 4, 6, 8, 12, 16, 24, 32, 48$ members and interpolated linearly between grid sizes; DocSweep computes them exactly at every integer. On hidden-profile tasks the draw is stratified by seat, so an ensemble of size $n$ contains seats in the same proportion as the real group, and sizes between multiples of the group size are drawn as evenly as possible. The stratified $T$ is exact, and $V$ is exact by enumeration or estimated from at least 2{,}000 draws per question and size. A strict pool never reads a reference beyond its own size. A cell whose budget exceeds its pool's largest size is graded at levels 1 and 2 against the references at that size; its third level is recorded as not gradable in that unit, shown as the level reached and marked capped.

\paragraph{Pool size.} From the budget-cap experiments on, a small smoke run measures each cell's budget before the main cells run. The pool then holds $S\lceil 1.5\,\max n_B/S\rceil$ members per question for groups of $S$ members, with at least 64 members; the budget-cap experiment with four members received 68 members per question. The HiddenBench ladders of Table~\ref*{tab:ladder} use the earlier pool of 32 members per task, and DocSweep's pool has 16 members per question.

\paragraph{Tests.} Every level is a one-sided test at 0.025 of the paired difference between $Y$ and the reference, with question-clustered standard errors, and levels are tested in order: a cell reaches level $k$ when this test confirms levels 1 to $k$. An unconfirmed positive margin keeps the cell on the level below, and a cell below one member sits at level 0. Within a task, the Benjamini and Hochberg procedure over the task's cells marks a pass as not robust when its within-task $q$ exceeds 0.025; it never changes the level. Equivalence to the vote is tested with two one-sided tests at a fixed margin.

\paragraph{A/A tests.} Each experiment specifies its own A/A check. The pool may be split into two halves by seed or by member position, with ensembles from one half graded against the references of the other half; a pseudo-group of pool runs may be graded against the pool without them; or new same-batch seeds may be graded against the existing pool. The checks run at the vote and pass@$n$ levels, with a bound between a 95\% interval that contains zero and $|z|\le2.58$.

\section{Hidden-Profile Protocols and Scarce Attention}
\label{app:protocols}

\input{tables/methods}

\paragraph{Method names.}\label{app:names} Table~\ref{tab:methods} groups the protocols by what members exchange, and the tables use the following short names.
\begin{itemize}[leftmargin=*,itemsep=1pt,topsep=2pt]
  \item \textbf{Independent references} (\famtag{Solo}). \textit{Single} is one independent member, $V(1)$. \textit{Vote@B} and \textit{Pass@B} are majority vote and pass@$n$ over equally costly independent members, $V(n_B)$ and $T(n_B)$. \textit{Sequential-4} is one agent that answers four times in sequence.
  \item \textbf{Interaction baselines} (\famtag{Argue}). \textit{BB} is the blackboard, and \textit{Debate-2R} and \textit{Debate-3R} are debates of two and three rounds~\citep{du2024debate}. \textit{Self-Refine-2R} is a two-round self-correction in which each member reads only its own text, and the suffix \textit{+H} adds a hint that facts may be distributed.
  \item \textbf{Evidence protocols} (\famtag{Fact}). \textit{Fact forwarding} lets members forward only their given facts, and \textit{Shared context} attaches every member's facts with free comments; \textit{SC+Forwarding} combines the two. \textit{Facts-only BB} is a blackboard that accepts only forwarded facts, and \textit{Gated BB} is a blackboard in which only a complete reader may conclude.
  \item \textbf{Scarce attention.} \textit{No transfer} gives each member one reading unit, \textit{Self-organized} allows budget transfers and \textit{+ planning round} adds a two-step planning round. \textit{Budget cap} adds a cap that makes a single hub infeasible and \textit{no relay} forbids relaying under the cap (all \famtag{Self}); a \textit{self-written prompt} is written from the rules or evolved over generations. \textit{Hub} and \textit{Tree} are the designed controls (\famtag{Design}), and groups that follow a self-written prompt, with or without the cap, are \famtag{Notes}.
  \item \textbf{DocSweep.} \textit{Self-organized} (\famtag{Self}), \textit{Self-written prompt} (\famtag{Notes}) and \textit{Aggregator} (\famtag{Design}) are groups of four that organize themselves, follow a self-written prompt or follow a designed aggregator.
  \item \textbf{Recursion} (\famtag{Train}). \textit{G0} is the untrained model and \textit{G1} the generation trained on its protocol-verified first turns. \textit{G2} is the second generation, trained on the shortest verified first turn of each seat sampled from G1, and \textit{control} is its control, trained with the same operator from the same parent on first turns sampled from G0.
\end{itemize}

\paragraph{Tasks.} Each HiddenBench task is a decision among three options. The shared facts favour a decoy, and the private facts together favour another option~\citep{li2025hiddenbench}. Fifty-eight tasks have exactly four private facts. We exclude one task whose facts are table rows. Two offline checks then remove tasks in which one private fact contains another fact's key words or a merged message would be credited with a missing fact, leaving 56. Generated tasks follow the same template: Qwen3.8-27B writes them with a drawn decoy and answer, and a task is admitted when members given every fact agree, members given only the shared facts are misled, and members given their own facts mostly fail. The writer's answer key is kept out of the task file and used only to report label quality.

\paragraph{Messages and budgets.} Members post messages to the blackboard with a sub-task title and a body. A forward tool copies facts word for word, and the host checks every forwarded line against the task's facts and rejects lines that are not facts. The headers of unread messages state whose private facts a message holds, as computed by the host from the content. A member spends reading budget when it opens a message by another member. A transfer moves budget units to another member, and the host rejects a transfer that would exceed the cap and states the cap. Only a member that holds every private fact may post an analysis, and analyses are pushed to all members. A timing rule holds submissions until step 16 unless an analysis has arrived.

\paragraph{Designed controls and self-written prompts.} The designed hub receives the whole reading budget. The designed tree for four members gives the hub two units and the relay one, opens reading for the relay after two steps and for the hub after four, and gives each member a role card that states whom to read and what to forward. The model writes a prompt from a text that states the protocol's rules. Evolved prompts are revised over four generations of 16 generated tasks each, with feedback that describes the group's process, transfers, relays and refusals and never its answers.

\paragraph{Structure reader.} For every record the reader finds the first member to hold every private fact without relying on a pushed analysis. Each fact it holds has a depth: one for a fact forwarded by its owner, and one more than its source message for a relayed fact. A message is necessary when removing it from the reader's read set leaves the reader without every fact at the step it completed. A record is a star when the reader completed through depth-one messages only, a two-level tree when the deepest necessary message has depth two, and a deep relay when it has depth three or more. The null policy keeps or randomly gives its budget, forwards its own facts in the first two steps and then reads uniformly at random; we simulate it 10{,}000 times per question under each protocol.

\section{Coordination Token Accounting}
\input{tables/channels}
\label{app:ledger}

Every member turn is matched to its model call and classified by its tool calls. We count a turn as waiting when it makes no tool call, only waits, or holds a submission. Transfers, and posts or reads before a member's reading opens, count as negotiation. All other turns are work. Completion tokens and newly read prompt tokens are summed per class. The replay without waiting removes every waiting turn and keeps every other turn as recorded, with accuracy held fixed, so it gives an upper bound on the saving of a protocol that leaves waiting members idle. The thinking inside waiting turns is bounded from below by subtracting an upper bound on the visible output from each turn's completion tokens. The token accounting is descriptive, and its per-cell tests are uncorrected.

\section{Recursion Experiments}
\label{app:recursion}

\paragraph{Setup.} Generation 0 is the base model. Generation 1 is trained on first turns of its parent group on SplitBench training tasks that write every private fact of the member and contain no stance, with two training seeds that form two lineages. The second generation of each lineage samples four first turns per seat from its own generation 1 and trains on the shortest one that meets the criterion. Training is full-parameter supervised fine-tuning with batch 8 and learning rate $2\times10^{-6}$. First generations and the second generations of the 3-step debate and of 27B train for 240 steps on the verified first turns alone. The second generation of the 9B standard-debate run trains for 320 steps and adds commit turns, rows in which only the skeleton of a member's submission call is trained, weighted to one quarter of the loss; the 4B run uses an operator of its own (below). Every generation is graded on the 196 held-out SplitBench tasks against a frozen pool of the base model, 32 independent members per question, and on the 65 HiddenBench tasks as a secondary set. Contrasts are paired by question, with standard errors from 2{,}000 bootstrap draws over questions that recompute $n_B$ in every draw. In one lineage, the share of members whose first turn covers every private fact triples in the first generation.

\paragraph{9B in the 3-step debate.} A member takes at most three steps and is then asked once for its answer. A contrast holds when it exceeds zero by 1.96 standard errors and no budget exceeds the pool; the control contrast uses the multiplier of Appendix~\ref{app:prereg}. The second step adds $+0.187$ to the margin and halves the first call of a member. The control repeats the second generation's operator with the same parent, steps and seeds, but samples its first turns from the base model. The second generation is $+0.192$ above it, so all three contrasts hold and the naming rule calls the result weight-level recursion at 9B. Both steps also hold on HiddenBench.

\paragraph{Replication over training seeds.} The second generation and its control are retrained from the same parents and row files on four further training seeds. Pooled over them, the contrast is $+0.115\pm0.008$ and positive in all eight pairs, also when its error includes training randomness, and accuracy carries most of it.

\paragraph{27B in the 3-step debate.} The 9B recipe is repeated at 27B. The first generation trains on the verified first turns of the base group, with the standard-debate prefix of the same question and seat as context, and raises the margin from $-0.009$ to $+0.201$, a step that also holds on HiddenBench and when the error includes training randomness. A second generation, trained on the standard-debate rows of the first generation's verified turns, raises the margin to $+0.284$ at a lower budget and is $+0.268$ above a control trained with the same recipe on base-model turns. Table~\ref{tab:recursion} lists these margins.

\paragraph{4B run.} A 4B run repeats the design in the standard debate with an operator of its own: it keeps the shortest verified first turn of each seat and adds commit turns, of which only the skeleton of the submission call is trained. The first step raises the margin by $+0.421$, holds under the frozen base pool and on HiddenBench, and keeps its sign under each model's own pool. The second generation adds $+0.194$ and reaches the third level.

\paragraph{9B in the standard debate.} The same 9B design in the standard debate gives two steps of $+0.112$ and $+0.179$ on evaluation seeds 10 to 13. On seeds 14 to 17, which Table~\ref{tab:recursion} reports, the two steps are $+0.138$ and $+0.188$, the second with four more training seeds per lineage, and both raise the margin on the 65 official HiddenBench tasks. The second generation is $+0.116$ above a control trained like it on G0's turns, but the naming rule leaves this contrast undecided because the harness closes the control's members more often.

\input{tables/collective}
\paragraph{The gain belongs to the group.} A group can rise because its members learn to solve the task or because they learn to pool what each holds. Every model of the standard-debate run also plays alone with every fact on eight seeds of the 65 HiddenBench tasks. In the first step this solo accuracy stays unchanged within a bound of 0.03, while the group's accuracy under split facts, on the tasks that members of both generations solve alone, rises by $+0.177$; the second step adds $+0.064$ (Table~\ref{tab:collective}a). Each generation of the 3-step debate is also graded against a pool of its own members working independently (Table~\ref{tab:collective}b). At 9B both steps hold under each generation's own pool, the second generation stays above pass@$n$ of its own members, and a member working alone barely improves, so both 9B steps are collective. At 27B the first generation rises far above its own members, while the trained members working alone are no better than base-model members, so the first step is collective. The second generation stays above its own members; its step on the frozen base pool keeps the parent's accuracy at a lower budget, so by Equation~(\ref*{eq:channels}) it comes from the cost term.

\paragraph{Supervisor rewriting the 27B team.} A supervisor with the members' weights reads the unlabeled records of the 27B first-generation group on SplitBench training tasks and rewrites the team's roles and instructions, version by version, without an answer key. On the 196 held-out tasks, the final rewritten team is $+0.111$ above the fixed team and $+0.210$ above a team that the same weights write from the rules alone.

\section{Naming Rule for Recursion}
\label{app:prereg}

The naming rule fixes how a result of the recursion experiments is named. A result is recursion when three contrasts on the equal-budget margin hold: the first generation over the base model, the second generation over its parent, and the second generation over a control with the same parent, steps, sampling budget and verification, whose training turns come from the generation before the parent. The recursion endpoints of the three model scales form one family at the one-sided level 0.025, tested by a Holm step-down whose multipliers 2.39, 2.24 and 1.96 are the normal quantiles at 0.025/3, 0.025/2 and 0.025. The 9B control is judged with the multiplier 2.24. Its ratio, 12.0, exceeds the largest multiplier of the family, 2.39, so it holds at any place of the order. The replication on four further seeds uses 2.50, and every other endpoint uses 1.96.

\section{Shared Blackboard and Agent Harness}
\label{app:org}

\paragraph{Blackboard.} The blackboard is an append-only sequence of messages. A message has four fields: the sub-task it addresses, phrased the way a later searcher would phrase it; the result; a short rationale; and its references to earlier messages or to task resources such as snippets, documents, files and Mathlib declarations. A message receives the next address when written and is never edited or deleted, so references form a directed acyclic graph that grows only at its newest end. Sub-task and result together are limited to 256 tokens on audits and retrieval and to 1{,}024 tokens on proofs and code, and a message has at most 32 references. A publish is accepted only if every reference exists and the rationale names each reference and states its contribution; otherwise nothing is written and the member learns which reference failed.

\paragraph{Search and ranking.} \emph{Search} takes free text and returns the eight best messages, and \emph{visit} reads up to eight messages by address. Both return the four fields, the messages a message cites and up to eight of the newest messages that cite it. Authors and ranking weights stay hidden. Every message carries a weight $\omega_m\in[0,1)$ that starts at zero, grows as $\omega_m\leftarrow\omega_m+0.3(1-\omega_m)$ each time a new message cites it, and decays as $\omega_m\leftarrow0.98\,\omega_m$ at every group step. A query $q$ ranks the messages with positive BM25 relevance by $\omega_m+\mathrm{Rel}(q,m)/\max_{m'}\mathrm{Rel}(q,m')$, with ties going to older messages, so the weights reorder relevant messages and leave irrelevant ones out.

\paragraph{Harness.} All members receive the same prompt and tools and run asynchronously on a global step clock: in each step every active member takes one turn, which is one model call with its tool calls. Every tool call counts as one action, including rejected ones. When a member reaches its action limit or the run reaches its step limit, the member submits its best answer from the work so far. The histories of audit, retrieval, proof and document members are summarized by the same model once they reach ten turns, keeping the latest three turns, and code-repair members keep their full history. Sampling uses temperature 1 and native tool calling. A call generates at most 16{,}384 tokens at evaluation and 8{,}192 in training, except the opening tool-free deliberation, which may use 32{,}768. Inference compute is the sum of prompt and completion tokens of all calls, including summaries. Table~\ref{tab:limits} lists the limits.
\begin{table*}[t]
\centering
\tsz
\caption{\textbf{Per-member limits of the harness} (evaluation; training uses 60 actions and steps on audits, 80 on retrieval).}
\label{tab:limits}
\begin{tabular}{@{}lcccccc@{}}
\toprule
& Audit & Retrieval & Proofs & Code repair & PDFs and long documents & Multiple choice \\
\midrule
actions per member & 250 & 300 & 100 & 300 & 100 & 16 \\
steps per run & 250 & 400 & 120 & 400 & 120 & 24 \\
summarization & yes & yes & yes & no & yes & no \\
\bottomrule
\end{tabular}
\end{table*}

\section{SplitBench, Contradiction Audit and DocSweep}
\label{app:audit}

\paragraph{SplitBench.} SplitBench follows the format of HiddenBench: a decision in which the shared facts point to a decoy and the facts that overturn it are spread across members. Qwen3.8-27B writes 1{,}500 candidate tasks, and none is close to a HiddenBench task in wording (largest word-set Jaccard similarity 0.143). Admission uses only the behaviour of the same model, with four independent members per condition. A task is dropped when members solve it without pooling their facts, when the shared facts alone do not lead members to the decoy, or when members given every fact disagree. The 697 admitted tasks are divided by a hash of their identifier into 501 training tasks and 196 held-out tasks, and both sets are frozen before any generation is trained. On the admitted tasks, the consensus of members given every fact matches the writer's answer key on 98.6\% of tasks.

\paragraph{Corpora.} The evaluation corpora are the Markdown and reStructuredText documentation of 16 projects, fetched at pinned commits: CPython, Docker, Godot, JAX, Kubernetes, the Linux kernel, LLVM, NumPy, pandas, PyTorch, Ray, SciPy, scikit-learn, SQLAlchemy, SymPy and Transformers. We keep reference and guide pages and exclude pages that describe releases, plans, templates or licensing. Paragraphs are packed greedily into snippets of about 1{,}024 tokens without splitting a paragraph, and the first line of every snippet names its file. Training uses the same construction on 35 further projects (21{,}430 snippets), with four of them held out for validation.

\paragraph{Planting contradictions.} Candidate snippet pairs share at least two rare words. A model names up to three shared topics and rewrites the natural-language text of one snippet so that it contradicts the other, and leaves code, tables, list items and headings untouched. A mechanical check bounds the edit to at most 30\% of the text and eight changes, requires more than formatting, and requires every changed sentence to be unique in the corpus. Four independent judges must all accept the pair as natural, non-trivial, self-consistent and a real conflict. Every corpus has 10 same-file and 10 cross-file contradictions over 40 distinct snippets. The level $\times n$ adds $40(n-1)$ distractor snippets drawn with a fixed seed, preferring whole files and keeping the original order; we evaluate at $\times4$.

\paragraph{Tools, answers and scores.} Members read snippets with \texttt{fetch\_snippet} and search them with \texttt{search\_snippet}, an exact, case-insensitive search that returns at most ten matches with context. A claimed contradiction is a snippet pair whose rationale quotes the conflicting text of both snippets verbatim. The group's deliverable is the union of its members' distinct pairs. Exact recall counts planted pairs found with both snippets right, precision is the fraction of committed pairs that are planted, and exact-pair F1 combines the two. The benchmark's original score credits a contradiction when any reported pair contains its edited snippet, which rewards long lists of pairs, so we report it only for comparability. In the distributed setting the snippets are split at random into $N$ shares and each member can open only its own share, so members find cross-share contradictions only by building on what others publish; the task prompt states this for every method alike.

\paragraph{DocSweep.} Each of the 128 questions requires combining facts that are scattered across several documents of a searchable corpus. One type of question joins facts across documents, and the other aggregates over many of them. A question is admitted when a reader that opens every relevant document answers it and the shared corpus offers no shortcut. Admission also requires that a single capped member answers at most one of four seeds. Because admission uses the 27B model's own behaviour, the question set is selected for this model and its external validity is limited. Each group member may take 12 actions, so no single member can open every relevant document. Self-organizing members claim documents on a shared board, plan during a window before reading and follow a timing rule. The designed aggregator assigns shares of the corpus and integrates the digests that members post, and the model writes a prompt from the rules once and freezes it.

\section{Evaluation Protocols and Baselines}
\label{app:eval}

\paragraph{BrowseComp-Plus.} We use two protocols over the fixed corpus of 100{,}195 documents~\citep{chen2025browsecompplus}. Members use search, open and find tools following the open deep-research setup of \citet{li2026openresearcher}, over a BM25 index whose retrieval metrics match the official Lucene index within 0.2 points on all 830 questions. Table~\ref{tab:prior} uses the first 100 questions, graded by Qwen3.6-35B-A3B with the grading prompt of the open deep-research setup.

\paragraph{SWE-bench Verified.} We use the first 100 instances~\citep{jimenez2024swebench}, each in a rootless sandbox of its official evaluation image pinned to its registry digest. Members have one shell tool, and the answer is the repository diff when the member stops. Every distinct patch is applied to a fresh sandbox and graded on the FAIL\_TO\_PASS and PASS\_TO\_PASS tests by the official harness.

\paragraph{LongBench v2.} We use all 503 questions~\citep{bai2025longbenchv2}. Contexts are split into sections of at most 6{,}000 characters that members search and read. A document of at most 65{,}536 tokens is also given whole, and every member first deliberates on it without tools. The single-model reference uses the benchmark's zero-shot template with middle truncation to 120{,}000 tokens.

\paragraph{GPQA-Diamond.} We use all 198 questions from the authors' release~\citep{rein2024gpqa}. Members first deliberate without tools under the task's own instructions, then may read the blackboard, and then submit with a stated confidence. The single-model reference samples eight answers per question with the model card's thinking settings.

\paragraph{MMLongBench-Doc.} The first 300 questions of a seeded shuffle of the 1{,}082 questions over 135 PDF documents~\citep{ma2024mmlongbenchdoc}. Members work page by page: they search the text layers of the pages, read the text of a page and view a page as an image, at most 40 images per member. Answers are short and may state that the question is not answerable. Grading runs the benchmark's answer extraction on a local judge model in place of GPT-4o and then the benchmark's rule scorer. The page-level protocol differs from the benchmark's whole-document protocol, so these scores stand apart from its leaderboard.

\paragraph{FATE-M.} The first 100 of the 150 problems~\citep{jiang2025fate}, each an algebra theorem to prove in Lean~4 against the Mathlib version pinned by the benchmark. Members search Mathlib declarations and compile complete Lean files with a checker tool, at most ten compilations per member, and a member answers with a complete Lean file. A group and an ensemble alike solve a problem when the checker accepts the proof of any member, so Vote@B and Pass@B coincide. Published pass rates use 64 samples on all 150 problems and stand apart from these cells.

\paragraph{Baselines.} All organizations share the harness, tools, task prompts, action limits and graders. Parallel ReAct runs $N$ members without communication. Multi-agent debate broadcasts every member's intermediate work to all others each round. DeLM gives all members one shared context with task decomposition and a queue of sub-tasks, and the choices needed to fit its design to an asynchronous, tool-using harness follow its first public version. AOrchestra lets a coordinator create and dispatch sub-agents and stops when the coordinator stops; on code repair its answer is the diff of the latest sub-agent's workspace. Three adaptations matter for reading Table~\ref{tab:prior}. The debate has no fixed number of rounds: members see the others' work at every step under stable agent labels and may submit before reading it, while the debates of the ladder hold every submission for one or two rounds of reading. DeLM members submit their own answers in place of one final synthesis step. AOrchestra runs every sub-agent and its trace summaries on the same model, without model selection. Parallel ReAct and multi-agent debate commit through a forced final step, while DeLM and AOrchestra commit the answer recorded so far, so a member that records no answer contributes nothing to the vote. All organizations are scored by the same commitment rule, which reduces to a plain vote where a run records no stated confidence. On SWE-bench Verified no rule pools patches, so the committed answer is the patch of one member and Vote@B stays within 0.01 of the accuracy of one independent member.

\section{Additional Results}
\label{app:results}

\subsection{Supporting results for the ladder}
\label{app:ladder-results}

\input{tables/prior}

The comparison of Section~\ref*{sec:exp-symmetric} covers 78 Qwen3.5-9B group cells over six benchmark subsets and the audit task. On the five subsets whose two references differ, no cell is above Pass@B in completion tokens, total tokens, members or calls, and Table~\ref{tab:prior} lists the prior-method cells one by one. Among untrained 9B groups of 16 on the audit task, parallel ReAct reports more distinct pairs per run than the blackboard, while a single blackboard member recalls more of the planted contradictions than a ReAct member. Members of the blackboard build on each other and converge on what was published first. Shapley credit (Section~\ref*{sec:credit}) removes the incentive to copy, because a duplicated finding is split between its reporters.

\paragraph{Other families on the 56-task subset.} With a planning window, on the 56 HiddenBench tasks with exactly four private facts and in another batch, GLM-4.5-Air members almost never transfer budget, and gpt-oss-120b members transfer in over a third of the records but rarely form hubs. Allowing transfers changes neither family's accuracy detectably. Protocol, task set and batch all differ from Table~\ref*{tab:emergence}b, so the two readings are not compared.

\subsection{Shapley credit in detail}
\label{app:credit-results}
\label{app:further}

\input{figures/example}

\paragraph{Training setup.} The credit study trains Qwen3.5-4B on the contradiction audit, eight questions per step, with exact-pair F1 of the group's union as the group score $s$; every member can open every snippet of the corpus. Figure~\ref{fig:example} illustrates the confidence-gated score with $c=0.25$, and the last row of Table~\ref{tab:reward} trains with $c=0.5$; both serve for comparison and differ from the score of Table~\ref*{tab:credit}. Members train in groups of four and are evaluated in groups of four and eight. Each question gets eight rollouts, a member takes at most 60 actions and steps, and a call generates at most 8{,}192 tokens within training sequences of at most 32{,}768. The objective is the clipped policy gradient with $\epsilon_{\mathrm{low}}=0.2$ and $\epsilon_{\mathrm{high}}=0.28$, without a KL term, with four mini-batch updates per rollout batch, learning rate $10^{-6}$ and gradient clipping at 1.0; importance weights truncated at 2 correct for rollouts that overlap the previous optimizer step. Shapley credit uses $\lambda=0.25$ and $\gamma=0.5$. Training runs on four H200 GPUs, two for training and two for inference, and every comparison is paired on 16 held-out corpora and two seeds with 95\% paired bootstrap intervals over 4{,}000 resamples.

\paragraph{Credit rules on one run.} Figure~\ref{fig:example} shows the three credit rules on one run. Own success pays both reporters of a shared finding in full. Team reward pays a member with only wrong claims as much as the members who found the contradictions. Shapley credit splits the shared finding and charges each wrong claim to its author. Evaluated alone on the corpora and seeds of Table~\ref*{tab:credit}, the agent trained on its own success more than doubles its F1 by step 40, while its groups fall below the untrained group.

\input{tables/ablation}

\input{figures/mechanism}

\paragraph{Gating claims by confidence.} Rewards that make wrong claims cheap drift toward more claims at lower precision, as Proposition~\ref*{prop:threshold} predicts (Table~\ref{tab:reward}, Figure~\ref{fig:mechanism}b). A large penalty reverses the drift and also suppresses correct findings, because a 4B group rarely reaches a confidence threshold of one third.

\paragraph{Shares and flow.} In the distributed setting, where each member can open only its own share of the snippets, Shapley shares alone reproduce the gain of the full credit (Table~\ref{tab:stages}), so the measured benefit at 4B comes from the shares themselves. Its differences from the untrained group therefore differ from those of Table~\ref*{tab:credit}, which uses the shared corpus.

\section{Recursive Training on PlanBench, HiddenBench and Hanabi}
\label{app:chains}

\paragraph{Common design.} Every run starts from four Qwen3.5-9B agents served with vLLM~\citep{kwon2023vllm}. Each generation continues from the full-precision weights of its predecessor and trains mainly on its predecessor's latest accepted outputs.

\paragraph{PlanBench.} The four agents write Blocksworld plans in four calls each: a draft, a self-check, a revision after reading the other drafts, and an anonymous review of the others' plans. The group submits the final plan with the highest share of positive reviews. Training, selection and confirmation problems come from our own generator, which excludes every official PlanBench problem up to renaming of the blocks, and a plan simulator accepts each plan. Problems have mostly five blocks. Each call may use 8{,}192 tokens for thinking and answer together. The selection set has 300 problems and every confirmation slice 800. A candidate is released when $d\ge1.645\,\mathrm{SE}$ and $d>0$; training stops after four released generations.

\paragraph{HiddenBench.} The four agents debate for three rounds and commit an answer weighted by their stated stakes. Training and selection use hidden-profile tasks that the base model writes in the format of HiddenBench~\citep{li2025hiddenbench}; a turn is accepted when its committed answer matches the answer that at least three of four base-model members give with every fact and it passes the protocol's guards. The selection set has 200 hidden-profile and 100 control tasks run with two seeds, and every round draws new confirmation seeds. A candidate is released when $d$ exceeds 0.026 and its accuracy on control tasks stays above a floor anchored at the base model. T3 adds a matched single-agent self-training run under the same caps. T3k starts again from the base model with four changes: control tasks kept in the training rows, candidates at half and full step, decision tokens counted once, and answer values weighted like the call structure. It has no self-training counterpart. Both runs stop after eight rounds.

\paragraph{Hanabi.} Four agents play the cooperative card game~\citep{bard2020hanabi} with a fixed move vocabulary, one call per move. For every training deal the group plays four games and keeps the best one above the deal's mean as an elite game. Every round draws 300 fresh selection deals and 600 fresh confirmation deals, and a candidate is released when $d-\mathrm{SE}>0$ on the challenge score. A non-recursive control run branches after the first release and trains on games of a fixed generation. Training stops after six released generations. Table~\ref{tab:rounds} lists the release decision of every round in the three runs.

\input{tables/rounds}

\paragraph{Disclosures.} T3, T3k and Hanabi were designed after earlier runs on other benchmarks had been read, and the HiddenBench tasks used for reference in Figure~\ref{fig:bench-hiddenbench} had been read before. The design of T3k and the PlanBench domain and token cap were chosen after earlier measurements of the same lines. PlanBench lacks a single-agent self-training run, so its results leave collective and individual learning combined. On PlanBench the fraction of truncated calls falls across generations, and the tokens of every generation are reported beside its accuracy.

\section{Compute and Reproducibility}
\label{app:repro}

Inference for the ladder experiments ran on H200 GPUs of a cluster and on A800 GPUs of two eight-GPU nodes, with vLLM servers at tensor parallelism one or two. A cell of 448 records of four 27B members takes about 9 to 26 hours on one four-GPU group, depending on the protocol, and the 27B GPQA-Diamond batch of ten cells and pools took seven and a half hours on eight H200 GPUs. Supervised fine-tuning of a 27B generation takes about 75 minutes on eight A800 GPUs, and of a 9B generation about 17 to 27 minutes on four. The Shapley-credit runs used H200 GPUs: a 4B training step takes 10 to 14 minutes on four GPUs. Every run executes from a frozen snapshot of the code at a recorded commit. Benchmark data and model weights are pinned to revisions and verified against content hashes before use, and every pool is drawn on the platform and in the batch of the cells it grades, except where a table says otherwise: the descriptive cross-batch readings of the test-time self-written-prompt cells.

%% file: tables/methods.tex
\begin{table}[t]
\centering
\tsz
\setlength{\tabcolsep}{5pt}
\renewcommand{\arraystretch}{1.12}
\caption{\textbf{The protocols we place on the ladder}, grouped by what members exchange and by who decides the organization, with the family codes of Figure~\ref*{fig:ladder}a. Appendix~\ref{app:names} defines each short name.}
\label{tab:methods}
\begin{adjustbox}{max width=\linewidth}
\begin{tabular}{@{}llll@{}}
\toprule
\textbf{Family} & \textbf{Exchange} & \textbf{Organized by} & \textbf{Protocols} \\
\midrule
\refrow \famtag{Solo} Independent references & nothing & none & Single, Vote@B, Pass@B, Sequential-4 \\
\famtag{Argue} Argument exchange & answers, reasons & fixed rounds & BB, Debate-2R, Debate-3R, Self-Refine-2R, +H \\
\oursrow \famtag{Fact} Evidence exchange & facts as given & fixed protocol & Fact forwarding, Shared context, SC+Forwarding, Facts-only BB, Gated BB \\
\oursrow \famtag{Self} Self-built organization & reading budget, relays & the members & No transfer, self-organized, + planning round, budget cap, no relay \\
\oursrow \famtag{Notes} Written organization & self-written prompt & the model & From the rules, evolved, each with and without a budget cap \\
\refrow \famtag{Design} Designed organization & assigned roles & the designer & Hub, Tree, Aggregator \\
\oursrow \famtag{Train} Trained members & verified turns & training & G0, G1, G2, control \\
\bottomrule
\end{tabular}
\end{adjustbox}
\end{table}

%% file: tables/channels.tex
\begin{table}[tp]
\centering
\tsz
\setlength{\tabcolsep}{3.6pt}
\renewcommand{\arraystretch}{1.08}
\caption{\textbf{Accuracy, cost and coordination overhead.} \textbf{(a)} A protocol against a reference protocol on HiddenBench, both graded on the same pool: their budgets $n_B$, the protocol's margin $Y-T(n_B)$ over pass@$n$ and the reference's margin (Ref.), whose difference splits into the accuracy and cost terms, Margin $-$ Ref.\ $=$ Accuracy $+$ Cost (Eq.~\ref*{eq:channels}), the margin it would have at the reference's budget (At ref.), and what carries its level~3 (Via): accuracy when the margin at the reference's budget is confirmed above zero, accuracy and cost together when only the margin at its own budget is, and none when neither is. Debate-2R+H is the two-round debate with the hint, and Self-organized, 56 tasks, is the self-organized group under the free budget on the 56 tasks (accuracy 0.690, $n_B$ 23.4), a different cell from the capped group of Table~\ref*{tab:emergence}; rows 1--7 and 11 use the 65 tasks, the other rows the 56 tasks with four private facts; \#: point in Figure~\ref*{fig:channels}. \textbf{(b)} Token accounting on the 56 tasks: share of tokens spent waiting (Wait) and on all coordination (Coord.), and budget and margin now and with every waiting turn removed (Free wait). $\pm$: standard error.}
\label{tab:channels}
\label{tab:price}
\textbf{(a)}\par\smallskip
\begin{adjustbox}{max width=\linewidth}
\begin{tabular}{@{}rllccccccl@{}}
\toprule
\# & \textbf{Protocol} & \textbf{Reference} & Budgets & \textbf{Margin} & Ref. & Accuracy & Cost & At ref. & Via \\
\midrule
\tgroup{10}{\famtag{Fact} protocols} \\
1 & Fact forwarding, 9B & Debate-2R+H & 9.5/7.2 & $+0.160$\se{0.057} & $-0.345$ & $+0.550$\se{0.039} & $-0.045$\se{0.005} & $+0.205$\se{0.057} & \viaacc \\
2 & Fact forwarding, 27B & Debate-2R+H & 6.2/8.8 & $+0.310$\se{0.049} & $+0.079$ & $+0.177$\se{0.033} & $+0.054$\se{0.007} & $+0.256$\se{0.049} & \viaacc \\
3 & Fact forwarding, 35B & Debate-2R+H & 10.6/9.5 & $+0.181$\se{0.049} & $-0.176$ & $+0.373$\se{0.036} & $-0.016$\se{0.002} & $+0.197$\se{0.049} & \viaacc \\
4 & Shared context, 27B & Debate-2R+H & 7.9/8.8 & $+0.190$\se{0.048} & $+0.079$ & $+0.099$\se{0.025} & $+0.012$\se{0.002} & $+0.178$\se{0.048} & \viaacc \\
5 & SC+Forwarding, 27B & Shared context & 6.0/7.9 & $+0.328$\se{0.049} & $+0.190$ & $+0.091$\se{0.023} & $+0.047$\se{0.006} & $+0.282$\se{0.050} & \viaacc \\
\tgroup{10}{\famtag{Design} and \famtag{Notes} organizations, 27B} \\
6 & Designed hub & Self-organized & 11.3/21.3 & $+0.130$\se{0.056} & $-0.124$ & $+0.187$\se{0.034} & $+0.067$\se{0.014} & $+0.063$\se{0.055} & \viacost \\
7 & Designed hub & Self-organized + planning & 11.3/18.3 & $+0.130$\se{0.056} & $+0.037$ & $+0.041$\se{0.028} & $+0.052$\se{0.010} & $+0.078$\se{0.055} & \viacost \\
8 & Self-written prompt (rules) & Self-organized, 56 tasks & 9.2/23.4 & $+0.205$\se{0.055} & $-0.100$ & $+0.190$\se{0.036} & $+0.115$\se{0.019} & $+0.090$\se{0.050} & \viacost \\
9 & Self-written prompt (evolved) & Self-organized, 56 tasks & 9.7/23.4 & $+0.180$\se{0.055} & $-0.100$ & $+0.172$\se{0.037} & $+0.109$\se{0.018} & $+0.072$\se{0.050} & \viacost \\
10 & Designed tree & Self-organized + planning, cap & 10.8/44.4 & $+0.107$\se{0.051} & $-0.180$ & $+0.125$\se{0.035} & $+0.162$\se{0.030} & $-0.055$\se{0.041} & \viacost \\
\tgroup{10}{\famtag{Self} organizations and budget-capped \famtag{Notes}, 27B} \\
11 & Self-organized + planning & Self-organized & 18.3/21.3 & $+0.037$\se{0.047} & $-0.124$ & $+0.146$\se{0.033} & $+0.015$\se{0.004} & $+0.021$\se{0.047} & \vianone \\
12 & Cap + prompt (evolved) & Self-organized + planning, cap & 25.0/44.4 & $+0.043$\se{0.041} & $-0.180$ & $+0.163$\se{0.027} & $+0.060$\se{0.014} & $-0.017$\se{0.039} & \vianone \\
13 & Self-organized + planning, cap & No relay, cap & 44.4/37.0 & $-0.180$\se{0.046} & $-0.267$ & $+0.106$\se{0.025} & $-0.019$\se{0.005} & $-0.161$\se{0.045} & \vianone \\
\bottomrule
\end{tabular}
\end{adjustbox}
\par\medskip\textbf{(b)}\par\smallskip
\begin{adjustbox}{max width=\linewidth}
\begin{tabular}{@{}lcccccc@{}}
\toprule
& \multicolumn{2}{c}{\textbf{Share}} & \multicolumn{2}{c}{\textbf{Budget}} & \multicolumn{2}{c}{$\boldsymbol{\Delta}$\textbf{Pass@}$\boldsymbol{n}$} \\
\cmidrule(lr){2-3}\cmidrule(lr){4-5}\cmidrule(lr){6-7}
\textbf{Method} & Wait & Coord. & Now & Free wait & Now & Free wait \\
\midrule
Self-organized + planning, 56 tasks & 0.258 & 0.67 & 22.7 & 16.9 & $+0.066$ & $+0.100$\se{0.047} \\
Self-organized + planning, cap & 0.377 & 0.69 & 44.4 & 27.7 & $-0.180$ & $-0.131$\se{0.045} \\
Cap + prompt (rules) & 0.404 & 0.67 & 40.2 & 23.9 & $-0.115$ & $-0.061$\se{0.047} \\
Cap + prompt (evolved) & 0.332 & 0.64 & 25.0 & 16.7 & $+0.043$ & $+0.090$\se{0.044} \\
\refrow Designed tree, cap & 0.308 & 0.50 & 10.8 & 7.5 & $+0.107$ & $+0.155$\se{0.052} \\
\bottomrule
\end{tabular}
\end{adjustbox}
\end{table}

%% file: tables/collective.tex
\begin{table}[t]
\centering
\tsz
\setlength{\tabcolsep}{3.5pt}
\caption{\textbf{Gains at the group level.} \textbf{(a)} 9B on the 65 HiddenBench tasks: accuracy of one member given every fact (Solo), its change (Individual), and the change in the group's accuracy under split facts on the tasks that members of both generations solve alone (Team). \textbf{(b)} The 3-step debate, each generation graded against pass@$n$ of base-model members (base pass@$n$) or of its own members' independent runs (own pass@$n$); 9B over four training seeds per lineage; values without an error are point values. Each 9B row is a separate measurement on its own records, so the two gains need not add up to the G2 margin over its own pass@$n$.}
\label{tab:collective}
\begin{adjustbox}{max width=\linewidth}
\begin{tabular}{@{}lccc@{}}
\toprule
\textbf{(a) Step} & Solo & Individual & Team \\
\midrule
G0 $\to$ G1 & $0.927\to0.926$ & $-0.001$\se{0.009} & $+0.177$\se{0.038} \\
G1 $\to$ G2 & $0.926\to0.909$ & $-0.017$\se{0.008} & $+0.064$\se{0.028} \\
\midrule
\textbf{(b) Reading} & \multicolumn{3}{c}{Value} \\
\midrule
\multicolumn{4}{@{}l}{\textit{Qwen3.5-9B}} \\
G0 over base pass@$n$ & \multicolumn{3}{c}{$-0.253$\se{0.027}} \\
G2 over own pass@$n$ & \multicolumn{3}{c}{$+0.082$\se{0.025}} \\
G1 gain over G0, own pass@$n$ & \multicolumn{3}{c}{$+0.197$\se{0.026}} \\
G2 gain over G1, own pass@$n$ & \multicolumn{3}{c}{$+0.127$\se{0.017}} \\
Member alone, G2 minus G1 & \multicolumn{3}{c}{$+0.005$\se{0.003}} \\
\multicolumn{4}{@{}l}{\textit{Qwen3.8-27B}} \\
G1 over own pass@$n$ & \multicolumn{3}{c}{$+0.238$\se{0.032}} \\
G1 gain over G0, own pass@$n$ & \multicolumn{3}{c}{$+0.245$\se{0.027}} \\
G1 own pass@$n$ minus base pass@$n$ & \multicolumn{3}{c}{$-0.037$} \\
G2 over own pass@$n$ & \multicolumn{3}{c}{$+0.234$\se{0.032}} \\
\bottomrule
\end{tabular}
\end{adjustbox}
\end{table}

%% file: tables/prior.tex
\begin{table*}[t]
\centering
\tsz
\setlength{\tabcolsep}{4pt}
\renewcommand{\arraystretch}{1.06}
\caption{\textbf{Prior multi-agent methods against the equal-budget vote.} Committed-answer accuracy minus Vote@B in completion tokens with Qwen3.5-9B members, for multi-agent debate~\citep{du2024debate}, DeLM~\citep{mao2026delm} and AOrchestra~\citep{ruan2026aorchestra}, which sets its own group size, and for the blackboard with a reviewer that selects among its members' answers. $^{\downarrow}$ and $^{\uparrow}$: the 95\% interval lies below or above zero; $^{\dagger}$: evaluated on part of the subset.}
\label{tab:prior}
\begin{adjustbox}{max width=\textwidth}
\begin{tabular}[t]{@{}lcl@{}}
\toprule
\multicolumn{3}{@{}l}{\textbf{Multi-agent debate}} \\
Benchmark & $N$ & $\Delta$Vote \\
\midrule
GPQA-Diamond & 4 & $-0.005$\se{0.013} \\
GPQA-Diamond & 8 & $+0.008$\se{0.015} \\
GPQA-Diamond & 16 & $+0.022$\se{0.016} \\
BrowseComp-Plus & 4 & $-0.055^{\dagger}$\se{0.036} \\
BrowseComp-Plus & 8 & $-0.178^{\downarrow}$\se{0.032} \\
BrowseComp-Plus & 16 & $-0.163^{\downarrow}$\se{0.031} \\
LongBench v2 & 8 & $-0.016$\se{0.014} \\
MMLongBench-Doc & 16 & $-0.016$\se{0.017} \\
SWE-bench Verified & 8 & $-0.173^{\downarrow}$\se{0.025} \\
FATE-M & 4 & $-0.057$\se{0.029} \\
FATE-M & 8 & $-0.045$\se{0.028} \\
\bottomrule
\end{tabular}%
\hspace{1.4em}%
\begin{tabular}[t]{@{}lcl@{}}
\toprule
\multicolumn{3}{@{}l}{\textbf{DeLM}} \\
Benchmark & $N$ & $\Delta$Vote \\
\midrule
GPQA-Diamond & 4 & $+0.022$\se{0.014} \\
GPQA-Diamond & 8 & $+0.015$\se{0.011} \\
GPQA-Diamond & 16 & $+0.022$\se{0.015} \\
BrowseComp-Plus & 4 & $-0.002^{\dagger}$\se{0.044} \\
BrowseComp-Plus & 16 & $-0.243^{\downarrow}$\se{0.044} \\
MMLongBench-Doc & 16 & $-0.066^{\downarrow}$\se{0.022} \\
SWE-bench Verified & 8 & $-0.506^{\downarrow}$\se{0.034} \\
FATE-M & 4 & $-0.235^{\downarrow}$\se{0.035} \\
FATE-M & 8 & $-0.304^{\downarrow}$\se{0.041} \\
\bottomrule
\end{tabular}%
\hspace{1.4em}%
\begin{varwidth}[t]{\textwidth}
\begin{tabular}[t]{@{}ll@{}}
\toprule
\multicolumn{2}{@{}l}{\textbf{AOrchestra}} \\
Benchmark & $\Delta$Vote \\
\midrule
GPQA-Diamond & $+0.021$\se{0.020} \\
BrowseComp-Plus & $-0.075^{\dagger}$\se{0.046} \\
MMLongBench-Doc & $-0.112^{\downarrow}$\se{0.022} \\
SWE-bench Verified & $-0.360^{\downarrow}$\se{0.043} \\
\bottomrule
\end{tabular}\par\vspace{7pt}
\begin{tabular}[t]{@{}lclll@{}}
\toprule
\multicolumn{5}{@{}l}{\textbf{Blackboard with a reviewer (ours)}} \\
Benchmark & $N$ & BB & + 9B reviewer & + 27B reviewer \\
\midrule
GPQA-Diamond & 16 & $+0.013$\se{0.014} & $+0.010$\se{0.018} & $+0.050^{\uparrow}$\se{0.019} \\
BrowseComp-Plus & 16 & $+0.016$\se{0.026} & $+0.113^{\uparrow}$\se{0.036} & $+0.131^{\uparrow}$\se{0.038} \\
LongBench v2 & 8 & $-0.013$\se{0.012} & $+0.016$\se{0.015} & $+0.037^{\uparrow}$\se{0.015} \\
MMLongBench-Doc & 16 & $+0.013$\se{0.017} & $-0.025$\se{0.019} & $+0.044^{\uparrow}$\se{0.018} \\
\bottomrule
\end{tabular}
\end{varwidth}
\end{adjustbox}
\end{table*}

%% file: figures/example.tex
\begin{figure}[t]
\centering
\newcommand{\chipok}[1]{\tikz[baseline=-0.6ex]\node[draw=sage, fill=sage!15, rounded corners=1.5pt, inner sep=1.5pt, font=\scriptsize]{#1 \textcolor{sage}{\ding{51}}};}
\newcommand{\chipno}[1]{\tikz[baseline=-0.6ex]\node[draw=BrickRed!70, fill=BrickRed!8, rounded corners=1.5pt, inner sep=1.5pt, font=\scriptsize]{#1 \textcolor{BrickRed!80}{\ding{55}}};}
\footnotesize
\setlength{\tabcolsep}{6pt}
\renewcommand{\arraystretch}{1.25}
\begin{tabular}{@{}llccc@{}}
\toprule
\textbf{Member} & \textbf{Claims in its submission} & \textbf{Own success} $s_x(o_i)$ & \textbf{Team reward} $R$ & \textbf{Shapley credit} $\phi_i$ \\
\midrule
1 & \chipok{$p_1$}\; \chipok{$p_2$} & $+0.500$ & $+0.250$ & \ours{$+0.375$} \\
2 & \chipok{$p_1$}\; \chipno{$w_1$} & $+0.188$ & $+0.250$ & \ours{$+0.063$} \\
3 & \chipno{$w_2$}\; \chipno{$w_3$}\; \chipno{$w_4$} & $-0.188$ & $+0.250$ & \ours{$-0.188$} \\
\midrule
\multicolumn{2}{@{}l}{sum of credits (outcome of the group $R=0.250$)} & $+0.500$ & $+0.750$ & \ours{$+0.250$} \\
\bottomrule
\end{tabular}
\caption{\textbf{Credit for one run under three rules}, illustrated with the confidence-gated score (four true findings, $c=0.25$, group outcome $R=0.25$); Table~\ref*{tab:credit} trains on exact-pair F1. Own success pays members 1 and 2 in full for the same finding $p_1$; team reward pays member 3, who made only wrong claims, as much as the others; Shapley credit splits $p_1$, charges each wrong claim to its author, and sums to $R$.}
\label{fig:example}
\end{figure}

%% file: tables/ablation.tex
\begin{table}[t]
\centering
\tsz
\setlength{\tabcolsep}{4pt}
\renewcommand{\arraystretch}{1.08}
\caption{\textbf{What each reward pays for.} 4B training rollouts on the audit task: claims per run and their precision early and late in the run, measured on training rollouts rather than held-out corpora. $q^\ast$: the confidence above which one more claim raises the expected reward. $^\dagger$Held-out precision at steps 0 and 40, $N=4$.}
\label{tab:reward}
\begin{adjustbox}{max width=\linewidth}
\begin{tabular}{@{}llccccccc@{}}
\toprule
& & & \multicolumn{2}{c}{\textbf{Steps}} & \multicolumn{2}{c}{\textbf{Claims per run}} & \multicolumn{2}{c}{\textbf{Precision}} \\
\cmidrule(lr){4-5}\cmidrule(lr){6-7}\cmidrule(lr){8-9}
\textbf{Reward} & \textbf{Credit} & $q^\ast$ & early & late & early & late & early & late \\
\midrule
Snippet coverage & share + flow & $0$ & 1 & 41 & 2.1 & 17.7 & 0.33 & 0.12$^\dagger$ \\
Exact-pair recall & own success & $0$ & 1--5 & 27--31 & 0.8 & 36.0 & 0.34 & 0.05 \\
Exact-pair recall & share & $0$ & 1--3 & 8--10 & 3.0 & 4.1 & 0.30 & 0.24 \\
Exact-pair F1 & share + flow & F1$/2$ & 1--5 & 6--10 & 3.6 & 4.8 & 0.25 & 0.20 \\
\midrule
Gated claims, $c=0.5$ & share + flow & $1/3$ & 1--5 & 10--12 & 3.1 & 0.8--1.8 & 0.28 & 0.42--0.54 \\
\bottomrule
\end{tabular}
\end{adjustbox}
\end{table}

\begin{table}[t]
\centering
\tsz
\setlength{\tabcolsep}{5pt}
\renewcommand{\arraystretch}{1.08}
\caption{\textbf{Shapley shares and the full credit.} Qwen3.5-4B in the distributed setting, $N=4$; held-out differences from the untrained model in points with 95\% bootstrap intervals. \textbf{Bold}: the interval excludes zero.}
\label{tab:stages}
\label{tab:stages-ci}
\begin{adjustbox}{max width=\linewidth}
\begin{tabular}{@{}lccccc@{}}
\toprule
\textbf{Method} & \textbf{Step} & $\Delta$Recall & 95\% CI & $\Delta$F1 & 95\% CI \\
\midrule
\oursrow Shapley credit (share + flow) & 20 & \textbf{+1.9} & [+0.8, +3.1] & \textbf{+3.0} & [+1.2, +5.2] \\
Shapley credit w/o flow (share only) & 20 & \textbf{+2.2} & [+0.5, +4.1] & \textbf{+2.5} & [+0.2, +5.0] \\
\midrule
\oursrow Shapley credit (share + flow) & 40 & \textbf{+3.8} & [+1.2, +6.4] & \textbf{+4.6} & [+1.1, +8.4] \\
\bottomrule
\end{tabular}
\end{adjustbox}
\end{table}

%% file: figures/mechanism.tex
\begin{figure}[t]
\centering
\subfloat[Reporting thresholds (Proposition~\ref{prop:threshold}).\label{fig:threshold}]{\includegraphics[width=0.485\linewidth]{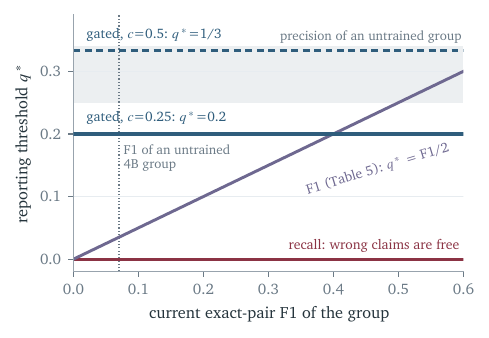}}\hfill
\subfloat[Training dynamics under five credit rules.\label{fig:dynamics}]{\includegraphics[width=0.485\linewidth]{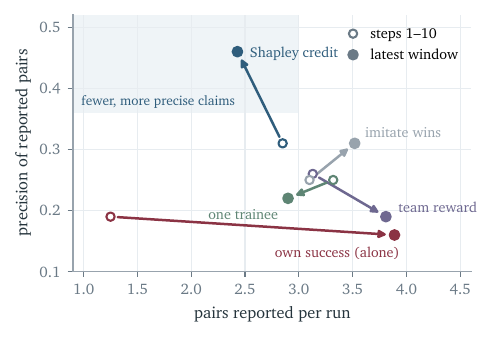}}
\caption{\textbf{Why the credit rule and the confidence threshold matter.} \textbf{(a)}~Smallest probability of being right at which one more claim raises the expected reward (Proposition~\ref{prop:threshold}). \textbf{(b)}~Claims per run and their precision during 4B training (open markers: first ten steps; filled: latest window).}
\label{fig:mechanism}
\end{figure}

%% file: tables/rounds.tex
\begin{table}[t]
\centering
\tsz
\setlength{\tabcolsep}{5pt}
\renewcommand{\arraystretch}{1.08}
\caption{\textbf{Release decisions of the three runs with at least four releases.} Each round compares the selected candidate with the incumbent on a fresh confirmation slice, paired by problem; $d$ is the paired difference with its standard error. PlanBench releases when $d\ge1.645\,\mathrm{SE}$, HiddenBench (T3k) when $d$ exceeds 0.026, and Hanabi when $d-\mathrm{SE}>0$. PlanBench and HiddenBench score the fraction of problems the group solves, and Hanabi the challenge score out of 25. Values not recorded in the run log are marked (\blank).}
\label{tab:rounds}
\begin{tabular}{@{}c cc c cc c cc@{}}
\toprule
& \multicolumn{2}{c}{\textbf{PlanBench}} & & \multicolumn{2}{c}{\textbf{HiddenBench (T3k)}} & & \multicolumn{2}{c}{\textbf{Hanabi}} \\
\cmidrule(lr){2-3}\cmidrule(lr){5-6}\cmidrule(l){8-9}
\textbf{Round} & $d$ (SE) & Decision & & $d$ (SE) & Decision & & $d$ (SE) & Decision \\
\midrule
1 & $+0.295$ (0.020) & G1 & & $+0.203$ (\blank) & G1 & & $+0.678$ (0.045) & G1 \\
2 & $+0.080$ (0.014) & G2 & & $+0.044$ (0.026) & G2 & & $+0.04$ (0.051) & flat \\
3 & $+0.024$ (0.012) & G3 & & $+0.001$ (0.022) & flat & & $-0.085$ (0.054) & flat \\
4 & $+0.021$ (0.010) & G4 & & $+0.044$ (\blank) & G3 & & $-0.012$ (0.052) & flat \\
5 & & & & $+0.065$ (0.020) & G4 & & \blank & flat \\
6 & & & & $+0.001$ (0.021) & flat & & $+2.363$ (0.075) & G2 \\
7 & & & & $+0.046$ (0.021) & G5 & & $+1.102$ (0.087) & G3 \\
8 & & & & $+0.015$ (0.016) & flat & & $+0.208$ (0.077) & G4 \\
9 & & & & & & & $+0.087$ (0.074) & G5 \\
10 & & & & & & & $+0.125$ (0.075) & G6 \\
\bottomrule
\end{tabular}
\end{table}

%% file: main.bbl
\begin{thebibliography}{77}
\providecommand{\natexlab}[1]{#1}
\providecommand{\url}[1]{\texttt{#1}}
\expandafter\ifx\csname urlstyle\endcsname\relax
  \providecommand{\doi}[1]{doi: #1}\else
  \providecommand{\doi}{doi: \begingroup \urlstyle{rm}\Url}\fi

\bibitem[Agarwal et~al.(2024)Agarwal, Vieillard, Zhou, Stanczyk, Ramos, Geist, and Bachem]{agarwal2024gkd}
Rishabh Agarwal, Nino Vieillard, Yongchao Zhou, Piotr Stanczyk, Sabela Ramos, Matthieu Geist, and Olivier Bachem.
\newblock {On-Policy Distillation of Language Models: Learning from Self-Generated Mistakes}.
\newblock In \emph{International Conference on Learning Representations (ICLR)}, 2024.

\bibitem[Aggarwal et~al.(2023)Aggarwal, Madaan, Yang, and Mausam]{aggarwal2023adaptive}
Pranjal Aggarwal, Aman Madaan, Yiming Yang, and Mausam.
\newblock {Let's Sample Step by Step: Adaptive-Consistency for Efficient Reasoning and Coding with LLMs}.
\newblock In \emph{Conference on Empirical Methods in Natural Language Processing (EMNLP)}, pages 12375--12396, 2023.

\bibitem[Bai et~al.(2025)Bai, Tu, Zhang, Peng, Wang, Lv, Cao, Xu, Hou, Dong, Tang, and Li]{bai2025longbenchv2}
Yushi Bai, Shangqing Tu, Jiajie Zhang, Hao Peng, Xiaozhi Wang, Xin Lv, Shulin Cao, Jiazheng Xu, Lei Hou, Yuxiao Dong, Jie Tang, and Juanzi Li.
\newblock {LongBench v2: Towards Deeper Understanding and Reasoning on Realistic Long-context Multitasks}.
\newblock In \emph{Annual Meeting of the Association for Computational Linguistics (ACL)}, pages 3639--3664, 2025.

\bibitem[Bard et~al.(2020)Bard, Foerster, Chandar, Burch, Lanctot, Song, Parisotto, Dumoulin, Moitra, Hughes, Dunning, Mourad, Larochelle, Bellemare, and Bowling]{bard2020hanabi}
Nolan Bard, Jakob~N. Foerster, Sarath Chandar, Neil Burch, Marc Lanctot, H.~Francis Song, Emilio Parisotto, Vincent Dumoulin, Subhodeep Moitra, Edward Hughes, Iain Dunning, Shibl Mourad, Hugo Larochelle, Marc~G. Bellemare, and Michael Bowling.
\newblock The {Hanabi} challenge: A new frontier for {AI} research.
\newblock \emph{Artificial Intelligence}, 280:\penalty0 103216, 2020.

\bibitem[Bikhchandani et~al.(1992)Bikhchandani, Hirshleifer, and Welch]{bikhchandani1992cascades}
Sushil Bikhchandani, David Hirshleifer, and Ivo Welch.
\newblock {A Theory of Fads, Fashion, Custom, and Cultural Change as Informational Cascades}.
\newblock \emph{Journal of Political Economy}, 100\penalty0 (5):\penalty0 992--1026, 1992.

\bibitem[Brown et~al.(2024)Brown, Juravsky, Ehrlich, Clark, Le, R{\'e}, and Mirhoseini]{brown2024monkeys}
Bradley Brown, Jordan Juravsky, Ryan Ehrlich, Ronald Clark, Quoc~V. Le, Christopher R{\'e}, and Azalia Mirhoseini.
\newblock {Large Language Monkeys: Scaling Inference Compute with Repeated Sampling}.
\newblock \emph{arXiv preprint arXiv:2407.21787}, 2024.

\bibitem[Cemri et~al.(2025)Cemri, Pan, Yang, Agrawal, Chopra, Tiwari, Keutzer, Parameswaran, Klein, Ramchandran, Zaharia, Gonzalez, and Stoica]{cemri2025fail}
Mert Cemri, Melissa~Z. Pan, Shuyi Yang, Lakshya~A. Agrawal, Bhavya Chopra, Rishabh Tiwari, Kurt Keutzer, Aditya Parameswaran, Dan Klein, Kannan Ramchandran, Matei Zaharia, Joseph~E. Gonzalez, and Ion Stoica.
\newblock {Why Do Multi-Agent LLM Systems Fail?}
\newblock In \emph{Advances in Neural Information Processing Systems (NeurIPS), Datasets and Benchmarks Track}, 2025.

\bibitem[Chen et~al.(2024)Chen, Davis, Hanin, Bailis, Stoica, Zaharia, and Zou]{chen2024morecalls}
Lingjiao Chen, Jared~Quincy Davis, Boris Hanin, Peter Bailis, Ion Stoica, Matei Zaharia, and James Zou.
\newblock {Are More LLM Calls All You Need? Towards the Scaling Properties of Compound AI Systems}.
\newblock In \emph{Advances in Neural Information Processing Systems (NeurIPS)}, 2024.

\bibitem[Chen et~al.(2025)Chen, Yuan, Qian, Yang, Liu, and Sun]{chen2024optima}
Weize Chen, Jiarui Yuan, Chen Qian, Cheng Yang, Zhiyuan Liu, and Maosong Sun.
\newblock {Optima: Optimizing Effectiveness and Efficiency for LLM-Based Multi-Agent System}.
\newblock In \emph{Findings of the Association for Computational Linguistics (ACL)}, 2025.

\bibitem[Chen et~al.(2026)Chen, Ma, Zhuang, Nie, Zou, Sharifymoghaddam, Liu, Green, Patel, Meng, Su, Li, Hong, Shi, Liu, Oyarhoseini, Thakur, Zhang, Gao, Chen, and Lin]{chen2025browsecompplus}
Zijian Chen, Xueguang Ma, Shengyao Zhuang, Ping Nie, Kai Zou, Sahel Sharifymoghaddam, Andrew Liu, Joshua Green, Kshama Patel, Ruoxi Meng, Mingyi Su, Yanxi Li, Haoran Hong, Xinyu Shi, Xuye Liu, Hosna Oyarhoseini, Nandan Thakur, Crystina Zhang, Luyu Gao, Wenhu Chen, and Jimmy Lin.
\newblock {BrowseComp-Plus: A Fair and Disentangled Evaluation Benchmark for Deep Search Agents}.
\newblock In \emph{Annual Meeting of the Association for Computational Linguistics (ACL)}, pages 22349--22370, 2026.

\bibitem[Choi et~al.(2025)Choi, Zhu, and Li]{choi2025debatevote}
Hyeong~Kyu Choi, Xiaojin Zhu, and Sharon Li.
\newblock {Debate or Vote: Which Yields Better Decisions in Multi-Agent Large Language Models?}
\newblock In \emph{Advances in Neural Information Processing Systems (NeurIPS)}, 2025.

\bibitem[Dang et~al.(2025)Dang, Qian, Luo, Fan, Xie, Shi, Chen, Yang, Che, Tian, Xiong, Han, Liu, and Sun]{dang2025puppeteer}
Yufan Dang, Chen Qian, Xueheng Luo, Jingru Fan, Zihao Xie, Ruijie Shi, Weize Chen, Cheng Yang, Xiaoyin Che, Ye~Tian, Xuantang Xiong, Lei Han, Zhiyuan Liu, and Maosong Sun.
\newblock {Multi-Agent Collaboration via Evolving Orchestration}.
\newblock In \emph{Advances in Neural Information Processing Systems (NeurIPS)}, 2025.

\bibitem[Dong et~al.(2023)Dong, Xiong, Goyal, Zhang, Chow, Pan, Diao, Zhang, Shum, and Zhang]{dong2023raft}
Hanze Dong, Wei Xiong, Deepanshu Goyal, Yihan Zhang, Winnie Chow, Rui Pan, Shizhe Diao, Jipeng Zhang, Kashun Shum, and Tong Zhang.
\newblock {RAFT: Reward rAnked FineTuning for Generative Foundation Model Alignment}.
\newblock \emph{Transactions on Machine Learning Research}, 2023.

\bibitem[Du et~al.(2024)Du, Li, Torralba, Tenenbaum, and Mordatch]{du2024debate}
Yilun Du, Shuang Li, Antonio Torralba, Joshua~B. Tenenbaum, and Igor Mordatch.
\newblock {Improving Factuality and Reasoning in Language Models through Multiagent Debate}.
\newblock In \emph{International Conference on Machine Learning (ICML)}, 2024.

\bibitem[Feng et~al.(2026{\natexlab{a}})Feng, Bai, Yang, Wang, Tan, Yan, Lei, Ding, Shi, Wang, Qi, Jiang, Wang, Huang, Fei, Yao, Du, Zettlemoyer, Choi, and Tsvetkov]{feng2026moco}
Shangbin Feng, Yuyang Bai, Ziyuan Yang, Yike Wang, Zhaoxuan Tan, Jiajie Yan, Zhenyu Lei, Wenxuan Ding, Weijia Shi, Haojin Wang, Zhenting Qi, Yuru Jiang, Heng Wang, Chengsong Huang, Yu~Fei, Jihan Yao, Yilun Du, Luke Zettlemoyer, Yejin Choi, and Yulia Tsvetkov.
\newblock {MoCo: A One-Stop Shop for Model Collaboration Research}.
\newblock \emph{arXiv preprint arXiv:2601.21257}, 2026{\natexlab{a}}.

\bibitem[Feng et~al.(2026{\natexlab{b}})Feng, Wang, Shi, Zettlemoyer, Choi, and Tsvetkov]{feng2026participation}
Shangbin Feng, Yike Wang, Weijia Shi, Luke Zettlemoyer, Yejin Choi, and Yulia Tsvetkov.
\newblock {Scaling Participation in Modular AI Systems}.
\newblock \emph{arXiv preprint arXiv:2606.07812}, 2026{\natexlab{b}}.

\bibitem[Fernando et~al.(2024)Fernando, Banarse, Michalewski, Osindero, and Rockt{\"a}schel]{fernando2024promptbreeder}
Chrisantha Fernando, Dylan~Sunil Banarse, Henryk Michalewski, Simon Osindero, and Tim Rockt{\"a}schel.
\newblock {Promptbreeder: Self-Referential Self-Improvement via Prompt Evolution}.
\newblock In \emph{International Conference on Machine Learning (ICML)}, pages 13481--13544, 2024.

\bibitem[Foerster et~al.(2018)Foerster, Farquhar, Afouras, Nardelli, and Whiteson]{foerster2018coma}
Jakob Foerster, Gregory Farquhar, Triantafyllos Afouras, Nantas Nardelli, and Shimon Whiteson.
\newblock {Counterfactual Multi-Agent Policy Gradients}.
\newblock In \emph{AAAI Conference on Artificial Intelligence}, pages 2974--2982, 2018.

\bibitem[{GLM-4.5 Team}(2025)]{glm2025glm45}
{GLM-4.5 Team}.
\newblock {GLM-4.5: Agentic, Reasoning, and Coding (ARC) Foundation Models}.
\newblock \emph{arXiv preprint arXiv:2508.06471}, 2025.

\bibitem[Gu et~al.(2024)Gu, Dong, Wei, and Huang]{gu2024minillm}
Yuxian Gu, Li~Dong, Furu Wei, and Minlie Huang.
\newblock {MiniLLM: Knowledge Distillation of Large Language Models}.
\newblock In \emph{International Conference on Learning Representations (ICLR)}, 2024.

\bibitem[Han and Zhang(2025)]{han2025blackboard}
Bochen Han and Songmao Zhang.
\newblock {Exploring Advanced LLM Multi-Agent Systems Based on Blackboard Architecture}.
\newblock \emph{arXiv preprint arXiv:2507.01701}, 2025.

\bibitem[Hayes-Roth(1985)]{hayesroth1985blackboard}
Barbara Hayes-Roth.
\newblock {A Blackboard Architecture for Control}.
\newblock \emph{Artificial Intelligence}, 26\penalty0 (3):\penalty0 251--321, 1985.

\bibitem[Hong et~al.(2025)Hong, Yin, Wang, Liu, Chen, Yu, Li, Ye, Xiao, Chen, Zhou, Yue, Yang, Guo, Liu, Wei, and Gu]{hong2025mgrpo}
Haoyang Hong, Jiajun Yin, Yuan Wang, Jingnan Liu, Zhe Chen, Ailing Yu, Ji~Li, Zhiling Ye, Hansong Xiao, Yefei Chen, Hualei Zhou, Yun Yue, Minghui Yang, Chunxiao Guo, Junwei Liu, Peng Wei, and Jinjie Gu.
\newblock {Multi-Agent Deep Research: Training Multi-Agent Systems with M-GRPO}.
\newblock \emph{arXiv preprint arXiv:2511.13288}, 2025.

\bibitem[Hong and Page(2004)]{hong2004diverse}
Lu~Hong and Scott~E. Page.
\newblock {Groups of Diverse Problem Solvers Can Outperform Groups of High-Ability Problem Solvers}.
\newblock \emph{Proceedings of the National Academy of Sciences}, 101\penalty0 (46):\penalty0 16385--16389, 2004.

\bibitem[Huang et~al.(2023)Huang, Gu, Hou, Wu, Wang, Yu, and Han]{huang2023selfimprove}
Jiaxin Huang, Shixiang Gu, Le~Hou, Yuexin Wu, Xuezhi Wang, Hongkun Yu, and Jiawei Han.
\newblock {Large Language Models Can Self-Improve}.
\newblock In \emph{Conference on Empirical Methods in Natural Language Processing (EMNLP)}, pages 1051--1068, 2023.

\bibitem[Jiang et~al.(2025)Jiang, He, Wang, Gao, Hu, Wang, Guan, Wu, Dai, Xiao, and Dong]{jiang2025fate}
Jiedong Jiang, Wanyi He, Yuefeng Wang, Guoxiong Gao, Yongle Hu, Jingting Wang, Nailin Guan, Peihao Wu, Chunbo Dai, Liang Xiao, and Bin Dong.
\newblock {FATE: A Formal Benchmark Series for Frontier Algebra of Multiple Difficulty Levels}.
\newblock \emph{arXiv preprint arXiv:2511.02872}, 2025.

\bibitem[Jimenez et~al.(2024)Jimenez, Yang, Wettig, Yao, Pei, Press, and Narasimhan]{jimenez2024swebench}
Carlos~E. Jimenez, John Yang, Alexander Wettig, Shunyu Yao, Kexin Pei, Ofir Press, and Karthik Narasimhan.
\newblock {SWE-bench: Can Language Models Resolve Real-World GitHub Issues?}
\newblock In \emph{International Conference on Learning Representations (ICLR)}, 2024.

\bibitem[Jin et~al.(2025)Jin, Zeng, Yue, Yoon, Arik, Wang, Zamani, and Han]{jin2025searchr1}
Bowen Jin, Hansi Zeng, Zhenrui Yue, Jinsung Yoon, Sercan Arik, Dong Wang, Hamed Zamani, and Jiawei Han.
\newblock {Search-R1: Training LLMs to Reason and Leverage Search Engines with Reinforcement Learning}.
\newblock In \emph{Conference on Language Modeling (COLM)}, 2025.

\bibitem[Kwon et~al.(2023)Kwon, Li, Zhuang, Sheng, Zheng, Yu, Gonzalez, Zhang, and Stoica]{kwon2023vllm}
Woosuk Kwon, Zhuohan Li, Siyuan Zhuang, Ying Sheng, Lianmin Zheng, Cody~Hao Yu, Joseph~E. Gonzalez, Hao Zhang, and Ion Stoica.
\newblock {Efficient Memory Management for Large Language Model Serving with PagedAttention}.
\newblock In \emph{ACM Symposium on Operating Systems Principles (SOSP)}, 2023.

\bibitem[Laughlin and Ellis(1986)]{laughlin1986demonstrability}
Patrick~R. Laughlin and Alan~L. Ellis.
\newblock {Demonstrability and Social Combination Processes on Mathematical Intellective Tasks}.
\newblock \emph{Journal of Experimental Social Psychology}, 22\penalty0 (3):\penalty0 177--189, 1986.

\bibitem[Li et~al.(2024{\natexlab{a}})Li, Zhang, Yu, Fu, and Ye]{li2024moreagents}
Junyou Li, Qin Zhang, Yangbin Yu, Qiang Fu, and Deheng Ye.
\newblock {More Agents Is All You Need}.
\newblock \emph{Transactions on Machine Learning Research}, 2024{\natexlab{a}}.

\bibitem[Li et~al.(2024{\natexlab{b}})Li, Yuan, Feng, Pan, Wang, Sun, Wang, and Li]{li2024esc}
Yiwei Li, Peiwen Yuan, Shaoxiong Feng, Boyuan Pan, Xinglin Wang, Bin Sun, Heda Wang, and Kan Li.
\newblock {Escape Sky-high Cost: Early-stopping Self-Consistency for Multi-step Reasoning}.
\newblock In \emph{International Conference on Learning Representations (ICLR)}, 2024{\natexlab{b}}.

\bibitem[Li et~al.(2026{\natexlab{a}})Li, Naito, and Shirado]{li2025hiddenbench}
Yuxuan Li, Aoi Naito, and Hirokazu Shirado.
\newblock {Systematic Failures in Collective Reasoning under Distributed Information in Multi-Agent LLMs}.
\newblock In \emph{International Conference on Machine Learning (ICML)}, 2026{\natexlab{a}}.

\bibitem[Li et~al.(2026{\natexlab{b}})Li, Jiang, Ma, Zhang, Nie, Zhang, Zou, Xie, Zhang, and Chen]{li2026openresearcher}
Zhuofeng Li, Dongfu Jiang, Xueguang Ma, Haoxiang Zhang, Ping Nie, Yuyu Zhang, Kai Zou, Jianwen Xie, Yu~Zhang, and Wenhu Chen.
\newblock {OpenResearcher: A Fully Open Pipeline for Long-Horizon Deep Research Trajectory Synthesis}.
\newblock \emph{arXiv preprint arXiv:2603.20278}, 2026{\natexlab{b}}.

\bibitem[Li et~al.(2026{\natexlab{c}})Li, Zhang, Han, Liu, Xie, Zhang, Choi, Zou, and Lu]{li2026agentflow}
Zhuofeng Li, Haoxiang Zhang, Seungju Han, Sheng Liu, Jianwen Xie, Yu~Zhang, Yejin Choi, James Zou, and Pan Lu.
\newblock {In-the-Flow Agentic System Optimization for Effective Planning and Tool Use}.
\newblock In \emph{International Conference on Learning Representations (ICLR)}, 2026{\natexlab{c}}.

\bibitem[Liu et~al.(2025)Liu, Chen, Liang, Lyu, and Amato]{liu2025magrpo}
Shuo Liu, Tianle Chen, Zeyu Liang, Xueguang Lyu, and Christopher Amato.
\newblock {LLM Collaboration With Multi-Agent Reinforcement Learning}.
\newblock \emph{arXiv preprint arXiv:2508.04652}, 2025.

\bibitem[Lorenz et~al.(2011)Lorenz, Rauhut, Schweitzer, and Helbing]{lorenz2011social}
Jan Lorenz, Heiko Rauhut, Frank Schweitzer, and Dirk Helbing.
\newblock {How Social Influence Can Undermine the Wisdom of Crowd Effect}.
\newblock \emph{Proceedings of the National Academy of Sciences}, 108\penalty0 (22):\penalty0 9020--9025, 2011.

\bibitem[Lorge and Solomon(1955)]{lorge1955twomodels}
Irving Lorge and Herbert Solomon.
\newblock {Two Models of Group Behavior in the Solution of Eureka-Type Problems}.
\newblock \emph{Psychometrika}, 20\penalty0 (2):\penalty0 139--148, 1955.

\bibitem[Ma et~al.(2024)Ma, Zang, Chen, Chen, Jiao, Li, Lu, Liu, Ma, Dong, Zhang, Pan, Jiang, Wang, Cao, and Sun]{ma2024mmlongbenchdoc}
Yubo Ma, Yuhang Zang, Liangyu Chen, Meiqi Chen, Yizhu Jiao, Xinze Li, Xinyuan Lu, Ziyu Liu, Yan Ma, Xiaoyi Dong, Pan Zhang, Liangming Pan, Yu-Gang Jiang, Jiaqi Wang, Yixin Cao, and Aixin Sun.
\newblock {MMLongBench-Doc: Benchmarking Long-context Document Understanding with Visualizations}.
\newblock In \emph{Advances in Neural Information Processing Systems (NeurIPS), Datasets and Benchmarks Track}, 2024.

\bibitem[Mao et~al.(2026)Mao, Gu, Chauhan, Zhang, Kang, and Mirhoseini]{mao2026delm}
Yuzhen Mao, Jerry Gu, Aadi Chauhan, Qizheng Zhang, Hangoo Kang, and Azalia Mirhoseini.
\newblock {Decentralized Multi-Agent Systems with Shared Context}.
\newblock \emph{arXiv preprint arXiv:2606.10662}, 2026.

\bibitem[Motwani et~al.(2025)Motwani, Smith, Das, Rafailov, Laptev, Torr, Pizzati, Clark, and de~Witt]{motwani2025malt}
Sumeet~Ramesh Motwani, Chandler Smith, Rocktim~Jyoti Das, Rafael Rafailov, Ivan Laptev, Philip H.~S. Torr, Fabio Pizzati, Ronald Clark, and Christian~Schroeder de~Witt.
\newblock {MALT: Improving Reasoning with Multi-Agent LLM Training}.
\newblock In \emph{Conference on Language Modeling (COLM)}, 2025.

\bibitem[{OpenAI}(2025)]{openai2025gptoss}
{OpenAI}.
\newblock {gpt-oss-120b \& gpt-oss-20b Model Card}.
\newblock \emph{arXiv preprint arXiv:2508.10925}, 2025.

\bibitem[Park et~al.(2025)Park, Han, Guo, Ozdaglar, Zhang, and Kim]{park2025maporl}
Chanwoo Park, Seungju Han, Xingzhi Guo, Asuman Ozdaglar, Kaiqing Zhang, and Joo-Kyung Kim.
\newblock {MAPoRL2: Multi-Agent Post-Co-Training for Collaborative Large Language Models with Reinforcement Learning}.
\newblock In \emph{Annual Meeting of the Association for Computational Linguistics (ACL)}, pages 30215--30248, 2025.

\bibitem[Qian et~al.(2025)Qian, Xie, Wang, Liu, Zhu, Xia, Dang, Du, Chen, Yang, Liu, and Sun]{qian2025macnet}
Chen Qian, Zihao Xie, YiFei Wang, Wei Liu, Kunlun Zhu, Hanchen Xia, Yufan Dang, Zhuoyun Du, Weize Chen, Cheng Yang, Zhiyuan Liu, and Maosong Sun.
\newblock {Scaling Large Language Model-based Multi-Agent Collaboration}.
\newblock In \emph{International Conference on Learning Representations (ICLR)}, 2025.

\bibitem[{Qwen Team}(2026)]{qwen2026qwen35}
{Qwen Team}.
\newblock {Qwen3.5} models.
\newblock \url{https://modelscope.cn/models/Qwen/Qwen3.5-9B}, 2026.
\newblock Model release, February 2026.

\bibitem[Rashid et~al.(2018)Rashid, Samvelyan, de~Witt, Farquhar, Foerster, and Whiteson]{rashid2018qmix}
Tabish Rashid, Mikayel Samvelyan, Christian~Schroeder de~Witt, Gregory Farquhar, Jakob Foerster, and Shimon Whiteson.
\newblock {QMIX: Monotonic Value Function Factorisation for Deep Multi-Agent Reinforcement Learning}.
\newblock In \emph{International Conference on Machine Learning (ICML)}, pages 4295--4304, 2018.

\bibitem[Rein et~al.(2024)Rein, Hou, Stickland, Petty, Pang, Dirani, Michael, and Bowman]{rein2024gpqa}
David Rein, Betty~Li Hou, Asa~Cooper Stickland, Jackson Petty, Richard~Yuanzhe Pang, Julien Dirani, Julian Michael, and Samuel~R. Bowman.
\newblock {GPQA: A Graduate-Level Google-Proof Q\&A Benchmark}.
\newblock In \emph{Conference on Language Modeling (COLM)}, 2024.

\bibitem[Ruan et~al.(2026)Ruan, Xu, Peng, Ren, Yu, Liang, Xiang, Chen, Liu, Wu, Luo, and Zhang]{ruan2026aorchestra}
Jianhao Ruan, Zhihao Xu, Yiran Peng, Fashen Ren, Zhaoyang Yu, Xinbing Liang, Jinyu Xiang, Yongru Chen, Bang Liu, Chenglin Wu, Yuyu Luo, and Jiayi Zhang.
\newblock {AOrchestra: Automating Sub-Agent Creation for Agentic Orchestration}.
\newblock In \emph{International Conference on Machine Learning (ICML)}, 2026.

\bibitem[Schmidhuber(2003)]{schmidhuber2003godel}
Juergen Schmidhuber.
\newblock {Goedel Machines: Self-Referential Universal Problem Solvers Making Provably Optimal Self-Improvements}.
\newblock \emph{arXiv preprint arXiv:cs/0309048}, 2003.

\bibitem[Schulman et~al.(2017)Schulman, Wolski, Dhariwal, Radford, and Klimov]{schulman2017ppo}
John Schulman, Filip Wolski, Prafulla Dhariwal, Alec Radford, and Oleg Klimov.
\newblock {Proximal Policy Optimization Algorithms}.
\newblock \emph{arXiv preprint arXiv:1707.06347}, 2017.

\bibitem[Shao et~al.(2024)Shao, Wang, Zhu, Xu, Song, Bi, Zhang, Zhang, Li, Wu, and Guo]{shao2024deepseekmath}
Zhihong Shao, Peiyi Wang, Qihao Zhu, Runxin Xu, Junxiao Song, Xiao Bi, Haowei Zhang, Mingchuan Zhang, Y.~K. Li, Y.~Wu, and Daya Guo.
\newblock {DeepSeekMath: Pushing the Limits of Mathematical Reasoning in Open Language Models}.
\newblock \emph{arXiv preprint arXiv:2402.03300}, 2024.

\bibitem[Shapley(1953)]{shapley1953value}
Lloyd~S. Shapley.
\newblock {A Value for n-Person Games}.
\newblock In Harold~W. Kuhn and Albert~W. Tucker, editors, \emph{Contributions to the Theory of Games, Volume II}, number~28 in Annals of Mathematics Studies, pages 307--317. Princeton University Press, 1953.

\bibitem[Shen et~al.(2026)Shen, Chang, Liang, Newman, Liang, Shao, Ma, Yin, Chuang, Shi, Choi, Zettlemoyer, Koh, Iyyer, Wu, and Sontag]{shen2026swecraft}
Shannon~Zejiang Shen, Yapei Chang, Qiyao Liang, Benjamin Newman, Jenny Liang, Rulin Shao, Ziqiao Ma, Junjie~Oscar Yin, Yung-Sung Chuang, Weijia Shi, Yejin Choi, Luke Zettlemoyer, Pang~Wei Koh, Mohit Iyyer, Tongshuang Wu, and David Sontag.
\newblock {World of SWECraft: Agent Collaboration at Scale via Simulated Software Worlds}.
\newblock \emph{arXiv preprint arXiv:2610.24061}, 2026.

\bibitem[Singh et~al.(2024)Singh, Co-Reyes, Agarwal, Anand, Patil, Garcia, Liu, Harrison, Lee, Xu, Parisi, Kumar, Alemi, Rizkowsky, Nova, Adlam, Bohnet, Elsayed, Sedghi, Mordatch, Simpson, Gur, Snoek, Pennington, Hron, Kenealy, Swersky, Mahajan, Culp, Xiao, Bileschi, Constant, Novak, Liu, Warkentin, Qian, Bansal, Dyer, Neyshabur, Sohl-Dickstein, and Fiedel]{singh2024beyond}
Avi Singh, John~D. Co-Reyes, Rishabh Agarwal, Ankesh Anand, Piyush Patil, Xavier Garcia, Peter~J. Liu, James Harrison, Jaehoon Lee, Kelvin Xu, Aaron Parisi, Abhishek Kumar, Alex Alemi, Alex Rizkowsky, Azade Nova, Ben Adlam, Bernd Bohnet, Gamaleldin Elsayed, Hanie Sedghi, Igor Mordatch, Isabelle Simpson, Izzeddin Gur, Jasper Snoek, Jeffrey Pennington, Jiri Hron, Kathleen Kenealy, Kevin Swersky, Kshiteej Mahajan, Laura Culp, Lechao Xiao, Maxwell~L. Bileschi, Noah Constant, Roman Novak, Rosanne Liu, Tris Warkentin, Yundi Qian, Yamini Bansal, Ethan Dyer, Behnam Neyshabur, Jascha Sohl-Dickstein, and Noah Fiedel.
\newblock {Beyond Human Data: Scaling Self-Training for Problem-Solving with Language Models}.
\newblock \emph{Transactions on Machine Learning Research}, 2024.

\bibitem[Smit et~al.(2024)Smit, Grinsztajn, Duckworth, Barrett, and Pretorius]{smit2024mad}
Andries~Petrus Smit, Nathan Grinsztajn, Paul Duckworth, Thomas~D. Barrett, and Arnu Pretorius.
\newblock {Should we be going MAD? A Look at Multi-Agent Debate Strategies for LLMs}.
\newblock In \emph{Proceedings of the 41st International Conference on Machine Learning (ICML)}, volume 235 of \emph{Proceedings of Machine Learning Research}, pages 45883--45905, 2024.

\bibitem[Stasser and Titus(1985)]{stasser1985pooling}
Garold Stasser and William Titus.
\newblock {Pooling of Unshared Information in Group Decision Making: Biased Information Sampling during Discussion}.
\newblock \emph{Journal of Personality and Social Psychology}, 48\penalty0 (6):\penalty0 1467--1478, 1985.

\bibitem[Steiner(1972)]{steiner1972group}
Ivan~D. Steiner.
\newblock \emph{{Group Process and Productivity}}.
\newblock Academic Press, New York, 1972.

\bibitem[Subramaniam et~al.(2025)Subramaniam, Du, Tenenbaum, Torralba, Li, and Mordatch]{subramaniam2025multiagentft}
Vighnesh Subramaniam, Yilun Du, Joshua~B. Tenenbaum, Antonio Torralba, Shuang Li, and Igor Mordatch.
\newblock {Multiagent Finetuning: Self Improvement with Diverse Reasoning Chains}.
\newblock In \emph{International Conference on Learning Representations (ICLR)}, 2025.

\bibitem[Sunehag et~al.(2018)Sunehag, Lever, Gruslys, Czarnecki, Zambaldi, Jaderberg, Lanctot, Sonnerat, Leibo, Tuyls, and Graepel]{sunehag2018vdn}
Peter Sunehag, Guy Lever, Audrunas Gruslys, Wojciech~Marian Czarnecki, Vinicius Zambaldi, Max Jaderberg, Marc Lanctot, Nicolas Sonnerat, Joel~Z. Leibo, Karl Tuyls, and Thore Graepel.
\newblock {Value-Decomposition Networks For Cooperative Multi-Agent Learning Based On Team Reward}.
\newblock In \emph{International Conference on Autonomous Agents and MultiAgent Systems (AAMAS)}, pages 2085--2087, 2018.

\bibitem[Valmeekam et~al.(2023)Valmeekam, Marquez, Olmo, Sreedharan, and Kambhampati]{valmeekam2023planbench}
Karthik Valmeekam, Matthew Marquez, Alberto Olmo, Sarath Sreedharan, and Subbarao Kambhampati.
\newblock {PlanBench}: An extensible benchmark for evaluating large language models on planning and reasoning about change.
\newblock In \emph{Advances in Neural Information Processing Systems, Datasets and Benchmarks Track}, 2023.

\bibitem[Wang et~al.(2020)Wang, Zhang, Kim, and Gu]{wang2020shapleyq}
Jianhong Wang, Yuan Zhang, Tae-Kyun Kim, and Yunjie Gu.
\newblock {Shapley Q-value: A Local Reward Approach to Solve Global Reward Games}.
\newblock In \emph{AAAI Conference on Artificial Intelligence}, 2020.

\bibitem[Wang et~al.(2025)Wang, Wang, Athiwaratkun, Zhang, and Zou]{wang2024moa}
Junlin Wang, Jue Wang, Ben Athiwaratkun, Ce~Zhang, and James Zou.
\newblock {Mixture-of-Agents Enhances Large Language Model Capabilities}.
\newblock In \emph{International Conference on Learning Representations (ICLR)}, 2025.

\bibitem[Wang et~al.(2024)Wang, Wang, Su, Tong, and Song]{wang2024rethinking}
Qineng Wang, Zihao Wang, Ying Su, Hanghang Tong, and Yangqiu Song.
\newblock {Rethinking the Bounds of LLM Reasoning: Are Multi-Agent Discussions the Key?}
\newblock In \emph{Proceedings of the 62nd Annual Meeting of the Association for Computational Linguistics (ACL)}, pages 6106--6131, 2024.

\bibitem[Wang et~al.(2023)Wang, Wei, Schuurmans, Le, Chi, Narang, Chowdhery, and Zhou]{wang2023selfconsistency}
Xuezhi Wang, Jason Wei, Dale Schuurmans, Quoc Le, Ed~Chi, Sharan Narang, Aakanksha Chowdhery, and Denny Zhou.
\newblock {Self-Consistency Improves Chain of Thought Reasoning in Language Models}.
\newblock In \emph{International Conference on Learning Representations (ICLR)}, 2023.

\bibitem[Wolpert and Tumer(2001)]{wolpert2001collectives}
David~H. Wolpert and Kagan Tumer.
\newblock {Optimal Payoff Functions for Members of Collectives}.
\newblock \emph{Advances in Complex Systems}, 4\penalty0 (2--3):\penalty0 265--279, 2001.

\bibitem[Woolley et~al.(2010)Woolley, Chabris, Pentland, Hashmi, and Malone]{woolley2010collective}
Anita~Williams Woolley, Christopher~F. Chabris, Alex Pentland, Nada Hashmi, and Thomas~W. Malone.
\newblock {Evidence for a Collective Intelligence Factor in the Performance of Human Groups}.
\newblock \emph{Science}, 330\penalty0 (6004):\penalty0 686--688, 2010.

\bibitem[Yang et~al.(2025)Yang, Li, Yang, Zhang, Hui, Zheng, Yu, Gao, Huang, Lv, Zheng, Liu, Zhou, Huang, Hu, Ge, Wei, Lin, Tang, Yang, Tu, Zhang, Yang, Yang, Zhou, Zhou, Lin, Dang, Bao, Yang, Yu, Deng, Li, Xue, Li, Zhang, Wang, Zhu, Men, Gao, Liu, Luo, Li, Tang, Yin, Ren, Wang, Zhang, Ren, Fan, Su, Zhang, Zhang, Wan, Liu, Wang, Cui, Zhang, Zhou, and Qiu]{yang2025qwen3}
An~Yang, Anfeng Li, Baosong Yang, Beichen Zhang, Binyuan Hui, Bo~Zheng, Bowen Yu, Chang Gao, Chengen Huang, Chenxu Lv, Chujie Zheng, Dayiheng Liu, Fan Zhou, Fei Huang, Feng Hu, Hao Ge, Haoran Wei, Huan Lin, Jialong Tang, Jian Yang, Jianhong Tu, Jianwei Zhang, Jianxin Yang, Jiaxi Yang, Jing Zhou, Jingren Zhou, Junyang Lin, Kai Dang, Keqin Bao, Kexin Yang, Le~Yu, Lianghao Deng, Mei Li, Mingfeng Xue, Mingze Li, Pei Zhang, Peng Wang, Qin Zhu, Rui Men, Ruize Gao, Shixuan Liu, Shuang Luo, Tianhao Li, Tianyi Tang, Wenbiao Yin, Xingzhang Ren, Xinyu Wang, Xinyu Zhang, Xuancheng Ren, Yang Fan, Yang Su, Yichang Zhang, Yinger Zhang, Yu~Wan, Yuqiong Liu, Zekun Wang, Zeyu Cui, Zhenru Zhang, Zhipeng Zhou, and Zihan Qiu.
\newblock {Qwen3 Technical Report}.
\newblock \emph{arXiv preprint arXiv:2505.09388}, 2025.

\bibitem[Yu et~al.(2025)Yu, Zhang, Zhu, Yuan, Zuo, Yue, Dai, Fan, Liu, Liu, Liu, Lin, Lin, Ma, Sheng, Tong, Zhang, Zhang, Zhang, Zhu, Zhu, Chen, Chen, Wang, Yu, Song, Wei, Zhou, Liu, Ma, Zhang, Yan, Qiao, Wu, and Wang]{yu2025dapo}
Qiying Yu, Zheng Zhang, Ruofei Zhu, Yufeng Yuan, Xiaochen Zuo, Yu~Yue, Weinan Dai, Tiantian Fan, Gaohong Liu, Lingjun Liu, Xin Liu, Haibin Lin, Zhiqi Lin, Bole Ma, Guangming Sheng, Yuxuan Tong, Chi Zhang, Mofan Zhang, Wang Zhang, Hang Zhu, Jinhua Zhu, Jiaze Chen, Jiangjie Chen, Chengyi Wang, Hongli Yu, Yuxuan Song, Xiangpeng Wei, Hao Zhou, Jingjing Liu, Wei-Ying Ma, Ya-Qin Zhang, Lin Yan, Mu~Qiao, Yonghui Wu, and Mingxuan Wang.
\newblock {DAPO: An Open-Source LLM Reinforcement Learning System at Scale}.
\newblock \emph{arXiv preprint arXiv:2503.14476}, 2025.

\bibitem[Yuan et~al.(2024)Yuan, Pang, Cho, Li, Sukhbaatar, Xu, and Weston]{yuan2024selfrewarding}
Weizhe Yuan, Richard~Yuanzhe Pang, Kyunghyun Cho, Xian Li, Sainbayar Sukhbaatar, Jing Xu, and Jason Weston.
\newblock {Self-Rewarding Language Models}.
\newblock In \emph{International Conference on Machine Learning (ICML)}, 2024.

\bibitem[Zelikman et~al.(2022)Zelikman, Wu, Mu, and Goodman]{zelikman2022star}
Eric Zelikman, Yuhuai Wu, Jesse Mu, and Noah~D. Goodman.
\newblock {STaR: Bootstrapping Reasoning With Reasoning}.
\newblock In \emph{Advances in Neural Information Processing Systems (NeurIPS)}, 2022.

\bibitem[Zelikman et~al.(2024)Zelikman, Lorch, Mackey, and Kalai]{zelikman2024stop}
Eric Zelikman, Eliana Lorch, Lester Mackey, and Adam~Tauman Kalai.
\newblock {Self-Taught Optimizer (STOP): Recursively Self-Improving Code Generation}.
\newblock In \emph{Conference on Language Modeling (COLM)}, 2024.

\bibitem[Zhang et~al.(2025{\natexlab{a}})Zhang, Geng, Yu, Yin, Zhang, Tan, Zhou, Li, Xue, Li, Zhou, Chen, Zhang, Fan, Wang, Huang, Piedrahita-Velez, Liao, Wang, Yang, Ji, Wang, Yan, Torr, and Bai]{zhang2025agenticrl}
Guibin Zhang, Hejia Geng, Xiaohang Yu, Zhenfei Yin, Zaibin Zhang, Zelin Tan, Heng Zhou, Zhongzhi Li, Xiangyuan Xue, Yijiang Li, Yifan Zhou, Yang Chen, Chen Zhang, Yutao Fan, Zihu Wang, Songtao Huang, Francisco Piedrahita-Velez, Yue Liao, Hongru Wang, Mengyue Yang, Heng Ji, Jun Wang, Shuicheng Yan, Philip Torr, and Lei Bai.
\newblock {The Landscape of Agentic Reinforcement Learning for LLMs: A Survey}.
\newblock \emph{Transactions on Machine Learning Research}, 2025{\natexlab{a}}.

\bibitem[Zhang et~al.(2025{\natexlab{b}})Zhang, Cui, Chen, Wang, Zhang, Wang, Wu, and Hu]{zhang2025stopmad}
Hangfan Zhang, Zhiyao Cui, Jianhao Chen, Xinrun Wang, Qiaosheng Zhang, Zhen Wang, Dinghao Wu, and Shuyue Hu.
\newblock {Stop Overvaluing Multi-Agent Debate -- We Must Rethink Evaluation and Embrace Model Heterogeneity}.
\newblock \emph{arXiv preprint arXiv:2502.08788}, 2025{\natexlab{b}}.

\bibitem[Zhang et~al.(2026)Zhang, Hu, Lu, Lange, and Clune]{zhang2026dgm}
Jenny Zhang, Shengran Hu, Cong Lu, Robert~Tjarko Lange, and Jeff Clune.
\newblock {Darwin G{\"o}del Machine: Open-Ended Evolution of Self-Improving Agents}.
\newblock In \emph{International Conference on Learning Representations (ICLR)}, 2026.

\bibitem[Zhao et~al.(2025)Zhao, Yuksekgonul, Wu, and Zou]{zhao2025sirius}
Wanjia Zhao, Mert Yuksekgonul, Shirley Wu, and James Zou.
\newblock {SiriuS: Self-improving Multi-agent Systems via Bootstrapped Reasoning}.
\newblock In \emph{Advances in Neural Information Processing Systems (NeurIPS)}, 2025.

\bibitem[Zhao et~al.(2026)Zhao, Hu, Wang, Hou, Zhang, Ding, and Zhao]{zhao2025strongermas}
Yujie Zhao, Lanxiang Hu, Yang Wang, Minmin Hou, Hao Zhang, Ke~Ding, and Jishen Zhao.
\newblock {Stronger-MAS: Multi-Agent Reinforcement Learning for Collaborative LLMs}.
\newblock In \emph{International Conference on Learning Representations (ICLR)}, 2026.

\bibitem[Zuo et~al.(2025)Zuo, Zhang, Sheng, Qu, Cui, Zhu, Li, Zhang, Long, Hua, Qi, Sun, Ma, Yuan, Ding, and Zhou]{zuo2025ttrl}
Yuxin Zuo, Kaiyan Zhang, Li~Sheng, Shang Qu, Ganqu Cui, Xuekai Zhu, Haozhan Li, Yuchen Zhang, Xinwei Long, Ermo Hua, Biqing Qi, Youbang Sun, Zhiyuan Ma, Lifan Yuan, Ning Ding, and Bowen Zhou.
\newblock {TTRL: Test-Time Reinforcement Learning}.
\newblock In \emph{Advances in Neural Information Processing Systems (NeurIPS)}, 2025.

\end{thebibliography}
